\documentclass[preprint,preprintnumbers,nofootinbib]{revtex4-1}

\usepackage[normalem]{ulem}
\usepackage{graphicx}
\usepackage{dcolumn}
\usepackage{bm}
\usepackage{amsmath}
\usepackage{amsfonts} 
\usepackage{latexsym}
\usepackage{bbm}
\usepackage{color}
\usepackage{amssymb}
\usepackage{amsthm}
\usepackage{epsfig}
\usepackage{physics}
\usepackage{hyperref}
\usepackage{comment}
\usepackage{ulem}
\usepackage{tikz-feynhand}
\hypersetup{
setpagesize=false,
 bookmarksnumbered=true,
 bookmarksopen=true,
 colorlinks=true,
 linkcolor=blue,
 citecolor=red,
}

\begin{document}

\title{One-loop analysis of dark matter constraints in a complex scalar extension of the Standard Model}

\author{Chikako Idegawa$^{1}$}
\email{idegawa@mail.sysu.edu.cn}

\author{Koichi Funakubo$^2$}
\email{funakubo@cc.saga-u.ac.jp}

\affiliation{$^1$ MOE Key Laboratory of TianQin Mission,
TianQin Research Center for Gravitational Physics \& School of Physics and Astronomy,
Frontiers Science Center for TianQin,
Gravitational Wave Research Center of CNSA,
Sun Yat-sen University (Zhuhai Campus), Zhuhai 519082, China}

\affiliation{$^2$ Department of Physics, Saga University, Saga 840-8502, Japan}

\bigskip

\date{\today}
\begin{abstract}

We investigate the complex singlet extension of the Standard Model, which provides a scalar dark matter candidate. We impose the constraints from the observed relic abundance together with the most stringent limits from direct detection experiments on the model. The counterterms are determined so that the vacuum and mass conditions are consistently satisfied at one-loop order, and one-loop corrections to the scalar interactions relevant to dark matter processes are incorporated into the analysis. As a result, the allowed parameter region shows clear deviations from the tree-level analysis, demonstrating the impact of quantum corrections on the phenomenology of dark matter.

\end{abstract}

\maketitle

%%%%%%%%%%%%%%%%%%%%%%%%%%%%%%%%%%%%%%%%%%%%%%%%
%		          Introduction
%%%%%%%%%%%%%%%%%%%%%%%%%%%%%%%%%%%%%%%%%%%%%%%%

\section{Introduction}\label{sec:intro}

The discovery of the Higgs boson at the LHC~\cite{ATLAS:2012yve,CMS:2012qbp} marked a significant milestone in particle physics, completing the Standard Model (SM) in its minimal form. Nevertheless, compelling evidence from astrophysical and cosmological observations, such as the precise measurements of the cosmic microwave background, suggests the existence of dark matter (DM), indicating the necessity for new physics beyond the SM (BSM). One of the well-motivated DM candidates is the weakly interacting massive particle (WIMP), which arises naturally in many BSM scenarios. Extensive searches for BSM physics have been carried out at the LHC~\cite{ATLAS:2019nkf,CMS:2020gsy} and in DM direct detection experiments~\cite{LZ:2024zvo}, though no conclusive signals have been observed, thereby imposing stringent constraints on the allowed parameter space.

Among the simplest extensions of the SM that provide a viable WIMP candidate is the complex singlet extension of the SM (CxSM)~\cite{Barger:2008jx,Barger:2010yn,Gonderinger:2012rd,Coimbra:2013qq,Jiang:2015cwa,Chiang:2017nmu,Cheng:2018ajh,Grzadkowski:2018nbc,Chen:2019ebq,Cho:2021itv,Cho:2022our,Egle:2022wmq,Cho:2022zfg,Idegawa:2023bkh,Cho:2023oad}, in which a complex scalar singlet $S$ is added to the SM field content. The imaginary component of the singlet field, $\chi$, acts as a scalar DM candidate. In contrast, the real component mixes with the SM Higgs field, resulting in two mass eigenstates, $h_1$ and $h_2$. Two main approaches have been considered to satisfy the increasingly stringent bounds from the DM direct detection experiments. The first is to suppress the coupling between DM and the SM particles. For example, by taking a small scalar mixing angle $\alpha$. The second is to utilize built-in cancellation mechanisms within the model. In the CxSM, the degenerate scalar scenario provides a built-in suppression mechanism, where the two Higgs mass eigenstates are nearly equal in mass, i.e., $m_{h_1}\simeq m_{h_2}$. This near-degeneracy leads to an effective cancellation in their contributions to DM-nucleon interactions, thereby suppressing the spin-independent scattering cross section~\cite{Abe:2021nih,Cho:2023hek}. Furthermore, the DM relic density provides a crucial constraint on the model. The observed value of the relic abundance is $\Omega_{\mathrm{DM}} h^2 = 0.1200 \pm 0.0012$~\cite{Planck:2018vyg}, and any viable model must reproduce this within the observational uncertainty.

In addition to the DM-related constraints discussed above, we also impose theoretical consistency conditions , such as the lower-boundedness of the scalar potential, perturbative unitarity, and vacuum stability. Particular attention is paid to the impact of the one-loop corrections to the effective potential together with the counterterms, and how this affects the shape and size of the allowed regions. 
When the couplings related to $\chi$ are strong as in the case where the scalar masses originate mainly from the VEVs, the importance of the loop-level corrections becomes particularly important.

This paper is organized as follows. In Sec.\ref{sec:model}, we introduce the CxSM and describe the theoretical constraints at the tree and the one-loop levels. We also study the condition that $\chi$ remains a viable DM candidate by not acquiring a VEV at the tree level. Sec.~\ref{sec:DM} discusses the constraints from the DM direct detection experiments and introduces the method used to evaluate the DM relic abundance. The numerical analysis, which aims to identify viable regions of the parameter space consistent with all these constraints, is presented in Sec.\ref{sec:num}. Finally, we summarize our findings and comment on future studies in Sec.~\ref{sec:sum}.

%%%%%%%%%%%%%%%%%%%%%%%%%%%%%%%%%%%%%%%%%%%%%%%%
%		        Model
%%%%%%%%%%%%%%%%%%%%%%%%%%%%%%%%%%%%%%%%%%%%%%%%

\section{The Model}\label{sec:model} 
The CxSM extends the SM by incorporating a complex scalar field that is a singlet and neutral under the SM gauge group~\cite{Barger:2008jx}. In this study, we consider the following scalar potential:
\begin{align}
V_{0}(H, S)=\frac{m^{2}}{2} H^{\dagger} H + \frac{\lambda}{4} \left(H^{\dagger} H\right)^{2} + \frac{\delta_{2}}{2} H^{\dagger} H |S|^{2} + \frac{b_{2}}{2} |S|^{2} + \frac{d_{2}}{4} |S|^{4} + \left(a_{1} S + \frac{b_{1}}{4} S^{2} + \text{H.c.} \right),
\label{tree}
\end{align}
Where all parameters, including $a_1$ and $b_1$, which can in general be complex, are taken to be real. First, when $a_1 = 0$ and $b_1$ is real and nonzero, the global $U(1)$ symmetry of the singlet field $S$ is explicitly broken down to a $\mathbb{Z}_2 \times \mathbb{Z}_2$ symmetry. This residual symmetry corresponds to independent sign flips of $s = \mathrm{Re}S$ and $\chi = \mathrm{Im}S$.
The linear term proportional to $a_1$ explicitly breaks the $\mathbb{Z}_2$ symmetry acting on $s$, so when $a_1 \ne 0$, a nonzero vacuum expectation value (VEV) for $\mathrm{Re}S$ is generally expected. This explicit breaking of the $\mathbb{Z}_2$ symmetry prevents the formation of domain walls. On the other hand, the $\mathbb{Z}_2$ symmetry acting on $\chi$ remains intact, which guarantees the stability of the scalar particle associated with $\chi$ and allows it to serve as a DM candidate.

The lower-boundedness of the potential requires
\begin{align}
\lambda>0, \quad d_2>0, \quad\left(\left[\delta_2 \geq 0\right], \quad \text { or } \quad\left[\delta_2<0 \text { and } \lambda d_2 \geq \delta_2^2\right]\right).
\label{BFB}
\end{align}
These conditions ensure that the scalar potential remains positive in all directions in field space at large field values.
In addition to the lower-boundedness condition, we impose the tree-level perturbative unitarity constraints.
Requiring the zeroth partial-wave amplitudes for scalar two-body scattering to satisfy
\begin{align}
\left| \mathrm{Re}\, a_0 \right| \leq \frac{1}{2},
\end{align}
leads to
\begin{align}
|\lambda| \leq \frac{16\pi}{3}, \qquad
|d_2| \leq \frac{16\pi}{3}, \qquad
|\delta_2| \leq 16\pi .
\label{unitarity}
\end{align}

%%%

%%%
\subsection{Tree level}\label{subsec:tree}
We denote the classical background fields of the Higgs doublet and singlet as
\begin{align}
\langle H \rangle &=\left(\begin{array}{c}
0 \\
\frac{1}{\sqrt{2}}\varphi
\end{array}\right), \label{Hcomponent}\\
\langle S \rangle &=\frac{1}{\sqrt{2}}\left(\varphi_{S}+i \varphi_\chi\right),
\label{Scomponent}
\end{align}
where $\langle\cdots\rangle$ is defined as taking all fluctuation fields to zero. 
The first derivatives of $V_0$ for $\varphi$,$\varphi_{S}$ and $\varphi_{\chi}$ are respectively given by
\begin{align}
\frac{\partial V_0}{\partial \varphi} &=\frac{\varphi}{2}\left[m^2+\frac{\lambda}{2} \varphi^2+\frac{\delta_2}{2}\left(\varphi_S^2+\varphi_\chi^2\right)\right]=0, \label{tadpole1}\\
\frac{\partial V_0}{\partial \varphi_S} &=\sqrt{2} a_1+\frac{\varphi_S}{2}\left[b_2+b_1+\frac{d_2}{2}\left(\varphi_S^2+\varphi_\chi^2\right)+\frac{\delta_2}{2} \varphi^2\right]=0,  \label{tadpole2}\\
\frac{\partial V_0}{\partial \varphi_\chi} &=\frac{\varphi_\chi}{2}\left[b_2-b_1+\frac{d_2}{2}\left(\varphi_S^2+\varphi_\chi^2\right)+\frac{\delta_2}{2} \varphi^2\right]=0, \label{tadpole3}
\end{align}
Note that a nonzero singlet VEV, $\varphi_S\neq 0$, is enforced by the presence of the linear term, i.e., $a_1 \neq 0$.
In this study, we focus on the case $\varphi_\chi= 0$, i,e., our vacuum is located at $(\varphi,\varphi_S,\varphi_\chi) = (v, v_S, 0)$, where $v\simeq246$ GeV. This choice ensures the stability of the $\chi$ particle by preserving the $\mathbb{Z}_2$ symmetry acting on the imaginary part of the singlet field, and at the same time it avoids the domain wall problem associated with the spontaneous breaking of that discrete symmetry.

%%%

%%%
In this vacuum, the two scalar fields are parametrized as
\begin{align}
H=\binom{0}{\frac{1}{\sqrt{2}}\left(v+h\right)}, \quad S=\frac{1}{\sqrt{2}}\left(v_S+s+i \chi\right).
\end{align}
The mass matrix for the CP-even scalar states $(h, s)$ is given by
\begin{align}
\mathcal{M}_S^2=\left(\begin{array}{cc}
\lambda v^2 / 2 & \delta_2 v v_S / 2 \\
\delta_2 v v_S / 2 & \Lambda^2 
\end{array}\right),\quad\Lambda^2 \equiv \frac{d_2}{2} v_S^2-\sqrt{2}\frac{a_1}{v_S}. \label{MM}
\end{align}
This matrix is diagonalized by an orthogonal transformation $O(\alpha)$ as
\begin{align}
O(\alpha)^\top \mathcal{M}_S^2 O(\alpha)=\left(\begin{array}{cc}
m_{h_1}^2 & 0 \\
0 & m_{h_2}^2
\end{array}\right), \quad O(\alpha)=\left(\begin{array}{cc}
\cos \alpha & -\sin \alpha \\
\sin \alpha & \cos \alpha
\end{array}\right), 
\label{masseigenstate}
\end{align}
where $\alpha$ is the mixing angle between the interaction eigenstates.
The mass eigenstates $(h_1, h_2)$ are related to the gauge eigenstates by
\begin{align}
\left(\begin{array}{l}
h \\
s
\end{array}\right)=\left(\begin{array}{cc}
\cos \alpha & \sin \alpha \\
-\sin \alpha & \cos \alpha
\end{array}\right)\left(\begin{array}{l}
h_1 \\
h_2
\end{array}\right).
\end{align}
We emphasize that the limit $\alpha \to 0$ corresponds to the SM-like scenario, where $h_1 \to h$ and $h_2 \to s$.
The mass eigenvalues are given by
\begin{align}
m_{h_1, h_2}^2 &=\frac{1}{2}\left(\frac{\lambda}{2} v^2+\Lambda^2 \mp \frac{\frac{\lambda}{2} v^2-\Lambda^2}{\cos 2 \alpha}\right) \\
&=\frac{1}{2}\left(\frac{\lambda}{2} v^2+\Lambda^2 \mp \sqrt{\left(\frac{\lambda}{2} v^2-\Lambda^2\right)^2+4\left(\frac{\delta_2}{2} v v_S\right)^2}\right), \label{eigenvalue} \\
\cos 2 \alpha&=\frac{\frac{\lambda}{2} v^2-\Lambda^2}{m_{h_1}^2-m_{h_2}^2}
\end{align}
We fix $h_1$ as the observed Higgs boson with mass $m_{h_1} = 125.2$~GeV, in accordance with LHC measurements. 
The mass of $\chi$ is determined by the soft breaking parameters $a_1$ and $b_1$ as follows: 
\begin{align}
m_\chi^2
 &=
 \frac{b_2}{2}-\frac{b_1}{2}+\frac{\delta_2}{4} v^2+\frac{d_2}{4} v_S^2 \nonumber \\
 &=
 -\frac{\sqrt{2} a_1}{v_S}-b_1, \label{DMmass}
\end{align}
where the tadpole condition \eqref{tadpole2} is used in the second equality.
%%%

%%%
For later convenience, we summarize the relationship between the input parameters and the parameters appearing in the Lagrangian.
In the following analysis, we take the set $\{v, v_S, m_{h_1}, m_{h_2}, \alpha, m_\chi, a_1\}$ as input parameters.
Correspondingly, the Lagrangian parameters $\{m^2, b_2, \lambda, d_2, \delta_2, b_1\}$ can be expressed as functions of these inputs.
Among them, the parameters $m^2$ and $b_2$ are determined via the tadpole conditions given in Eqs.(\ref{tadpole1}) and (\ref{tadpole2}), and are thus eliminated in terms of the independent parameters as
\begin{align}
m^2 &=-\frac{\lambda}{2} v^2-\frac{\delta_2}{2}v_S^2  \label{msquare}\\
b_2 &=-\frac{\delta_2}{2} v^2-\frac{d_2}{2}v_S^2-\sqrt{2}\frac{a_1}{v_S}-b_1. \label{b2}
\end{align}
The remaining four parameters among the six Lagrangian parameters are given in terms of the input parameters as follows:
\begin{align}
\lambda&=\frac{2}{v^2}\left(m_{h_1}^2\cos^2\alpha+m_{h_2}^2\sin^2\alpha\right),\label{lambda}\\
\delta_2&=\frac{1}{v v_S}\left(m_{h_1}^2-m_{h_2}^2\right)\sin{2\alpha}\label{del2CPC1},\\
d_2&=2\left(\frac{m_{h_1}}{v_S}\right)^2\sin^2\alpha+2\left(\frac{m_{h_2}}{v_S}\right)^2\cos^2\alpha+2\sqrt{2}\frac{a_1}{v_S^3}\label{d2},\\
b_1 &=-m_{\chi}^2-\frac{\sqrt{2}}{v_S}a_1. 
\end{align}

Lastly, we consider the condition for vacuum stability along the $\varphi_\chi$-direction at the tree level.
The origin $(\varphi,\varphi_S,\varphi_\chi)=(0,0,0)$ is not a stationary point of $V_0$, since
$\partial V_0/\partial \varphi_S|_{(0,0,0)}=\sqrt{2}\,a_1\neq 0$.
We ensure that the $\mathbb{Z}_2$ symmetry associated with $\mathrm{Im}\,S$ remains unbroken by working on the $\varphi_\chi=0$ branch and solving the tree-level stationarity equations.
On $\varphi_\chi=0$, the tadpole conditions \eqref{tadpole1} and \eqref{tadpole2} reduce to
\begin{align}
m^2 + \frac{\lambda}{2}\,v^2 + \frac{\delta_2}{2}\,v_S^2 &= 0, \qquad
d_2\,v_S^3 + \bigl[2(b_2+b_1)+\delta_2\,v^2\bigr]v_S + 4\sqrt{2}\,a_1 = 0,
\label{eq:tadpoles_v_vS}
\end{align}
which determine the desired vacuum $(\varphi,\varphi_S,\varphi_\chi)=(v, v_{S}, 0)$.

Potential stationary points can be written explicitly.
Besides the desired vacuum $(v, v_S, 0)$, three additional branches may appear:\\
(i) Point $(\varphi, \varphi_S, \varphi_\chi) = (0,\, v_{S1},\,0)$:
\begin{align}
d_2 v_{S 1}^3+2\left(b_2+b_1\right) v_{S 1}+4 \sqrt{2} a_1=0.
\label{tadpolefalse1}
\end{align}
(ii) Point $(\varphi, \varphi_S, \varphi_\chi) = (0,\, v_{S2},\, v_{\chi2})$:
\begin{align}
& b_2+b_1+\frac{d_2}{2}\left(v_{S 2}^2+v_{\chi 2}^2\right)+2 \sqrt{2} \frac{a_1}{v_{S 2}}=0, \nonumber \\
& b_2-b_1+\frac{d_2}{2}\left(v_{S 2}^2+v_{\chi 2}^2\right)=0.
\label{tadpolefalse2}
\end{align}
(iii) Point $(\varphi, \varphi_S, \varphi_\chi) = (v_3,\, v_{S3},\, v_{\chi3})$:
\begin{align}
& m^2+\frac{\lambda}{2} v_3^2+\frac{\delta_2}{2}\left(v_{s 3}^2+v_{\chi 3}^2\right)=0, \nonumber \\
& b_2+b_1+\frac{d_2}{2}\left(v_{S 3}^2+v_{\chi 3}^2\right)+\frac{\delta_2}{2} v_3^2+2 \sqrt{2} \frac{a_1}{v_{S 3}}=0,\nonumber \\
& b_2-b_1+\frac{d_2}{2}\left(v_{S 3}^2+v_{\chi 3}^2\right)+\frac{\delta_2}{2} v_3^2=0.
\label{tadpolefalse3}
\end{align}
If either Point (ii) or Point (iii) turns out to be the global minimum of the potential, the discrete symmetry that stabilizes the $\chi$ particle would be spontaneously broken, thereby invalidating our scenario. 
The tadpole conditions \eqref{tadpolefalse2} and \eqref{tadpolefalse3} can be solved for the coordinates of Points (ii) and (iii) as\\
(ii) Point $(\varphi, \varphi_S, \varphi_\chi) = (0,\, v_{S2},\, v_{\chi2})$:
\begin{align}
v_{S2} = -\frac{\sqrt{2}\,a_1}{b_1}, \qquad
v_{\chi2}^2 = -\frac{2(b_2-b_1)}{d_2} - \frac{2a_1^2}{b_1^2}.
\label{eq:branch_v0}
\end{align}
(iii) Point $(\varphi, \varphi_S, \varphi_\chi) = (v_3,\, v_{S3},\, v_{\chi3})$:
\begin{align}
v_3^{\,2} = -\frac{2\bigl[d_2 m^2 - \delta_2(b_2-b_1)\bigr]}{\lambda d_2 - \delta_2^{\,2}}, \qquad
v_{S3} = -\frac{\sqrt{2}\,a_1}{b_1}, \qquad
v_{\chi3}^2 = \frac{2\delta_2 m^2}{\lambda d_2 - \delta_2^{\,2}}
              - \frac{2\lambda(b_2-b_1)}{\lambda d_2 - \delta_2^{\,2}}
              - \frac{2a_1^2}{b_1^2}.
\label{eq:branch_vneq0}
\end{align}
These stationary points exist only if the right-hand sides of $v_{\chi2}^2$ in Eq.~\eqref{eq:branch_v0} and $v_3^2$, $v_{\chi3}^2$ in Eq.~\eqref{eq:branch_vneq0} are non-negative.
For instance, when $(b_2-b_1)/d_2>0$, $v_{\chi2}^2$ is negative and the $(0,v_{S2},v_{\chi2})$ branch is absent. Thus, when the square of these VEVs becomes negative, stationary points with $\varphi_\chi\neq 0$ do not appear and the vacuum preserving the $\mathbb{Z}_2$ symmetry. Even if the solution exists, as long as the desired vacuum $V_0(v,v_S,0)$ is the global minimum, i.e., $V_0(v,v_S,0)<V_0(0,\, v_{S2},\, v_{\chi2})$ and/or $V_0(v,v_S,0)<V_0(v_3,\, v_{S3},\, v_{\chi3})$, one can conclude that the discrete symmetry is not broken.

%%%
\subsection{One-loop level}\label{subsec:loop}
The one-loop effective potential is given by
\begin{align}
V_{\mathrm{eff}}\left(\varphi, \varphi_S, \varphi_\chi\right)=V_0\left(\varphi, \varphi_S, \varphi_\chi\right)+V_{\mathrm{CW}}\left(\varphi, \varphi_S, \varphi_\chi\right).
\end{align}
$V_0$ includes not only the tree-level potential but also the one-loop level counterterms, i.e.,
\begin{align}
V_0&=\sqrt{2}\left( a_1+a_1^{(1)}\right) \varphi_s+\frac{m^2+m^{(1) 2}}{4} \varphi^2+\frac{b_2+b_1+b_2^{(1)}+b_1^{(1)}}{4} \varphi_s^2 + \frac{b_2-b_1+b_2^{(1)}-b_1^{(1)}}{4} \varphi_\chi^2 \nonumber \\
& +\frac{\lambda+\lambda^{(1)}}{16} \varphi^4+\frac{d_2+d_2^{(1)}}{16} (\varphi_S^2+\varphi_\chi^2)^2+\frac{\delta_2+\delta_2^{(1)}}{8} \varphi^2 (\varphi_S^2+\varphi_\chi^2),
\end{align}
where the quantities labeled with the superscript (1) represent the coefficients of the finite counterterms.
The one-loop correction is
\begin{align}
V_{\mathrm{CW}}\left(\varphi, \varphi_s, \varphi_\chi\right)=\sum_i \frac{n_i \bar{m}_i^4}{64 \pi^2}\left(\log \frac{\bar{m}_i^2}{\bar{\mu}^2}-\gamma_i\right).
\end{align}
All parameters are taken at the tree level. In this context, $\bar{m}_i^2$ represents the field-dependent mass squared, and $n_i$ denotes the number of degrees of freedom for the 
$i$-th species. The factor $\gamma_i$ is 2/3 for scalars and fermions, and 5/6 for gauge bosons. $\bar{\mu}$ is the renormalization scale and we set $\bar{\mu}=250$ GeV as an example.

As the tree-level case, we assume that the global minimum of the effective potential lies within the subspace defined by $\varphi_\chi=0$. As long as the $\mathbb{Z}_2$-symmetry under $\varphi_\chi\to-\varphi_\chi$ is preserved, the effective potential remains stationary in the $\varphi_\chi$-direction at $\varphi_\chi=0$. Note that the first and second derivatives with respect to $\chi$ are evaluated at $\chi=0$ after differentiation. At the one-loop level, the tadpole conditions take the form:
\begin{align}
\frac{\partial V_{\mathrm{eff}}}{\partial \varphi}= & \frac{\varphi}{2}\left(m^2+m^{(1) 2}+\frac{\lambda+\lambda^{(1)}}{2} \varphi^2+\frac{\delta_2+\delta_2^{(1)}}{2} \varphi_{S}^2\right)+\frac{\partial V_{\mathrm{CW}}}{\partial \varphi}=0, \\
\frac{\partial V_{\mathrm{eff}}}{\partial \varphi_s}= & \sqrt{2}\left(a_1+a_1^{(1)}\right)+\frac{\varphi_{S}}{2}\left(b_2+b_1+b_2^{(1)}+b_1^{(1)}+\frac{d_2+d_2^{(1)}}{2} \varphi_{S}^2+\frac{\delta_2+\delta_2^{(1)}}{2} \varphi^2\right) \\
& \quad+\frac{\partial V_{\mathrm{CW}}}{\partial \varphi_S}=0, \\
\frac{\partial V_{\mathrm{eff}}}{\partial \varphi_\chi}=&\frac{\varphi_\chi}{2}\left[b_2+b_2^{(1)}-b_1-b_1^{(1)}+\frac{d_2+d_2^{(1)}}{2}\left(\varphi_S^2+\varphi_\chi^2\right)+\frac{\delta_2+\delta_2^{(1)}}{2} \varphi^2\right]=0.
\end{align}
The tadpole condition of $\chi$ boson is trivially satisfied when $\varphi_\chi=0$.
The elements of the mass matrix of the CP-even scalars and the mass of the CP-odd scalar are
\begin{align}
\left(\mathcal{M}_h^2\right)_{11}&=\frac{m^2+m^{(1) 2}}{2}+\frac{3\left(\lambda+\lambda^{(1)}\right)}{4} v^2+\frac{\delta_2+\delta_2^{(1)}}{4} v_{S}^2+\frac{\partial^2 V_{\mathrm{CW}}}{\partial \varphi^2} \\
\left(\mathcal{M}_h^2\right)_{22}&=\frac{b_2+b_1+b_2^{(1)}+b_1^{(1)}}{2}+\frac{\delta_2+\delta_2^{(1)}}{4} v^2+\frac{3\left(d_2+d_2^{(1)}\right)}{4} v_{S}^2+\frac{\partial^2 V_{\mathrm{CW}}}{\partial \varphi_S^2} \\
\left(\mathcal{M}_h^2\right)_{12}&=\frac{\delta_2+\delta_2^{(1)}}{2} v v_{S}+\frac{\partial^2 V_{\mathrm{CW}}}{\partial \varphi \partial \varphi_S} \\
m_\chi^2 & =\frac{b_2+b_2^{(1)}-b_1^{(1)}-b_1^{(1)}}{2}+\frac{d_2+d_2^{(1)}}{4}\left(\varphi_S^2+3 \varphi_\chi^2\right)+\frac{\delta_2+\delta_2^{(1)}}{4} \varphi^2+\frac{\partial^2 V_{\mathrm{CW}}}{\partial \varphi_\chi^2}.
\label{DMmassloop}
\end{align}
The first and second derivatives of the one-loop correction are given in Appendix.~\ref{app:deriv}.

By requiring that the tadpole and mass conditions above retain the same form as at the tree level, the coefficients of the finite counterterms are determined accordingly. We can get
\begin{align}
\delta_2^{(1)} & =-\frac{2}{v v_{S}}\frac{\partial^2 V_{\mathrm{CW}}}{\partial \varphi \partial \varphi_S},
\label{del2CT} \\
\lambda^{(1)} & =-\frac{2}{v^2}\left(\left\langle\frac{\partial^2 V_{\mathrm{CW}}}{\partial \varphi^2}\right\rangle-\frac{1}{v}\left\langle\frac{\partial V_{\mathrm{CW}}}{\partial \varphi}\right\rangle\right), \\
m^{2 (1)} & = -\frac{\lambda^{(1)}}{2} v^2-\frac{\delta_2^{(1)}}{2} v_{S}^2-\frac{2}{v}\left\langle\frac{\partial V_{\mathrm{CW}}}{\partial \varphi}\right\rangle,
\end{align}
In addition, we set $a_1^{(1)}=0$ without loss of generality generality, we also find
\begin{align}
d_2^{(1)} & =-\frac{2}{v_{S}^2}\left(\frac{\partial^2 V_{\mathrm{CW}}}{\partial \varphi_S^2}-\frac{1}{v_{S}}\frac{\partial V_{\mathrm{CW}}}{\partial \varphi_s}\right),\label{d2CT} \\
b_2^{(1)}+b_1^{(1)} & =-\frac{\delta_2^{(1)}}{2} v^2-\frac{d_2^{(1)}}{2} v_{S}^2-\frac{2}{v_{S}}\frac{\partial V_{\mathrm{CW}}}{\partial \varphi_S},\\
b_2^{(1)}-b_1^{(1)}&=-\frac{d_2^{(1)}}{2} v_{S}^2-\frac{\delta_2^{(1)}}{2} v^2-2\frac{\partial^2 V_{\mathrm{CW}}}{\partial \varphi_\chi^2}.\label{b2-b1CT}
\end{align}
We adopt the same set of the input parameters as the tree level, by which all the counterterms are determined.

%%%%%%%%%%%%%%%%%%%%%%%%%%%%%%%%%%%%%%%%%%%%%%%%
%				DM-quark scattering
%%%%%%%%%%%%%%%%%%%%%%%%%%%%%%%%%%%%%%%%%%%%%%%%

\section{Dark Matter Phenomenology}\label{sec:DM}

\subsection{DM-quark scattering}\label{sec:DD}
%----------------------------------------------------------------------------------------------------------------------------------
\begin{figure}[htpb]
\center
\includegraphics[width=5cm]{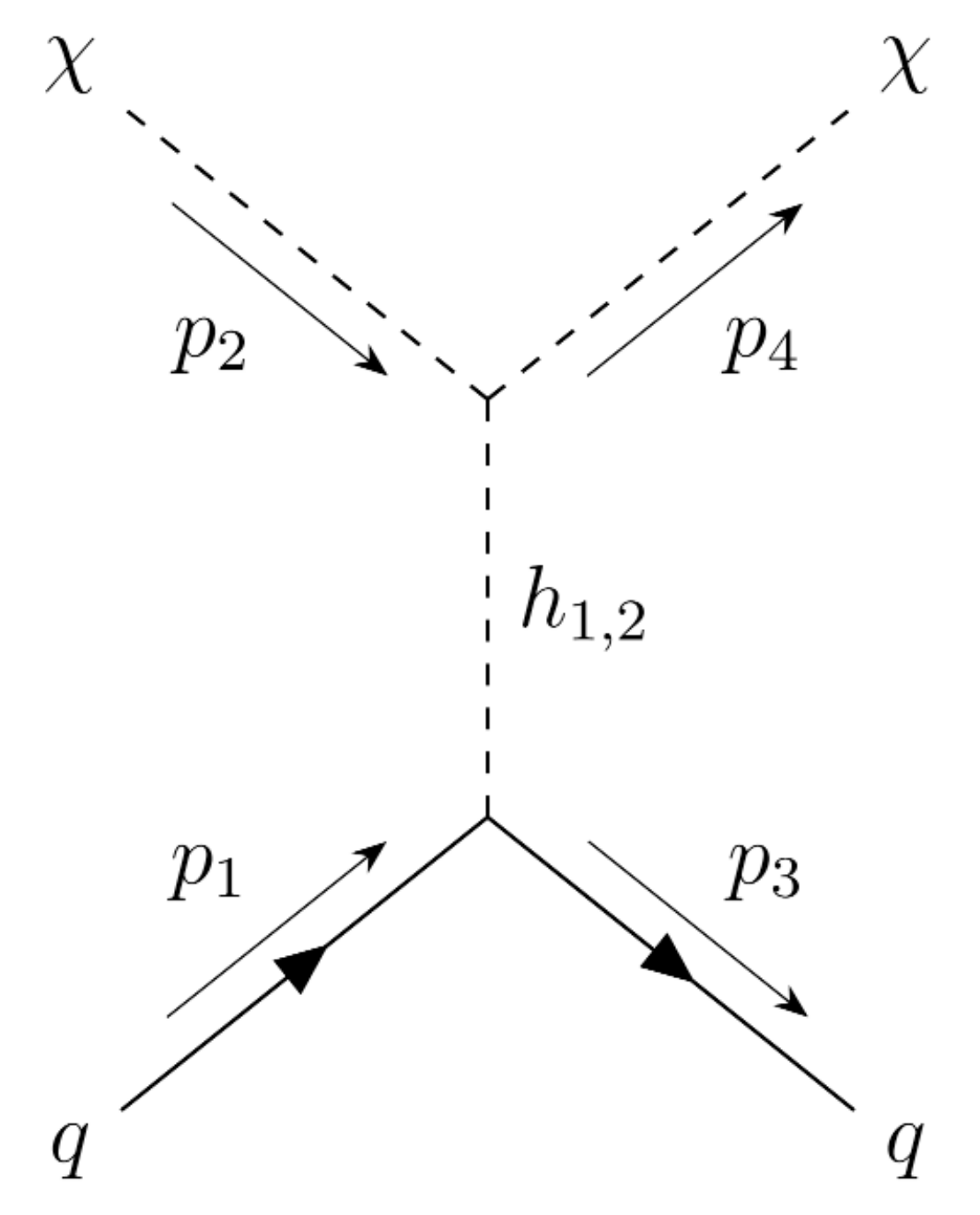}
\caption{Feynman diagram of the scattering process $\chi q \to \chi q $ mediated by $h_{1,2}$.}
\label{fig:scatt}
\end{figure}
%----------------------------------------------------------------------------------------------------------------------------------
In the CxSM, the scattering process of the DM $\chi$ off a quark $q$, $\chi q\to \chi q$ is described by the Feynman diagram in Fig.~\ref{fig:scatt}. The interaction Lagrangian of DM $\chi$ to the CP-even scalars $h_1$, $h_2$ is given by 
\begin{align}
\mathcal{L}_S
&=C_{\chi\chi h_1} \chi^2 h_1 +C_{\chi\chi h_2} \chi^2 h_2  \nonumber \\
&= \left(\frac{\delta_2}{4} v \cos \alpha+\frac{d_2}{4} v_S \sin \alpha \right)  \chi^2 h_1
+ \left(-\frac{\delta_2}{4} v \sin \alpha+\frac{d_2}{4} v_S \cos \alpha \right)  \chi^2 h_2  \nonumber \\
&= -\frac{m_{h_1}^2+\frac{\sqrt{2} a_1}{v_S}}{2 v_S} \sin \alpha~ \chi^2 h_1 +\frac{m_{h_2}^2+\frac{\sqrt{2} a_1}{v_S}}{2 v_S} \cos \alpha~ \chi^2 h_2  
\label{cch} 
\end{align}
while that of a quark $q$ to $h_1$ or $h_2$ is given by
\begin{align}
\mathcal{L}_Y&= C_{qq h_1}\bar{q}q h_1 + C_{qq h_2}\bar{q}q h_2 \nonumber \\
&=\frac{m_q}{v} \bar{q} q \left(h_1 \cos \alpha-h_2 \sin \alpha\right) \label{hff}, 
\end{align}
where $m_q$ denotes a mass of the quark $q$. Then, the scattering amplitude $\mathcal{M}$ is given by a sum of two amplitudes $\mathcal{M}_1$ and $\mathcal{M}_2$ mediated by $h_1$ and $h_2$, respectively;
\begin{align}
 i\mathcal{M} 
&=
i\qty(\mathcal{M}_1 + \mathcal{M}_2), 
\label{msumtr}
\\
 i\mathcal{M}_1 
&= 
-i 2 C_{\chi\chi h_1} C_{qq h_1}  \frac{1}{t-m_{h_1}^2} \bar{u}(p_3) u(p_1)\nonumber \\
&=-i \frac{m_q}{v v_S} \frac{m_{h_1}^2+\frac{\sqrt{2} a_1}{v_S}}{t-m_{h_1}^2} \sin \alpha \cos \alpha \bar{u}\left(p_3\right) u\left(p_1\right), 
\label{m1tr}
\\
 i\mathcal{M}_2 
&= 
-i 2 C_{\chi\chi h_2} C_{qq h_2}  \frac{1}{t-m_{h_2}^2} \bar{u}(p_3) u(p_1)\nonumber \\
&=-i \frac{m_q}{v v_S} \frac{m_{h_2}^2+\frac{\sqrt{2} a_1}{v_S}}{t-m_{h_2}^2} \sin \alpha \cos \alpha \bar{u}\left(p_3\right) u\left(p_1\right),
\label{m2tr}
\end{align}
where $t\equiv \qty(p_1 - p_3)^2$ is a momentum transfer and $u(p)~(\bar{u}(p))$ represents an incoming (outgoing) quark spinor with a momentum $p$. A factor 2 in the r.h.s of the first lines of \eqref{m1tr} and \eqref{m2tr} is a symmetry factor for the $\chi\chi h_i$ vertex. When $t\ll m_{h_1},m_{h_2}$, the sum of the two amplitudes can be expressed as
\begin{align}
i\mathcal{M}&=i \frac{m_f}{v v_S} \bar{u}\left(p_3\right) u\left(p_1\right) \sin \alpha \cos \alpha \nonumber \\
& \times\left\{\left(-\frac{m_{h_1}^2}{t-m_{h_1}^2}+\frac{m_{h_2}^2}{t-m_{h_2}^2}\right)+\frac{\sqrt{2} a_1}{v_S}\left(-\frac{1}{t-m_{h_1}^2}+\frac{1}{t-m_{h_2}^2}\right)\right\}.
\label{degesum}
\end{align}

LUX-ZEPLIN (LZ) experiment has placed strong constraints on the DM-quarks scattering cross section~\cite{LZ:2024zvo}. The main scenarios for satisfying this constraint in this CxSM are as follows:
\begin{enumerate}
  \item Considering a small mixing angle between CP-even scalars to realize a decoupling of the scalar particle from the SM,
  \item Considering a suppression mechanism such as the degenerate scalar scenario.
\end{enumerate}
In fact, when we consider the case $\alpha\to 0$, the scattering amplitude \eqref{degesum} vanishes. On the other hand, the amplitude also disappears when the masses of two scalar particles are degenerate ($m_{h_1} = m_{h_2}$), and we refer to this suppression mechanism as the degenerate scalar scenario. In numerical calculations, benchmark points corresponding to these two scenarios are used.

Before closing this section, we mention the consistency with the results of the collider experiments.
The Higgs signal strength, denoted by $\mu$, is used to compare the branching ratio of the 125 GeV Higgs boson observed at the LHC with the prediction from the SM.
The LHC Run-2 experiment gives constraints on $\mu$ as 
$0.92<\mu<1.20 $ at ATLAS~\cite{ATLAS:2019nkf} and  $0.90<\mu<1.16$ at CMS~\cite{CMS:2020gsy}. 
The total decay width of the Higgs boson is constrained as $\Gamma_{h}^{\mathrm{exp}}<$  14.4 MeV at ATLAS~\cite{ATLAS:2018jym} and $\Gamma_h^{\exp }=3.2_{-1.7}^{+2.4}$ MeV at CMS~\cite{CMS:2022ley}. 
If $h_2$ is heavier than $h_1$, consistency with the observed SM Higgs signal strength requires $|\alpha|<\pi / 14$ .

On the other hand, ref.~\cite{Abe:2021nih} has explored how collider experiments could probe the degenerate scalar scenario. The study notes that a mass splitting smaller than roughly 3 GeV between $h_1$ and $h_2$ is still allowed by current LHC data~\cite{CMS:2014afl}. In general, the mixing angle $\alpha$ is limited by the existence of an extra Higgs boson, but if the two Higgs masses are nearly identical, this constraint vanishes and $\alpha$ can reach its maximal value $\alpha=\pi/4$.
As shown in Eq.~(\ref{hff}), the couplings of $h_1~(h_2)$ to the SM particles are given by the SM Higgs couplings multiplied by $\cos{\alpha}$ (for $h_1$) and $-\sin{\alpha}$ (for $h_2$). Accordingly, the partial decay widths are expressed as:
\begin{align}
\Gamma_{h_1\to XX}&=\cos^2{\alpha}~\Gamma_{h\to XX}^{\mathrm{SM}}(m_{h_1}), \label{partialdecaywidth}\\
\Gamma_{h_2\to XX}&=\sin^2{\alpha}~\Gamma_{h\to XX}^{\mathrm{SM}}(m_{h_2}),
\end{align}
where $\Gamma_{h\to XX}^\mathrm{SM}(m_{h_{1,2}})$ denotes the SM Higgs partial decay width evaluated at the corresponding mass.
When the two Hiigs masses are nearly degenerate, it becomes experimentally difficult to distinguish the contributions from $h_1$ and $h_2$. As a result, the observed decay rate is effectively the sum of both:
\begin{align}
\Gamma_{h_1\to XX}+\Gamma_{h_2\to XX} \simeq \Gamma_{h\to XX}^{\mathrm{SM}}(m_h), 
\end{align}
which holds for any value of $\alpha$.
Therefore, in the degenerate mass limit, the predicted Higgs signal strength in the CxSM becomes identical to that in the SM.

\subsection{Dark matter abundance}\label{sec:relic}

To compute the DM relic abundance, we solve the Boltzmann equation governing the evolution of the DM number density $n_\chi(t)$. It is convenient to work with the comoving yield $Y_\chi(z) \equiv n_\chi(t)/s(T)$, where $s(T)$ is the entropy density and $z \equiv m_\chi / T$ is the dimensionless inverse temperature.
In terms of $Y_\chi(z)$, the Boltzmann equation takes the form:
\begin{align}
\frac{d Y_\chi(z)}{d z}=-\lambda(z)\left(Y_\chi(z)^2-Y_{\mathrm{eq}}(z)^2\right),
\end{align}
where
\begin{align}
Y_\chi^{\mathrm{eq}}(z)=\frac{n_\chi^{\mathrm{eq}}(T)}{s(T)}=\frac{45}{4 \pi^4} \frac{1}{h_{\mathrm{eff}}(T)} z^2 K_2\left( z\right).
\end{align}
$K_2(z)$ is the modified Bessel function of the second kind, which appears in the expression for the equilibrium number density of a massive particle and reflects the Boltzmann suppression in the non-relativistic regime. $h_{\mathrm{eff}}(T)$ denotes the effective relativistic degrees of freedom for the entropy. $\lambda(z)$ contains the effects of the thermally averaged annihilation cross section and the Hubble expansion rate defined as
\begin{align}
\lambda(z)=\frac{1}{8} \sqrt{\frac{\pi g_*(T)}{45}} \frac{m_{\mathrm{Pl}}}{m_\chi} \frac{\int_{4 z^2}^{\infty} d x \mathcal{F}\left(T^2 x\right) \sqrt{x-4 z^2} K_1(z)}{\left[z^2 K_2\left(z\right)\right]^2}.
\end{align}
Here, $\sqrt{g_*(T)}=\frac{h_{\mathrm{eff}}(T)}{\sqrt{g_{\mathrm{eff}}(T)}}\left(1+\frac{T}{3 h_{\mathrm{eff}}(T)} \frac{d h_{\mathrm{eff}}(T)}{d T}\right)$ and $g_{\mathrm{eff}}(T)$ denote the effective relativistic degrees of freedom for the energy. The relation between the function $\mathcal{F}(s)$ and the thermally averaged cross section $\langle \sigma v \rangle$ is given by
\begin{align}
\langle \sigma v \rangle = \frac{1}{2 s \sqrt{1 - 4 m_\chi^2 / s}} \, \mathcal{F}(s),
\end{align}
with $s$ being the Mandelstam variable representing the square of the center-of-mass energy.
The explicit expressions for $\mathcal{F}(s)$ can be found in Appendix.~\ref{app:crosssection}.

In Ref.~\cite{Cline:2013gha,Chakraborti:2025ced}, the DM relic abundance is approximately evaluated using the following expression:
\begin{align}
Y_\chi(\infty) \simeq \frac{Y_\chi\left(z_d\right)}{1+Y_\chi\left(z_d\right) A_d}=\frac{\left(1+\delta_d\right) Y_\chi^{\mathrm{eq}}\left(z_d\right)}{1+\left(1+\delta_d\right) Y_\chi^{\mathrm{eq}}\left(z_d\right) A_d},
\label{Yinf}
\end{align}
where 
\begin{align}
A_d=\int_{z_d}^{\infty} d z \lambda(z).
\end{align}
$z_d$ is the decoupling point defined by the condition $Y_\chi(z_d) = (1 + \delta_d) Y_\chi^{\mathrm{eq}}(z_d)$, and $\delta_d$ quantifies the deviation from equilibrium at freeze-out. In this analysis, we apply the following approximations for $z_d$ and $A_d$:
\begin{align}
z_d &\simeq \log \left[\frac{\delta_d\left(2+\delta_d\right)}{1+\delta_d} \frac{1}{8 \pi^3} \sqrt{\frac{45}{2 g_{\mathrm{eff}}(T)}} \frac{m_{\mathrm{Pl}}}{m_\chi} \mathcal{F}\left(4 m_\chi^2\right)\right] \nonumber \\
&-\frac{1}{2} \log \log \left[\frac{\delta_d\left(2+\delta_d\right)}{1+\delta_d} \frac{1}{8 \pi^3} \sqrt{\frac{45}{2 g_{\mathrm{eff}}(T)}} \frac{m_{\mathrm{Pl}}}{m_\chi} \mathcal{F}\left(4 m_\chi^2\right)\right], \\
A_d &\simeq \frac{1}{2} \sqrt{\frac{\pi g_*(T)}{45}} \frac{m_{\mathrm{Pl}}}{m_\chi} \cdot \mathcal{F}\left(4 m_\chi^2\right) \frac{1}{z_d}\left(1-\frac{27}{16 z_d}+\frac{319}{128 z_d^2}+\cdots\right) .
\end{align}
Here,  The function $\mathcal{F}(s)$, with $s$ being the squared center-of-mass energy of the initial state, is defined as the integration over the final-state phase space of the squared amplitudes, incorporating all contributions from the DM annihilation channels. For numerical evaluations, we take $\delta_d = 3$ to ensure the validity of the analytic approximation. For simplicity, we also set $g_{\mathrm{eff}} = h_{\mathrm{eff}} = g_*(T) = 80$ since the $T$-dependence of these quantities do not alter the numerical resutlts, and use the Planck mass $m_{\mathrm{Pl}} = 1.22 \times 10^{19}$ GeV.

The final DM abundance is obtained from the comoving yield $Y_\chi(\infty)$ \eqref{Yinf} via the relation
\begin{align}
\Omega_\chi h^2 = 2.7506 \times 10^8 \cdot m_\chi~\mathrm{GeV} \cdot Y_\chi(\infty).
\end{align}
This result can be directly compared with the observed value~\cite{Planck:2018vyg}:
\begin{align}
\Omega_{\mathrm{DM}} h^2 = 0.1200 \pm 0.0012.
\label{obsrelic}
\end{align}

%%%%%%%%%%%%%%%%%%%%%%%%%%%%%%%%%%%%%%%%%%%%%%%%
%				Numerical results
%%%%%%%%%%%%%%%%%%%%%%%%%%%%%%%%%%%%%%%%%%%%%%%%

\section{Numerical results}\label{sec:num}

We first impose the theoretical constraints discussed in Sec.~\ref{sec:model}, including the lower-boundedness condition and the tree-level perturbative unitarity constraints in Eqs.~\eqref{BFB} and \eqref{unitarity}.
We then explore the parameter space that satisfies two conditions:
\begin{enumerate}
  \item consistency with the latest bounds of the latest LZ experiment~\cite{LZ:2024zvo},
  \item reproduction of the observed DM relic abundance,
\end{enumerate}
To this end, we perform a parameter scan over the following ranges:
\begin{align}
& 10 \leq v_s~ \mathrm{GeV} \leq 1000 \nonumber \\
& -500^3 \leq a_1 / \mathrm{GeV}^3 \leq 0 \nonumber\\
& 125 \leq m_\chi~  \mathrm{GeV} \leq 5000.
\end{align}
As seen in Eq.~\eqref{tadpole2}, ensuring $\varphi_S > 0$ typically requires $a_1 < 0$. Therefore, we focus on negative values of $a_1$ in this analysis.
For the second scalar mass $m_{h_2}$, we consider four benchmark values: 126, 500, 1000, and 2000 GeV. When $m_{h_2}=126$ GeV, the degenerate scalar scenario described in Sec.~\ref{sec:DD} can be realized, while for other values of $m_{h_2}$, the small value of the mixing angle $\alpha$ is the key to satisfying the direct detection bound. The mixing angle $\alpha$ is fixed to $-\pi/14$, motivated by the Higgs search considerations discussed in Sec.~\ref{sec:DD}\footnote{The scalar degeneracy region does not restrict the mixing angle $\alpha$, but here we use the same $\alpha=-\pi/14$ as in other scenarios. The results do not vary greatly depending on the value of $\alpha$.}. Note that $\delta_2$ is invariant under the transformation $m_{h_1}^2-m_{h_2}^2\to -(m_{h_1}^2-m_{h_2}^2)$ and $\alpha\to -\alpha$ as can be seen from Eq.~\eqref{del2CPC1}. Here, since we consider the case where $m_{h_2} > m_{h_1}$, the mixing angle $\alpha$ is taken to be negative so that $\delta_2 > 0$.
We compare the allowed parameter regions with and without the one-loop corrections. The one-loop corrections include not only the corrections to the vertices involving the $\chi$-particle but also the finite counterterms in Eqs.~\eqref{del2CT} - \eqref{b2-b1CT}. In both direct detection and relic abundance, the relevant amplitudes are dominated by contributions with vanishing incoming momenta. The explicit expressions for the vertex functions are provided together with their numerical volumes for $m_{h_2}=$126 GeV and 2000 GeV in Appendix.~\ref{app:chivertex}.

%-------------------------------------------------------------%

\begin{figure}[h!]
  \centering
  \begin{minipage}{0.39\columnwidth}
    \centering
    
    \includegraphics[width=\columnwidth]{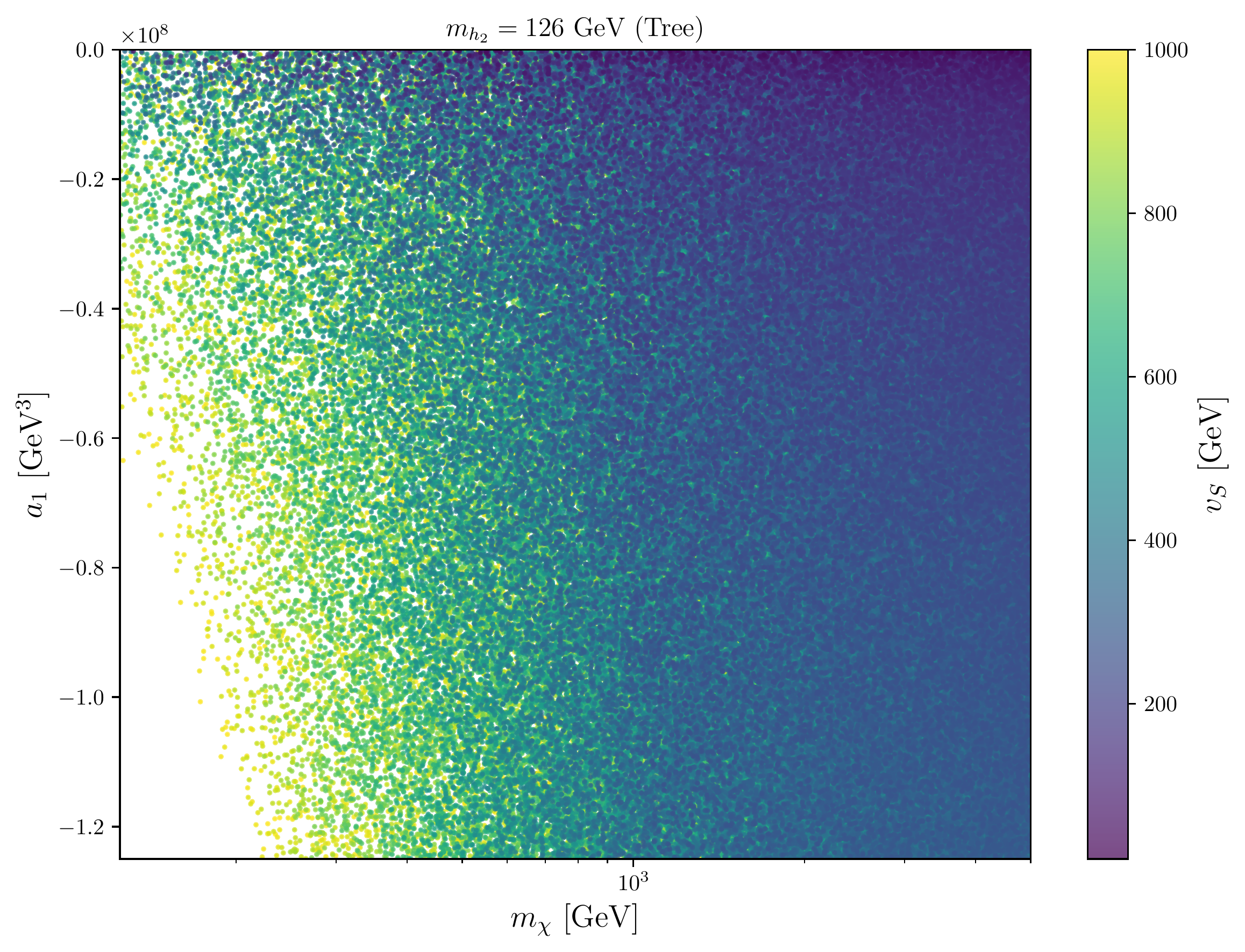}
  \end{minipage}
  \hspace{5mm}
  \begin{minipage}{0.39\columnwidth}
    \centering
    \includegraphics[width=\columnwidth]{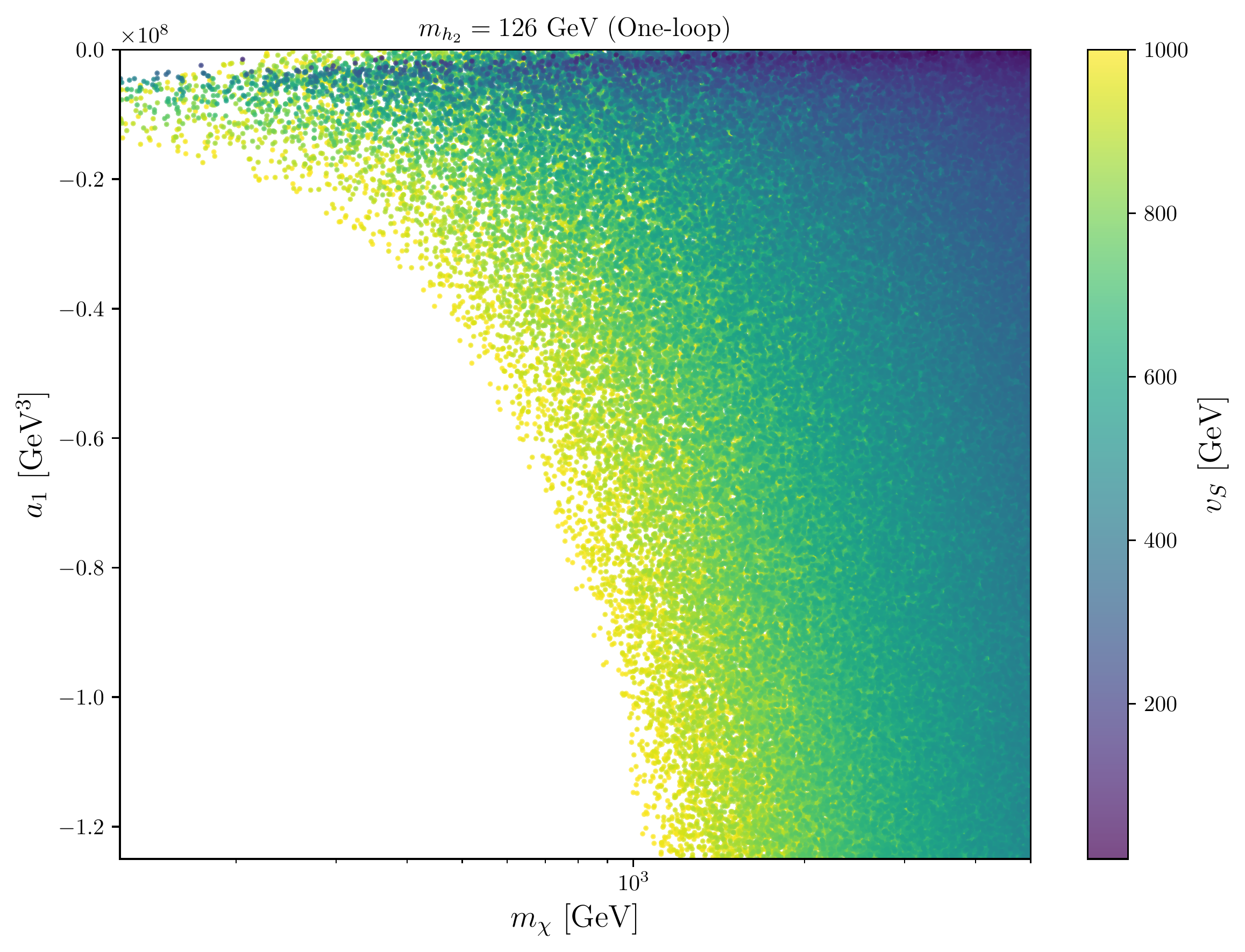}
  \end{minipage}
  
  \begin{minipage}{0.39\columnwidth}
    \centering
    \includegraphics[width=\columnwidth]{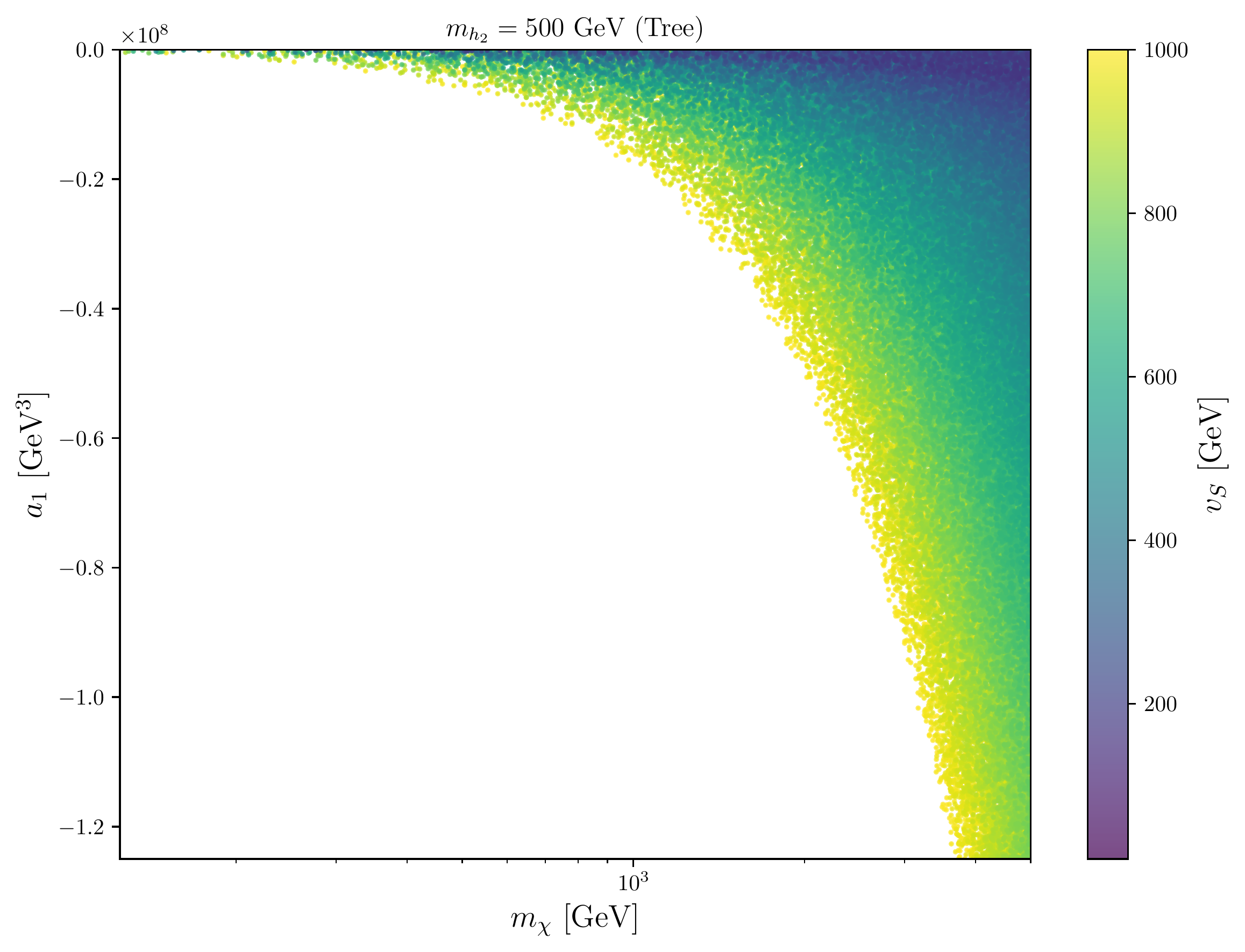}
  \end{minipage}
  \hspace{5mm}
  \begin{minipage}{0.39\columnwidth}
    \centering
    \includegraphics[width=\columnwidth]{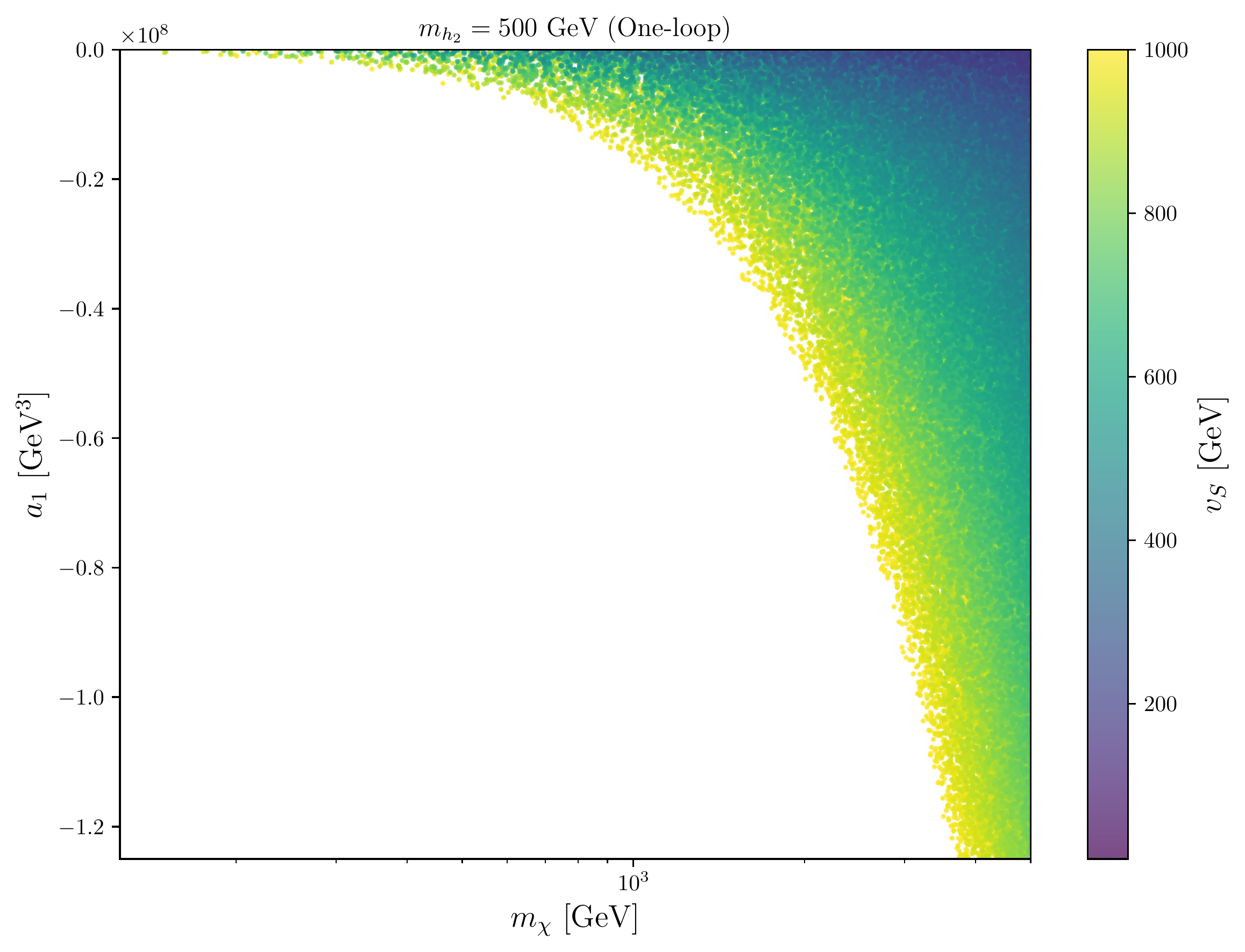}
  \end{minipage}
  
  \begin{minipage}{0.39\columnwidth}
    \centering
    \includegraphics[width=\columnwidth]{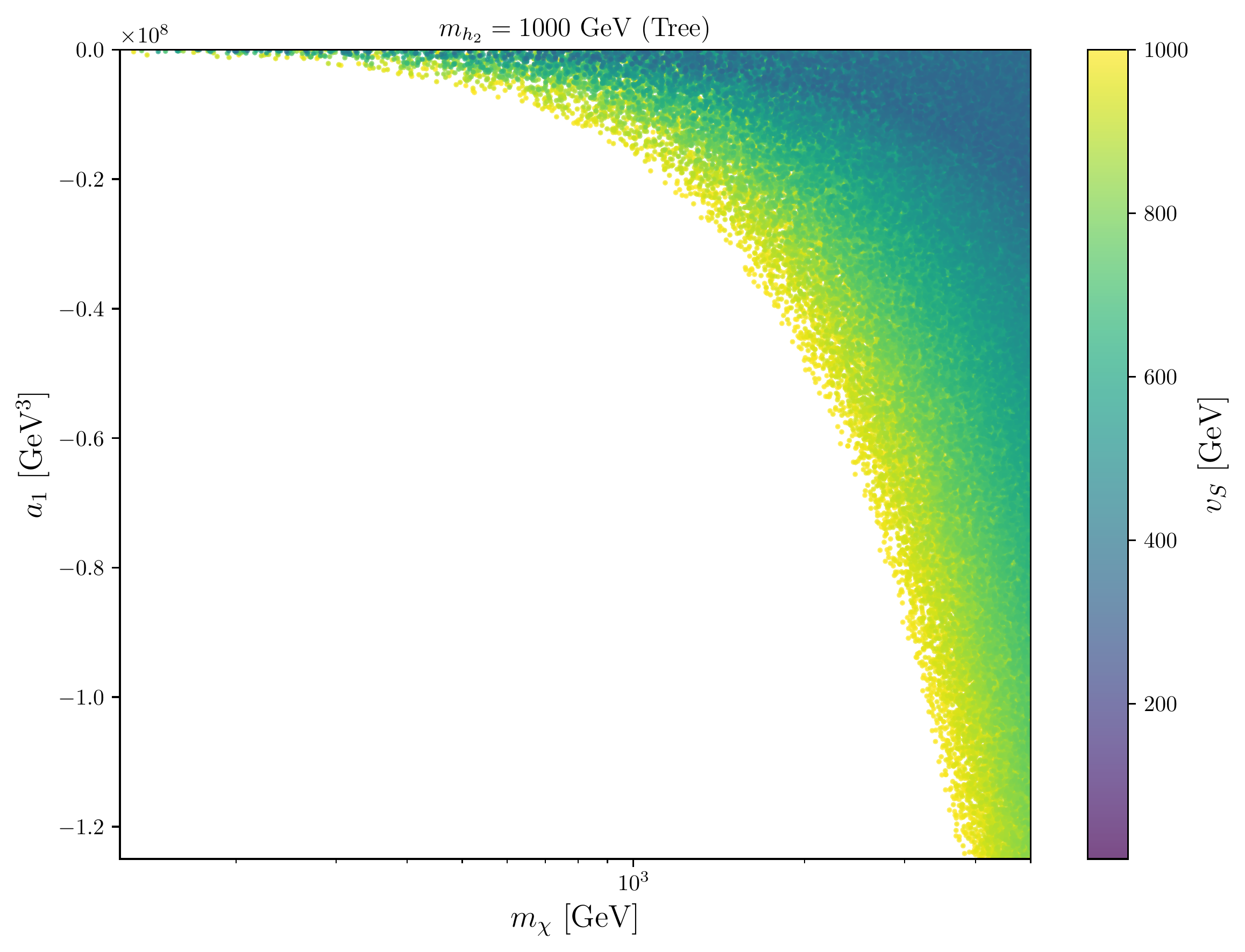}
  \end{minipage}
  \hspace{5mm}
  \begin{minipage}{0.39\columnwidth}
    \centering
    \includegraphics[width=\columnwidth]{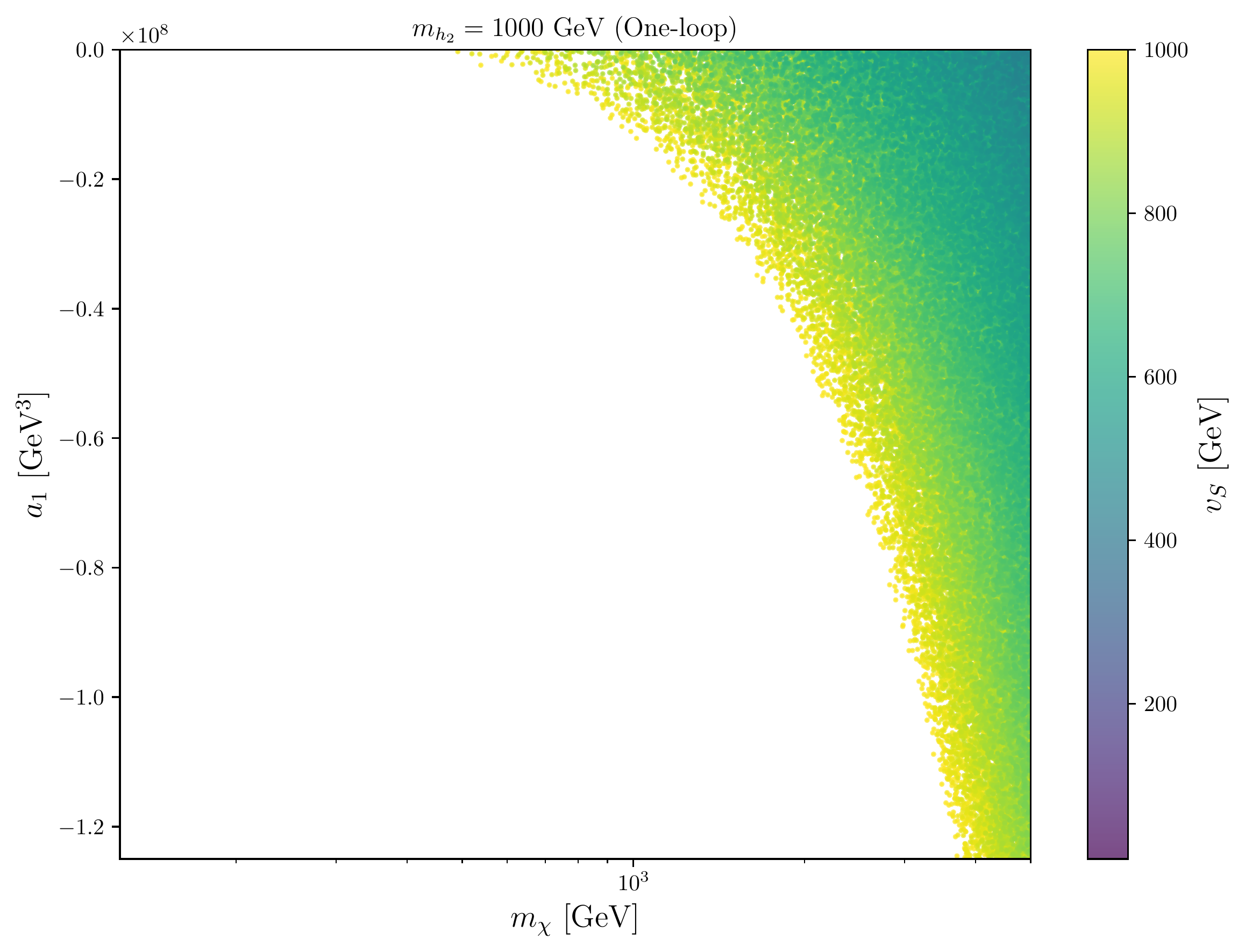}
  \end{minipage}
  
  \begin{minipage}{0.39\columnwidth}
    \centering
    \includegraphics[width=\columnwidth]{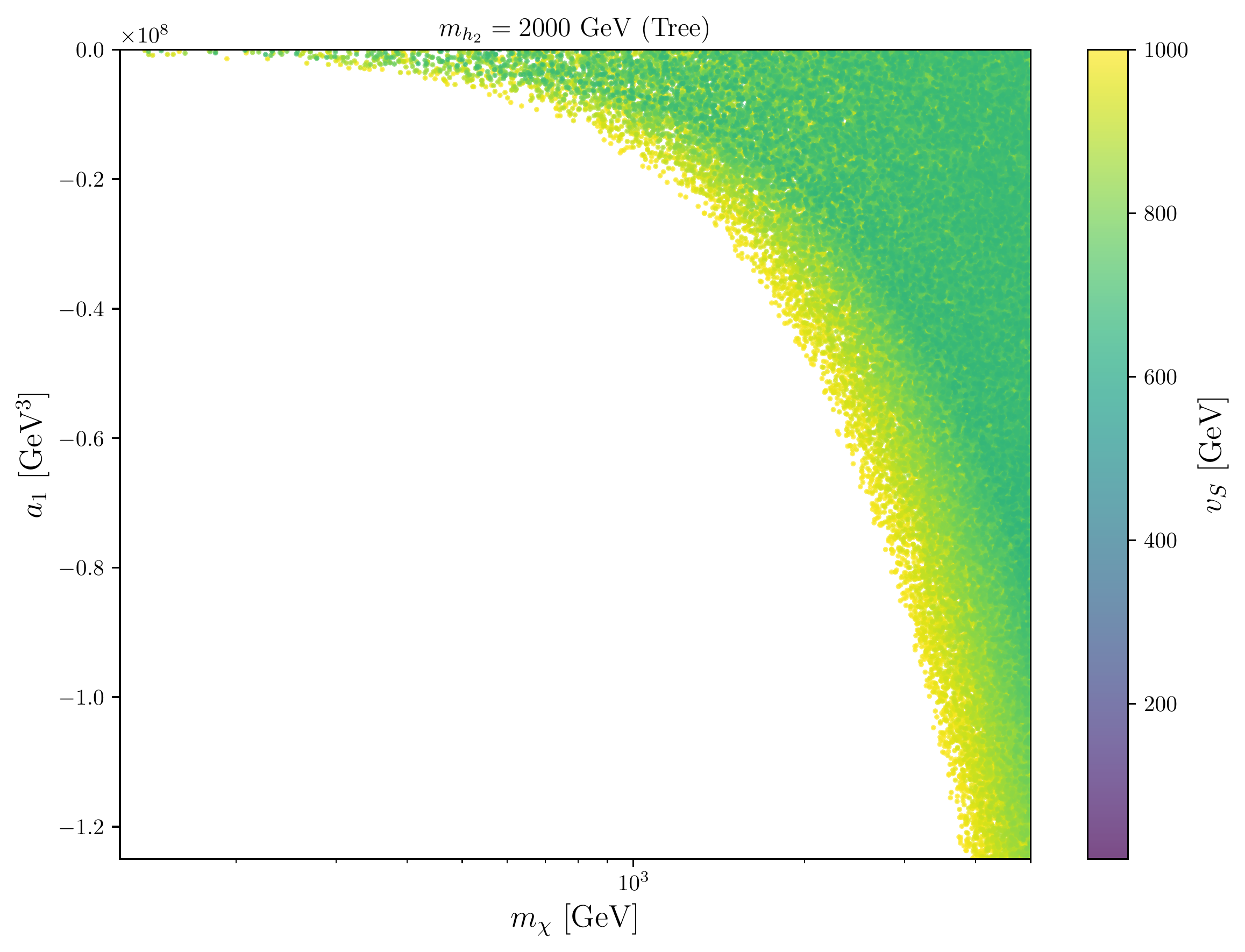}
  \end{minipage}
  \hspace{5mm}
  \begin{minipage}{0.39\columnwidth}
    \centering
    \includegraphics[width=\columnwidth]{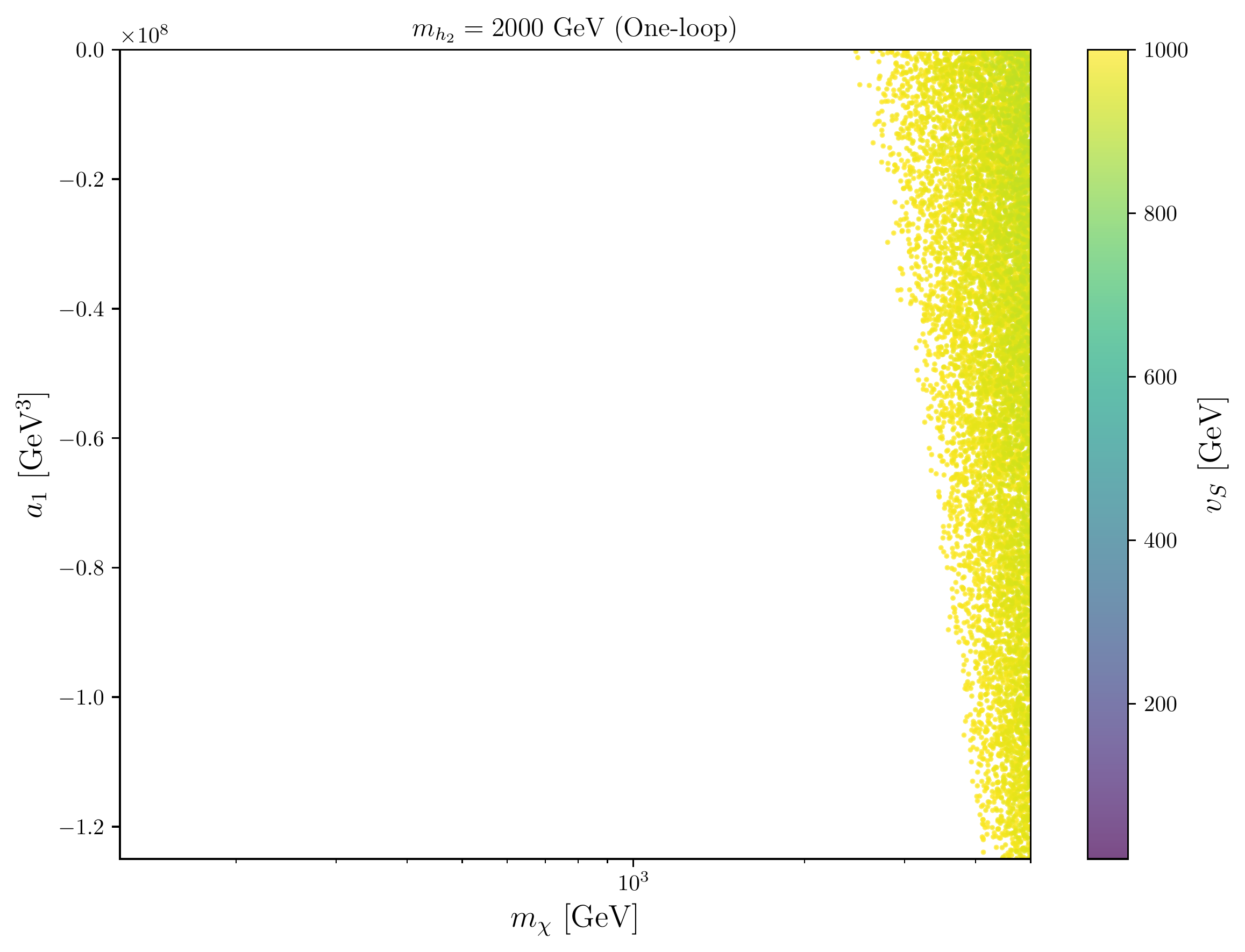}
  \end{minipage}

  \caption{%
   \setlength{\baselineskip}{14pt}
  The parameter region that satisfies the results of the DM direct detection experiment. We set $\alpha=-\pi/14$ and $m_{h_2}=126,500,1000,2000$ GeV. The left four panels show the allowed parameter regions from the tree-level analysis, while the right ones depict those obtained by including the one-loop corrections.}
    \label{fig:DD}
\end{figure}

%-------------------------------------------------------------%

Fig.~\ref{fig:DD} illustrates the parameter space consistent with the constraints from the LZ experiment. In the four tree-level panels on the left, one sees that a broad band of viable points survives whenever the two Higgs states are nearly degenerate.  Once the mass of $h_2$ departs from that of $h_1$, the size and shape of the allowed region depend only weakly on $m_{h_2}$; increasing $m_\chi$ always relaxes the bound. For small $v_S$, the magnitude of $d_2$ tends to increase, as can be seen from Eq.~\eqref{d2}. Consequently, the small-$v_S$ region is more strongly restricted by the theoretical constraints, in particular the tree-level unitarity bound on $d_2$.
If $a_1 = 0$, the DM-quark scattering amplitude is strongly suppressed in the low-energy limit ($t \to 0$ in Eq.~\eqref{degesum}). However in that case, the scalar potential has $\mathbb{Z}_2$ symmetry with respect to the scalar field $S$, causing the domain-wall problem. Thus, nonzero $a_1$ is needed in this model.  Even with $a_1\neq 0$, the second term in Eq.~\eqref{degesum} is proportional to $1 / v_S$, hence a larger singlet VEV lowers the cross section and widens the allowed band.
 
Comparing the left and right panels of Fig.~\ref{fig:DD}, we observe that the inclusion of the one-loop corrections generally reduces the allowed region. As $m_{h_2}$ increases, the discrepancy from the tree-level predictions becomes more significant. This behavior originates from the enhancement of the effective couplings associated with larger $m_{h_2}$. The strengthened couplings amplify the scattering processes, thereby increasing the likelihood that the parameter points are excluded by the LZ constraint. For $m_{h_2}=126$ GeV, in the tree-level calculations, the vertex relationships caused cancellations as mentioned in Sec.~\ref{sec:DD}, but in the one-loop calculations, the effective vertex shifts make it harder for the degenerate scalars to suppress the scattering cross section. The explicit form and behavior of the effective vertices can be found in Appendix.~\ref{app:chivertex}.
%-------------------------------------------------------------%

\begin{figure}[h!]
  \centering
  \begin{minipage}{0.39\columnwidth}
    \centering
    
    \includegraphics[width=\columnwidth]{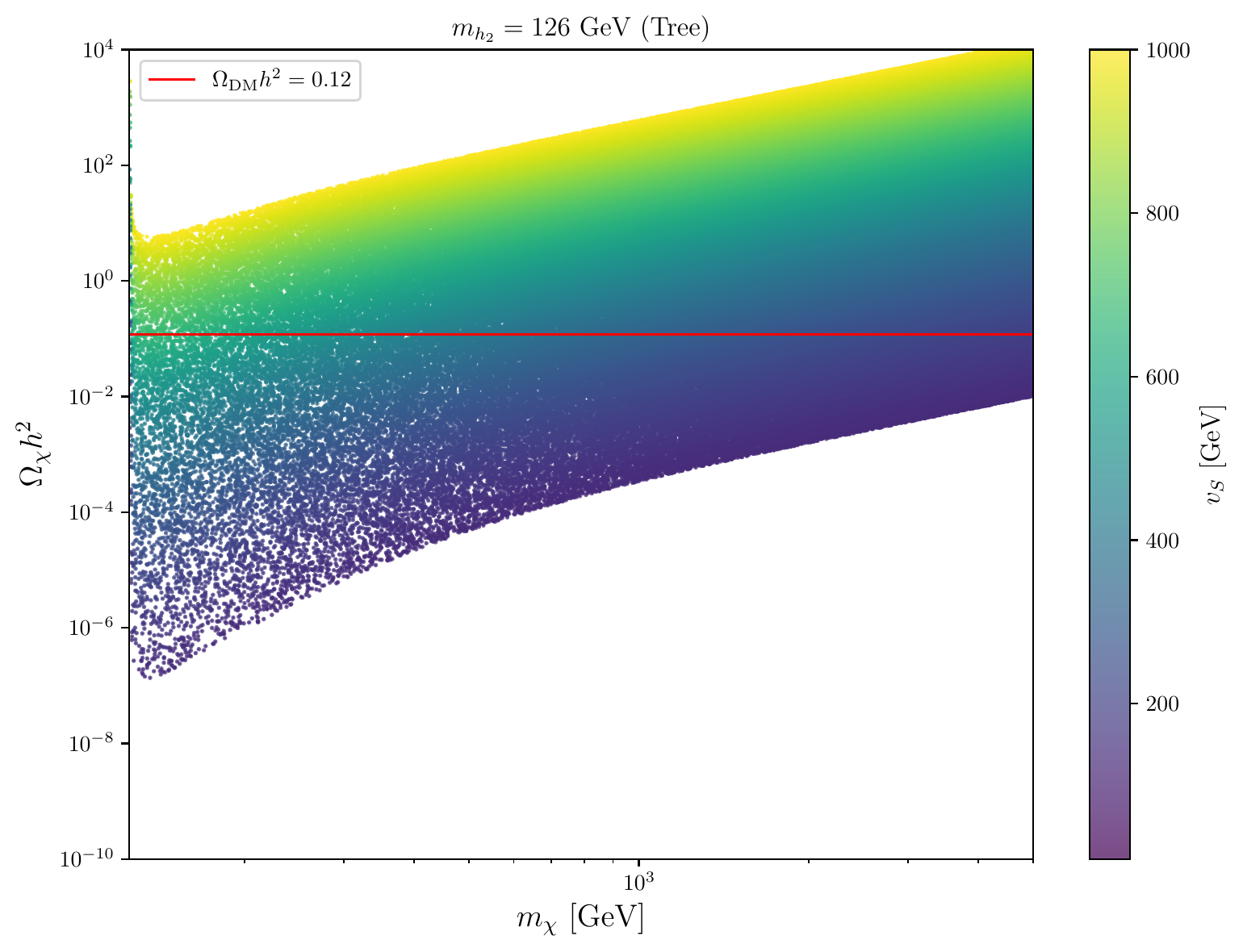}
  \end{minipage}
  \hspace{5mm}
  \begin{minipage}{0.39\columnwidth}
    \centering
    \includegraphics[width=\columnwidth]{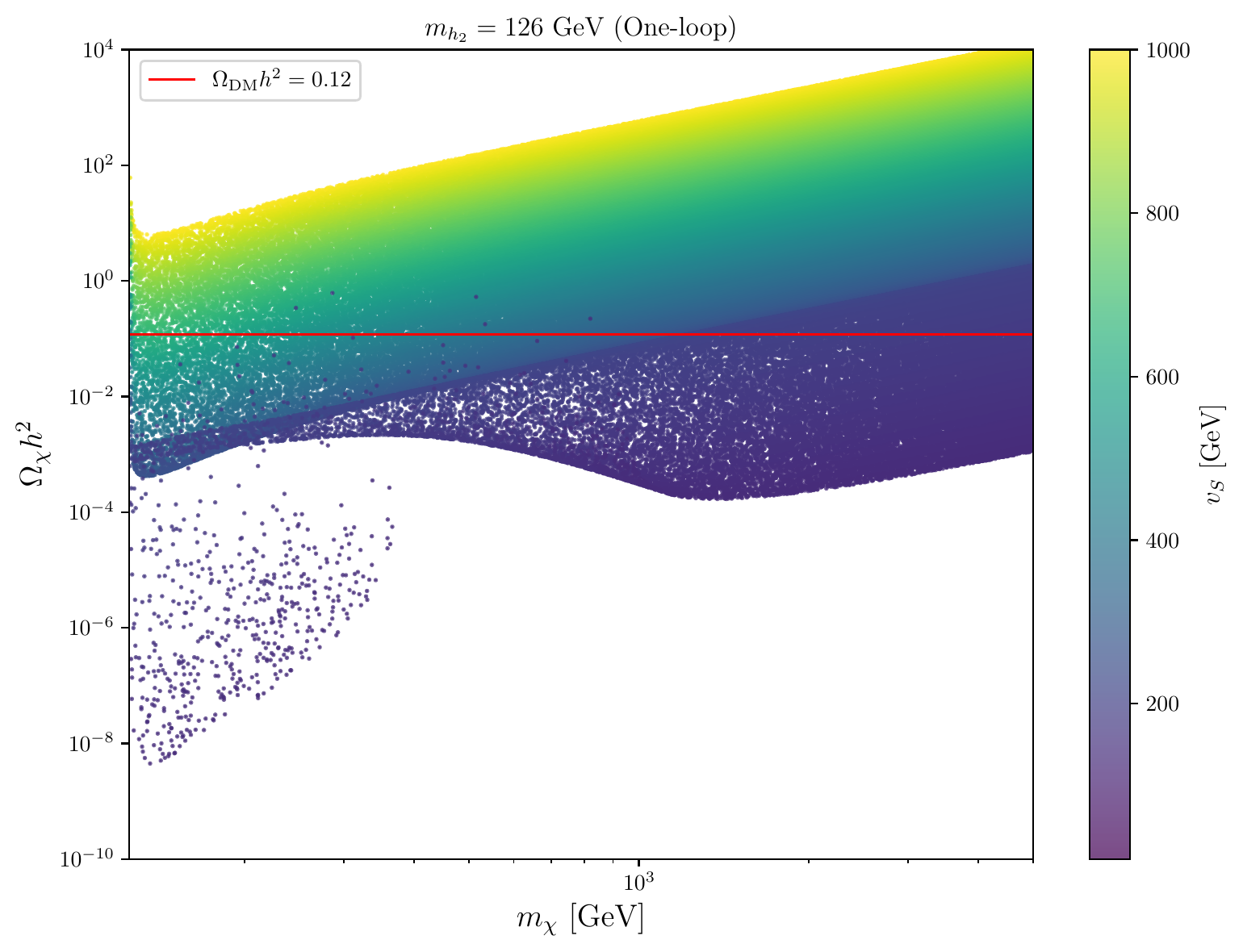}
  \end{minipage}
  
  \begin{minipage}{0.39\columnwidth}
    \centering
    \includegraphics[width=\columnwidth]{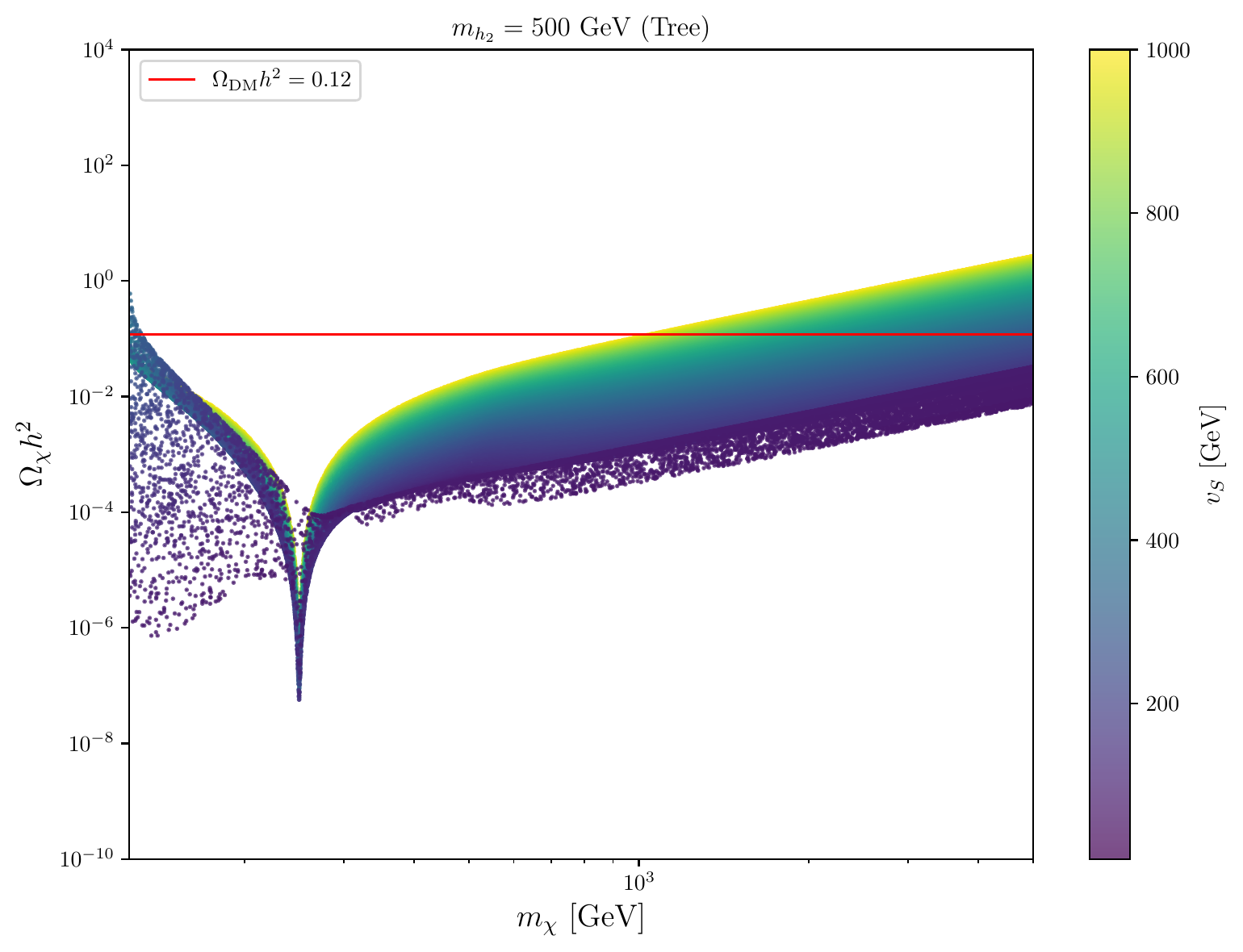}
  \end{minipage}
  \hspace{5mm}
  \begin{minipage}{0.39\columnwidth}
    \centering
    \includegraphics[width=\columnwidth]{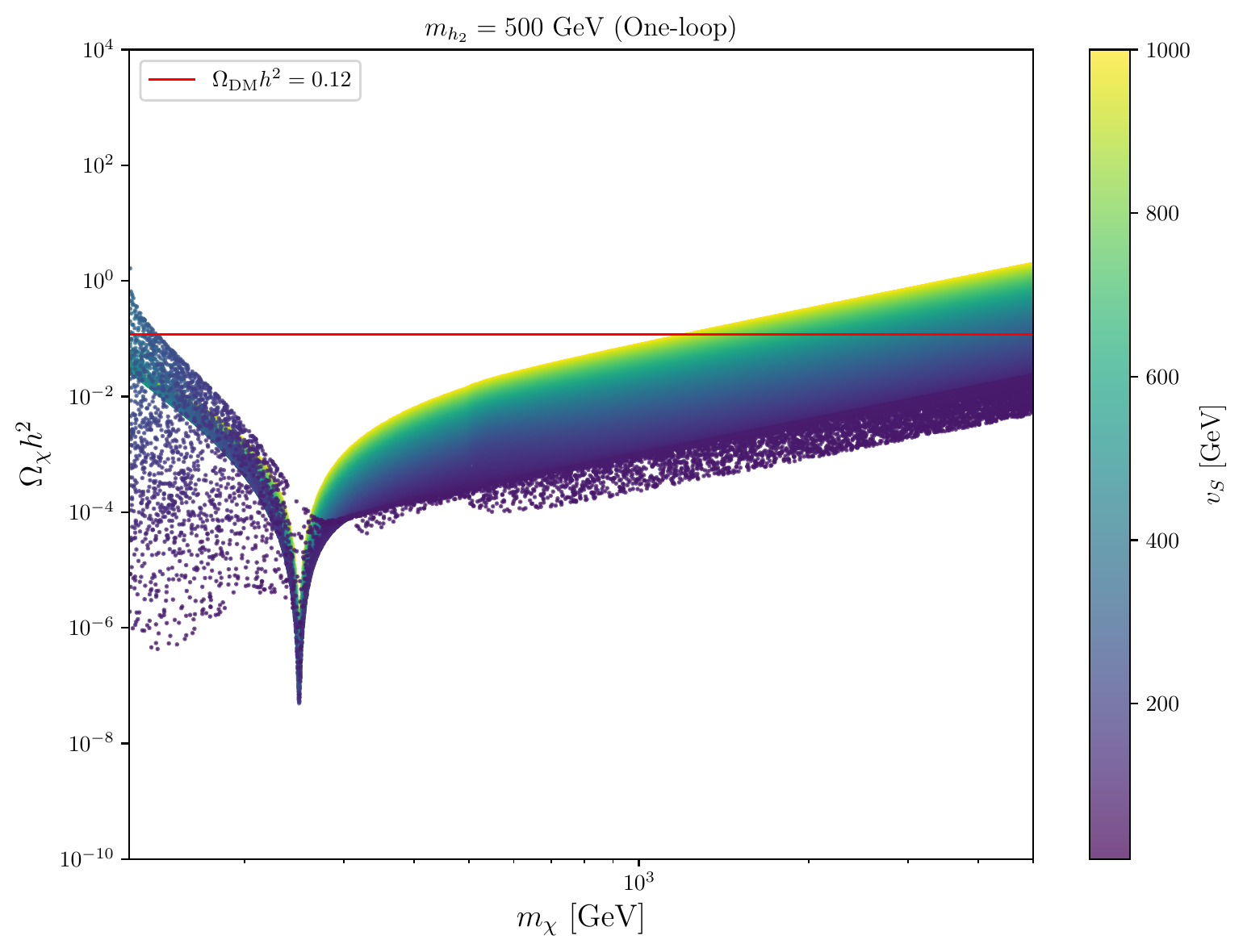}
  \end{minipage}
  
  \begin{minipage}{0.39\columnwidth}
    \centering
    \includegraphics[width=\columnwidth]{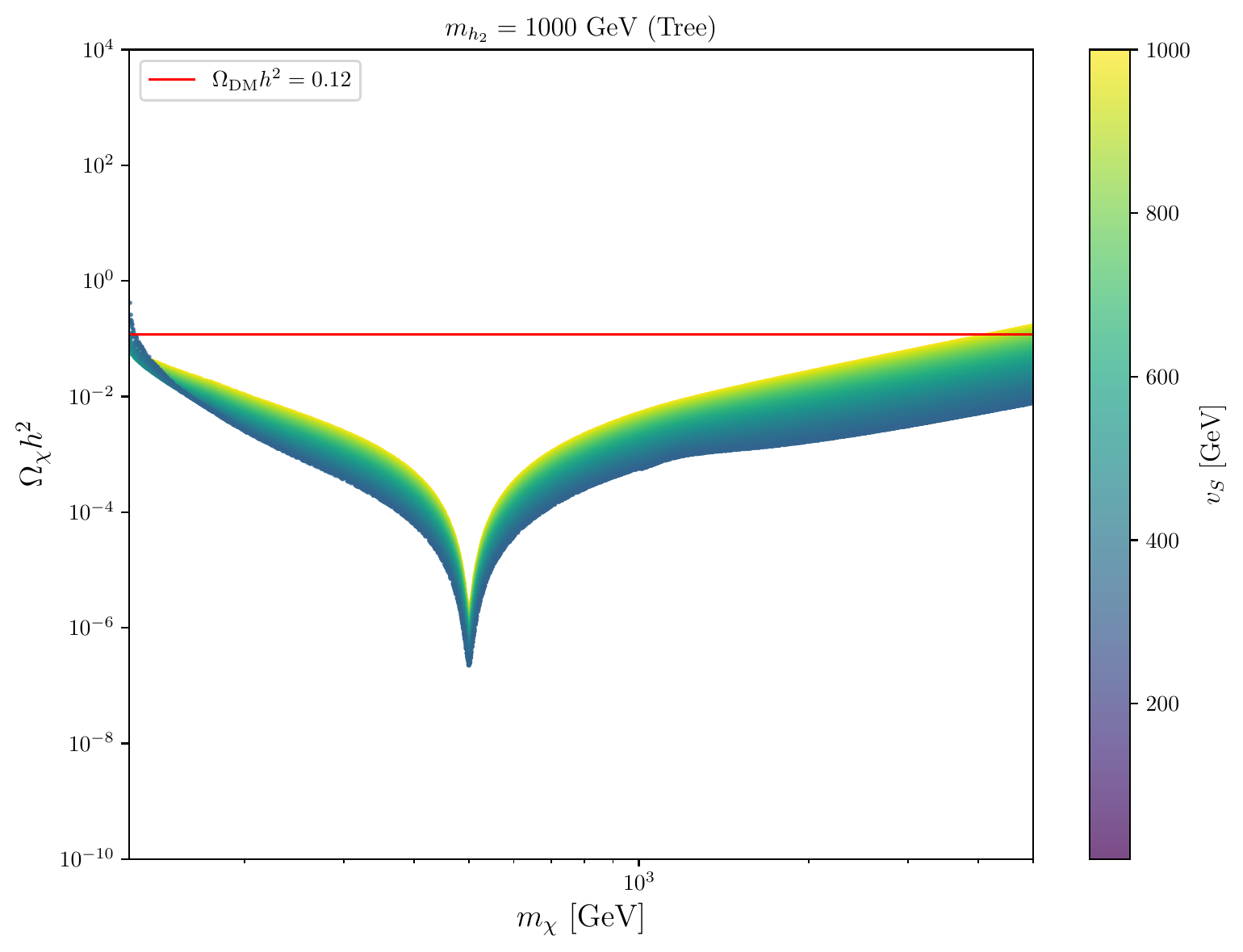}
  \end{minipage}
  \hspace{5mm}
  \begin{minipage}{0.39\columnwidth}
    \centering
    \includegraphics[width=\columnwidth]{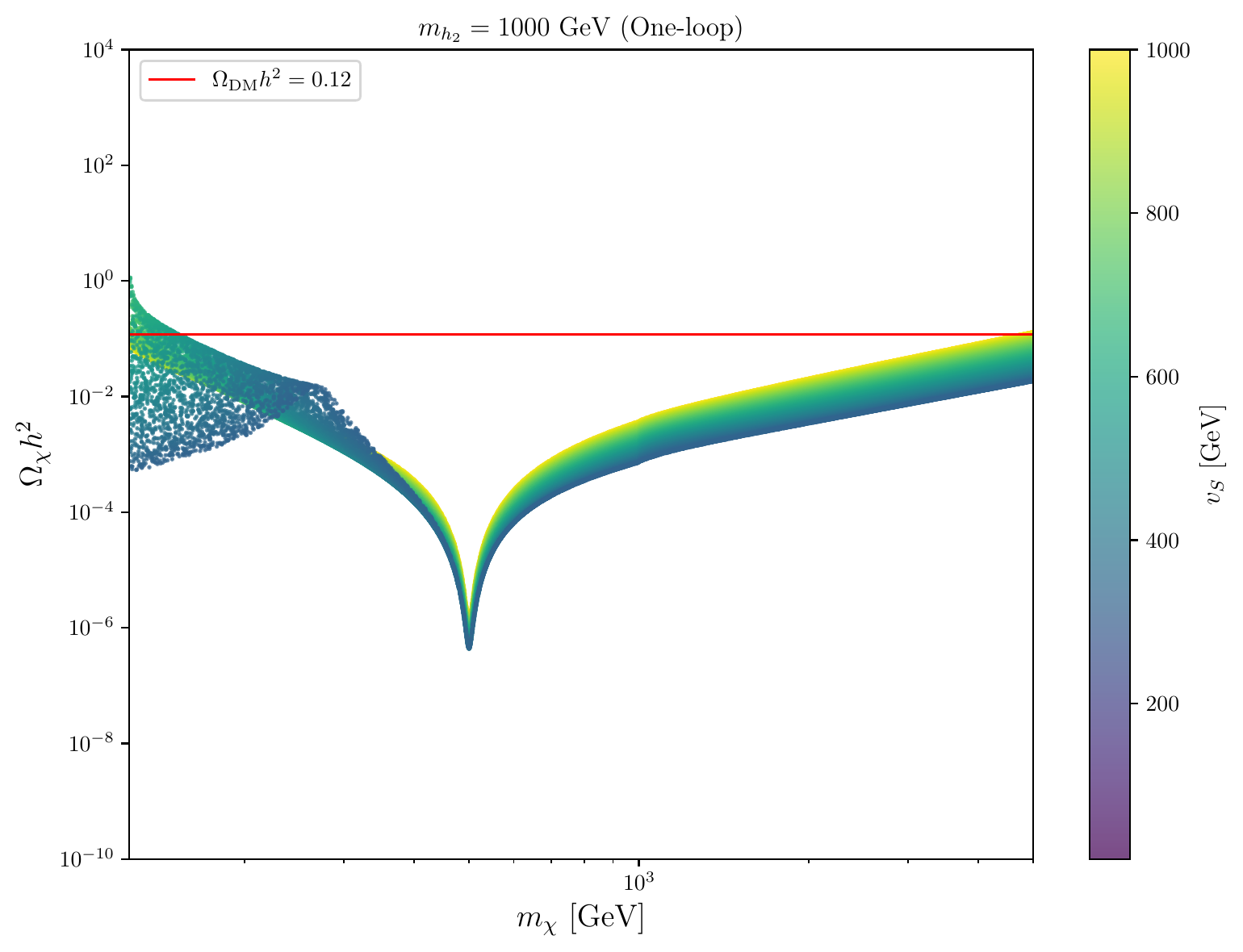}
  \end{minipage}
  
  \begin{minipage}{0.39\columnwidth}
    \centering
    \includegraphics[width=\columnwidth]{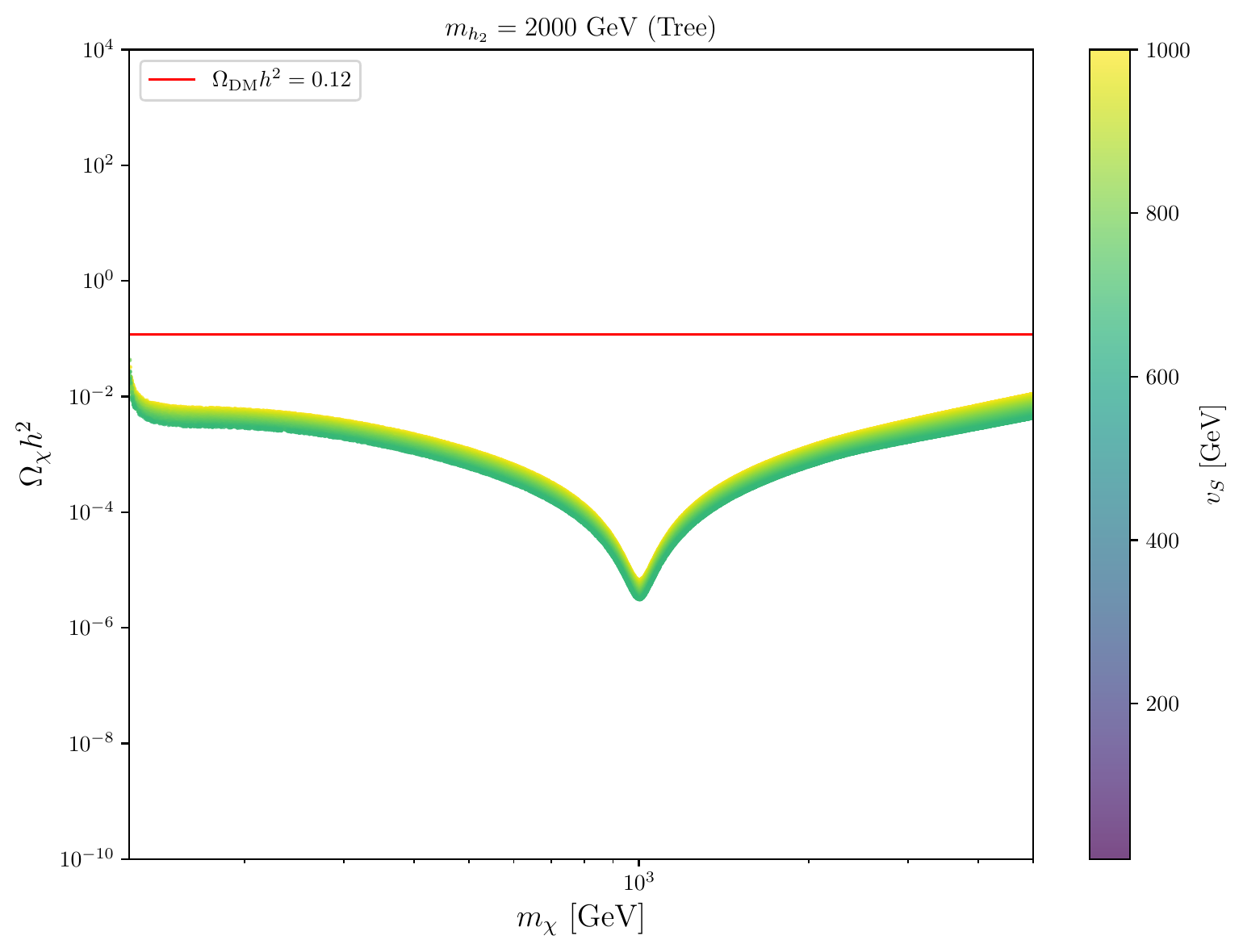}
  \end{minipage}
  \hspace{5mm}
  \begin{minipage}{0.39\columnwidth}
    \centering
    \includegraphics[width=\columnwidth]{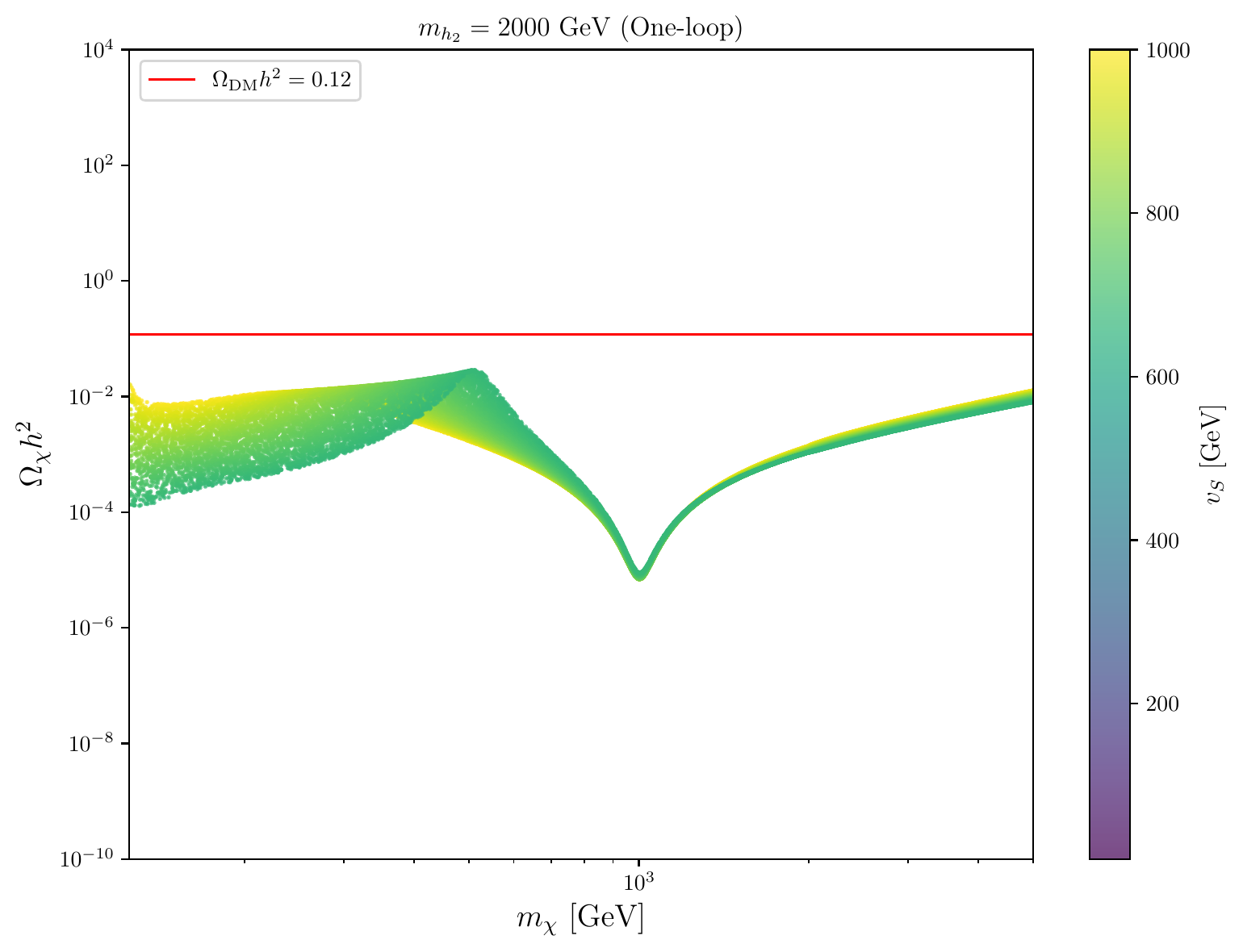}
  \end{minipage}
  \caption{%
   \setlength{\baselineskip}{14pt}
  The DM abundance as a function of the DM mass $m_\chi$. The color bar represents the change of $v_s$. We fix $a_1 = -246^3$ GeV$^3$. The horizontal red line expresses the center value of the observed DM relic density~\eqref{obsrelic}. We set $\alpha=-\pi/14$ and $m_{h_2}=126,500,1000,2000$ GeV. The left four panels show the allowed parameter regions from the tree-level analysis, while the right ones depict those obtained by including the one-loop corrections.}
    \label{fig:relic}
\end{figure}

%-------------------------------------------------------------%

In addition to the above results, we need to take into account the DM relic density Eq.~\eqref{obsrelic}~\cite{Planck:2018vyg}. We fix $a_1=-246^3$ GeV$^3$ as an example, and evaluate the relic abundance as functions of $m_\chi$ and $v_S$ using the method described in Sec.~\ref{sec:relic}, as shown in Fig.~\ref{fig:relic}. The parameter space shown here is also subject to the theoretical constraints discussed above. In particular, small values of $v_S$ are strongly restricted by the tree-level perturbative unitarity bound on $d_2$, since the magnitude of $d_2$ tends to increase as $v_S$ decreases.

A dip structure is observed when $m_\chi$ is approximately half of $m_{h_2}$, corresponding to the $s$-channel resonance in the annihilation process mediated by $h_2$. The one-loop corrections modify the effective $\chi\chi h_i$ couplings and hence the relative size of the $h_1$- and $h_2$-mediated contributions to DM annihilation into SM particles. Depending on the parameter point, these loop-induced shifts can either enhance or weaken the interference between the two mediator contributions. The loop effect can become particularly pronounced in the small-$m_\chi$ region, where the tree-level mediator contributions can partially cancel each other, making the total annihilation amplitude more sensitive to loop-induced shifts of the effective vertices.

%-------------------------------------------------------------%

\begin{figure}[h!]
  \centering
  \begin{minipage}{0.39\columnwidth}
    \centering
    
    \includegraphics[width=\columnwidth]{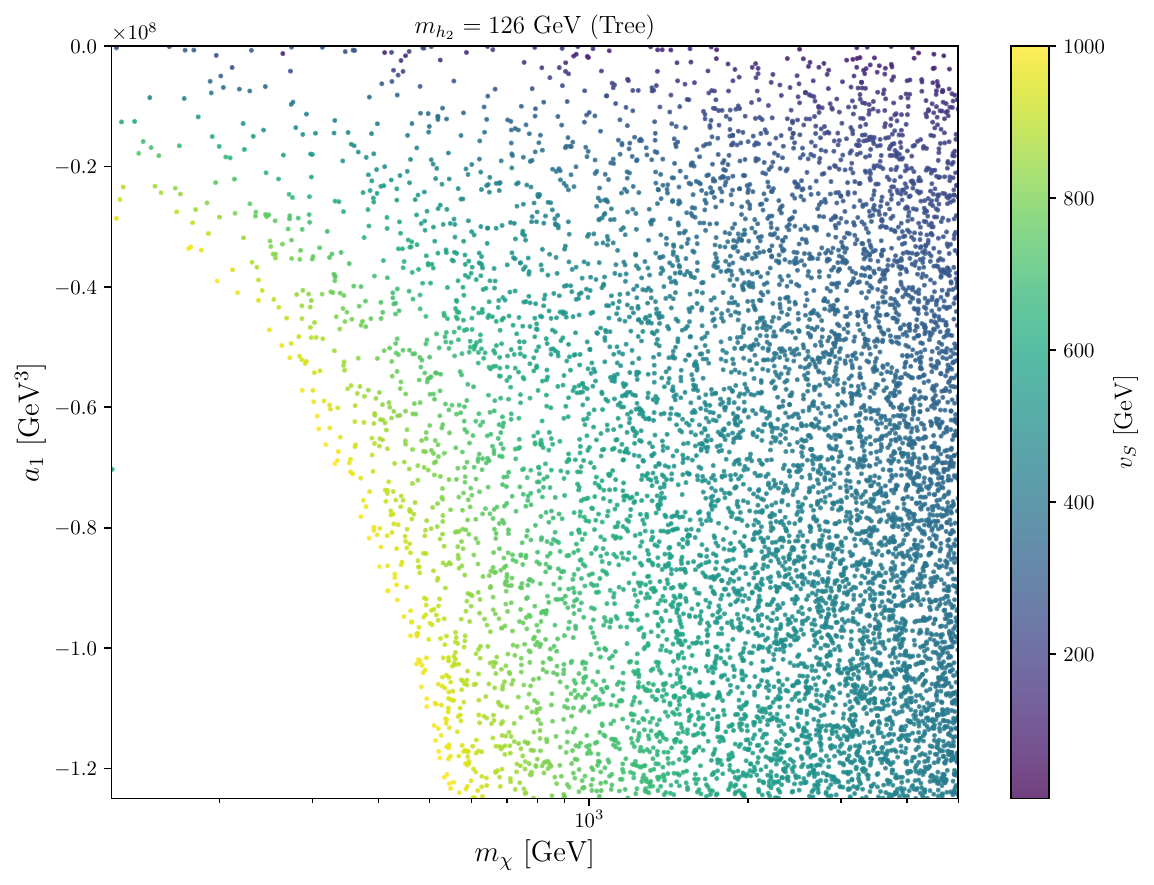}
  \end{minipage}
  \hspace{5mm}
  \begin{minipage}{0.39\columnwidth}
    \centering
    \includegraphics[width=\columnwidth]{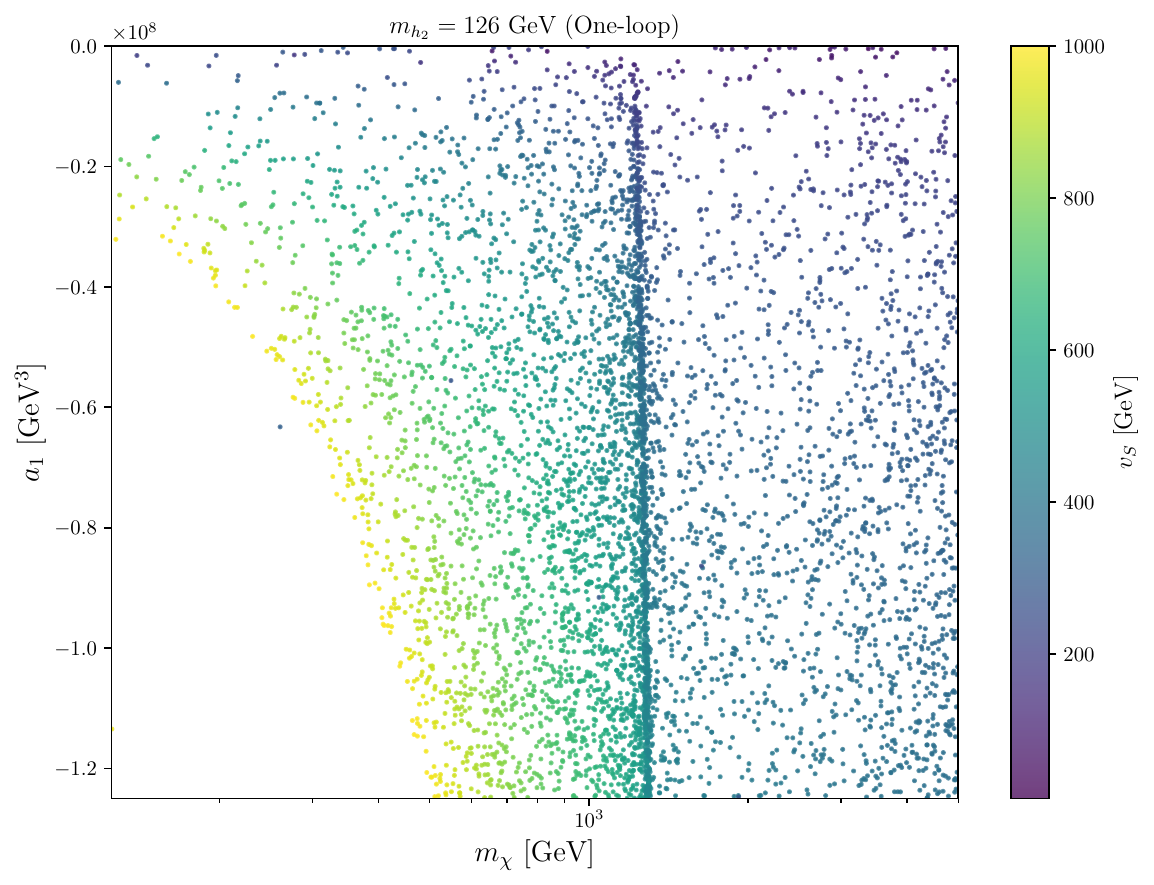}
  \end{minipage}
  
  \begin{minipage}{0.39\columnwidth}
    \centering
    \includegraphics[width=\columnwidth]{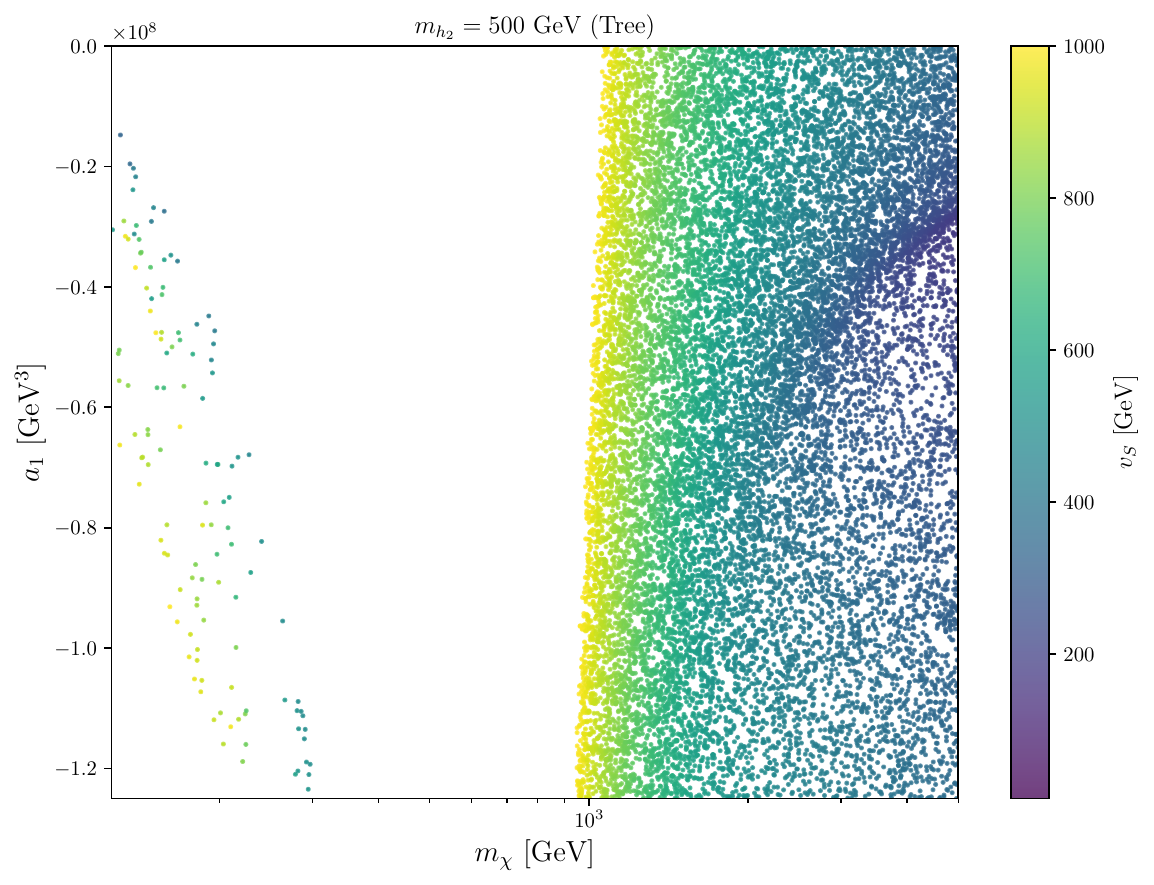}
  \end{minipage}
  \hspace{5mm}
  \begin{minipage}{0.39\columnwidth}
    \centering
    \includegraphics[width=\columnwidth]{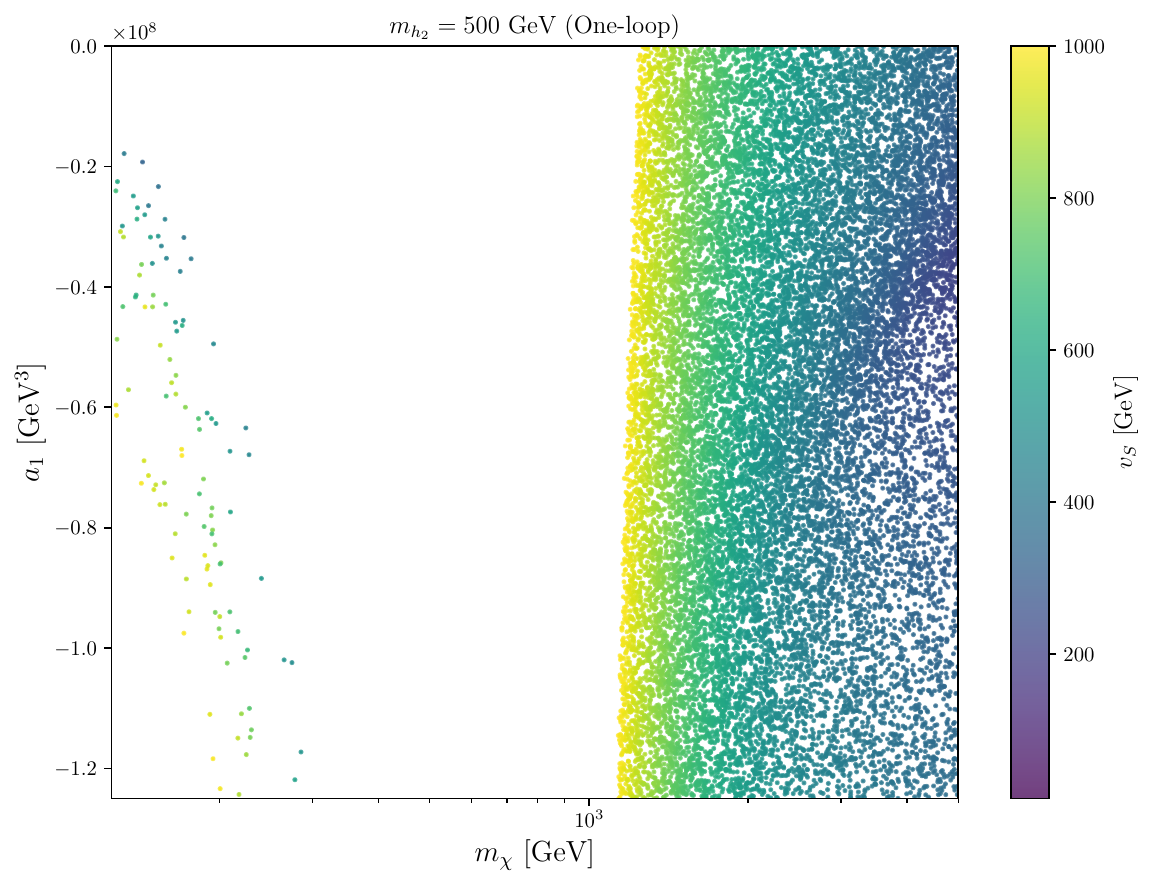}
  \end{minipage}
  
  \begin{minipage}{0.39\columnwidth}
    \centering
    \includegraphics[width=\columnwidth]{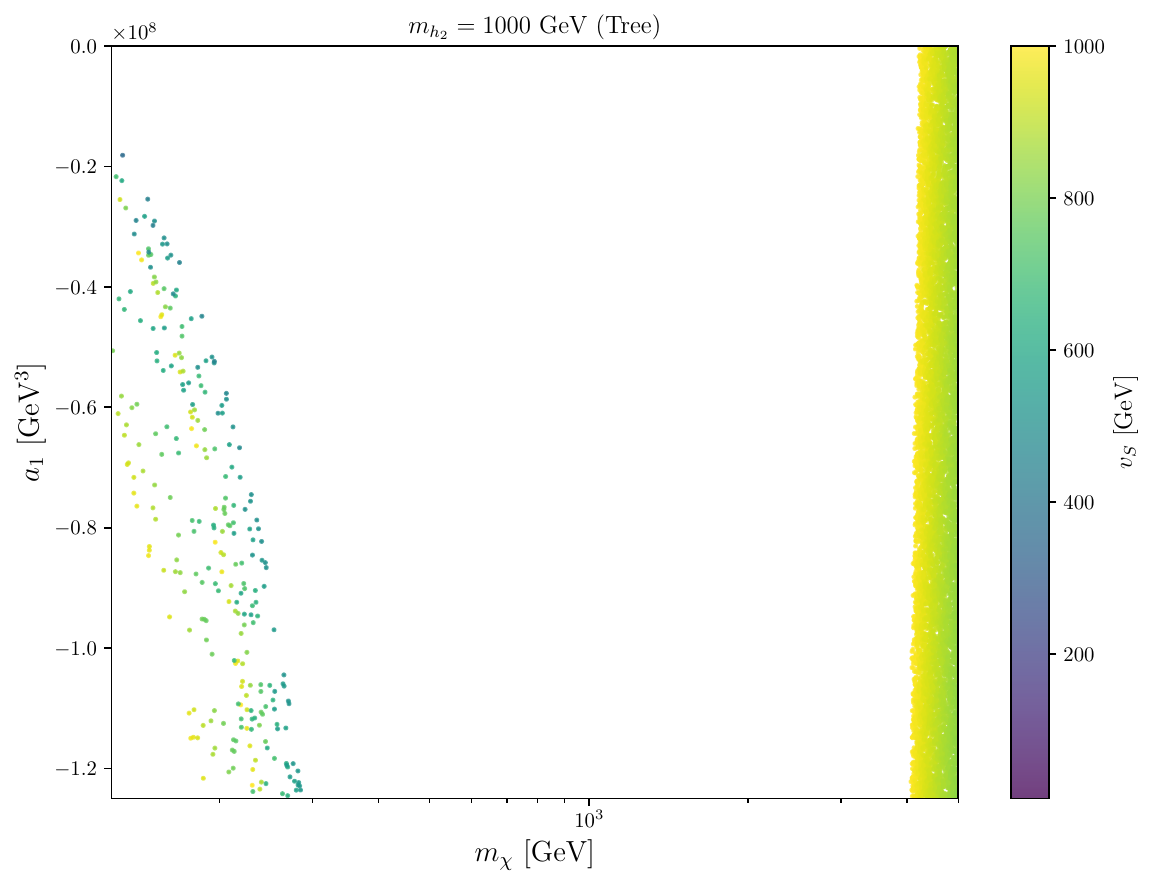}
  \end{minipage}
  \hspace{5mm}
  \begin{minipage}{0.39\columnwidth}
    \centering
    \includegraphics[width=\columnwidth]{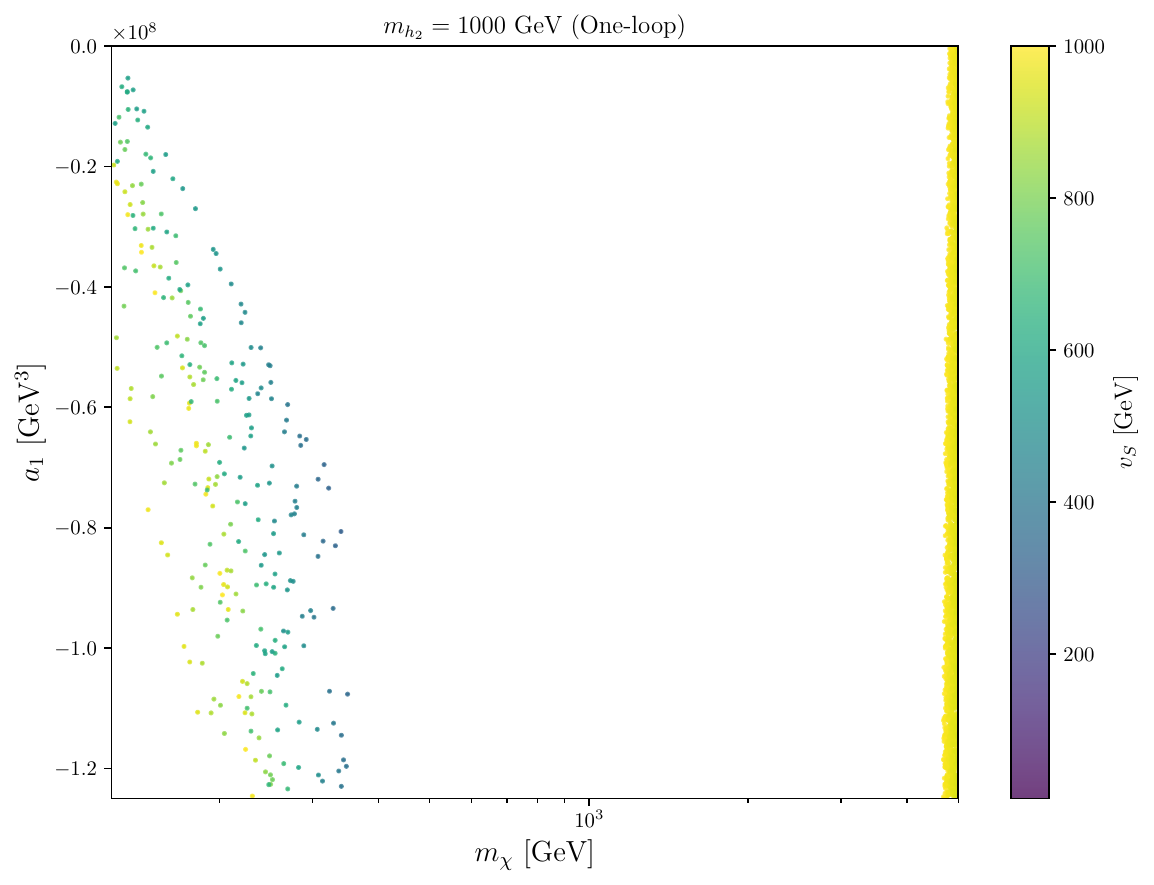}
  \end{minipage}
  
  \begin{minipage}{0.39\columnwidth}
    \centering
    \includegraphics[width=\columnwidth]{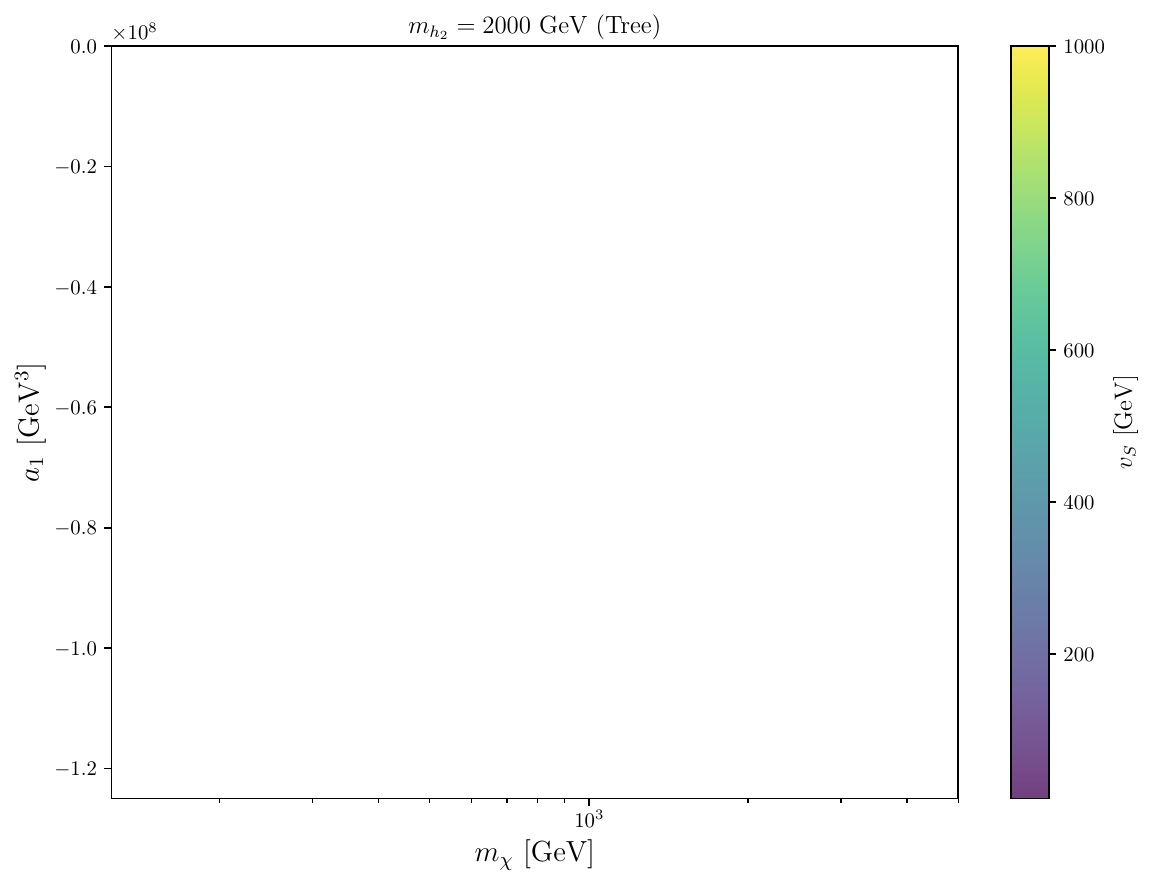}
  \end{minipage}
  \hspace{5mm}
  \begin{minipage}{0.39\columnwidth}
    \centering
    \includegraphics[width=\columnwidth]{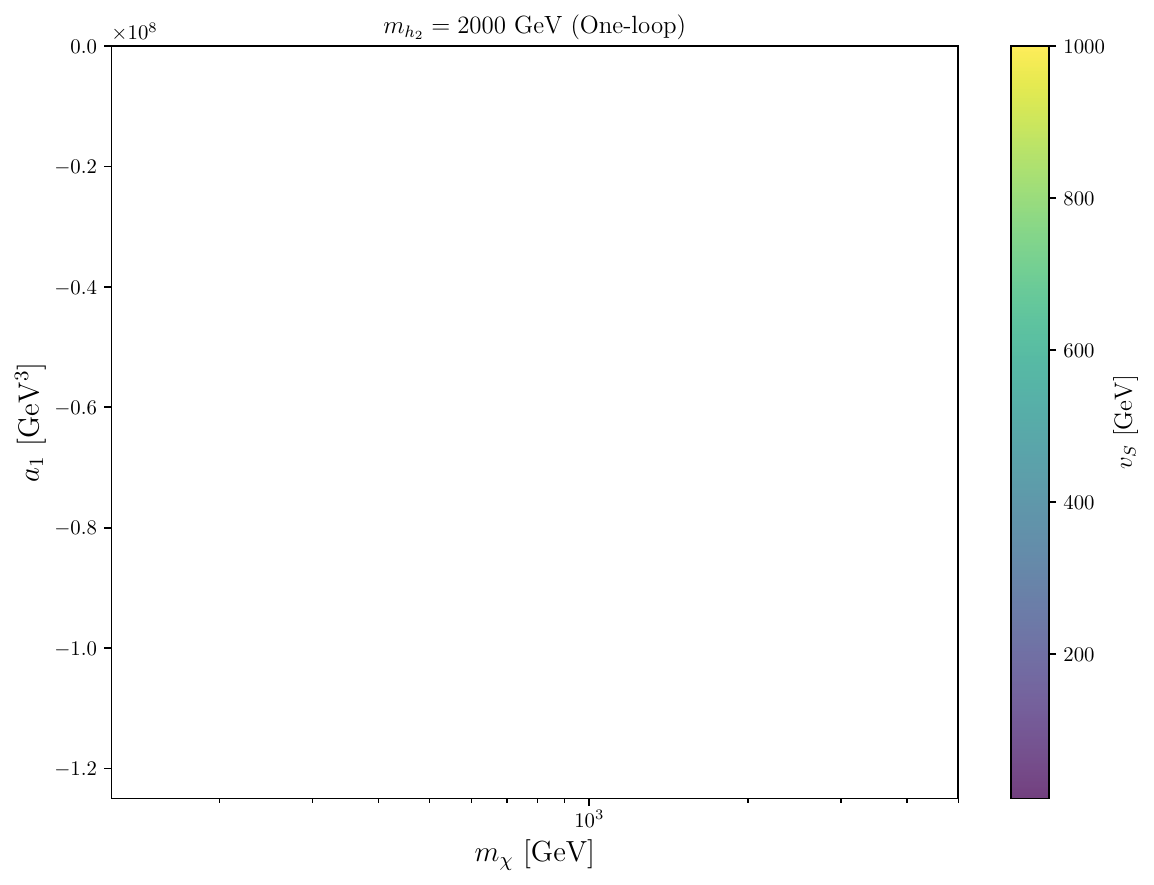}
  \end{minipage}
   \caption{%
   \setlength{\baselineskip}{14pt}
 The parameter region that satisfies the observed value of the DM relic abundance~\eqref{obsrelic} within the 2$\sigma$ level. We set $\alpha=-\pi/14$ and $m_{h_2}=126,500,1000,2000$ GeV. The left four panels show the allowed parameter regions from the tree-level analysis, while the right ones depict those obtained by including the one-loop corrections.}
    \label{fig:omega}
\end{figure}

%-------------------------------------------------------------%

Fig.~\ref{fig:omega} shows the points that satisfy the DM observed relic abundance within the 2$\sigma$ level.
The results in Fig.~\ref{fig:omega} largely reflect those in Fig.~\ref{fig:relic} and the absence of any points in the two lower panels ($m_{h_2} = 2000$ GeV) is also consistent, as the relic abundance falls below the observed value throughout the entire parameter region considered for that case.

%-------------------------------------------------------------%

\begin{figure}[h!]
  \centering
  \begin{minipage}{0.38\columnwidth}
    \centering
    
    \includegraphics[width=\columnwidth]{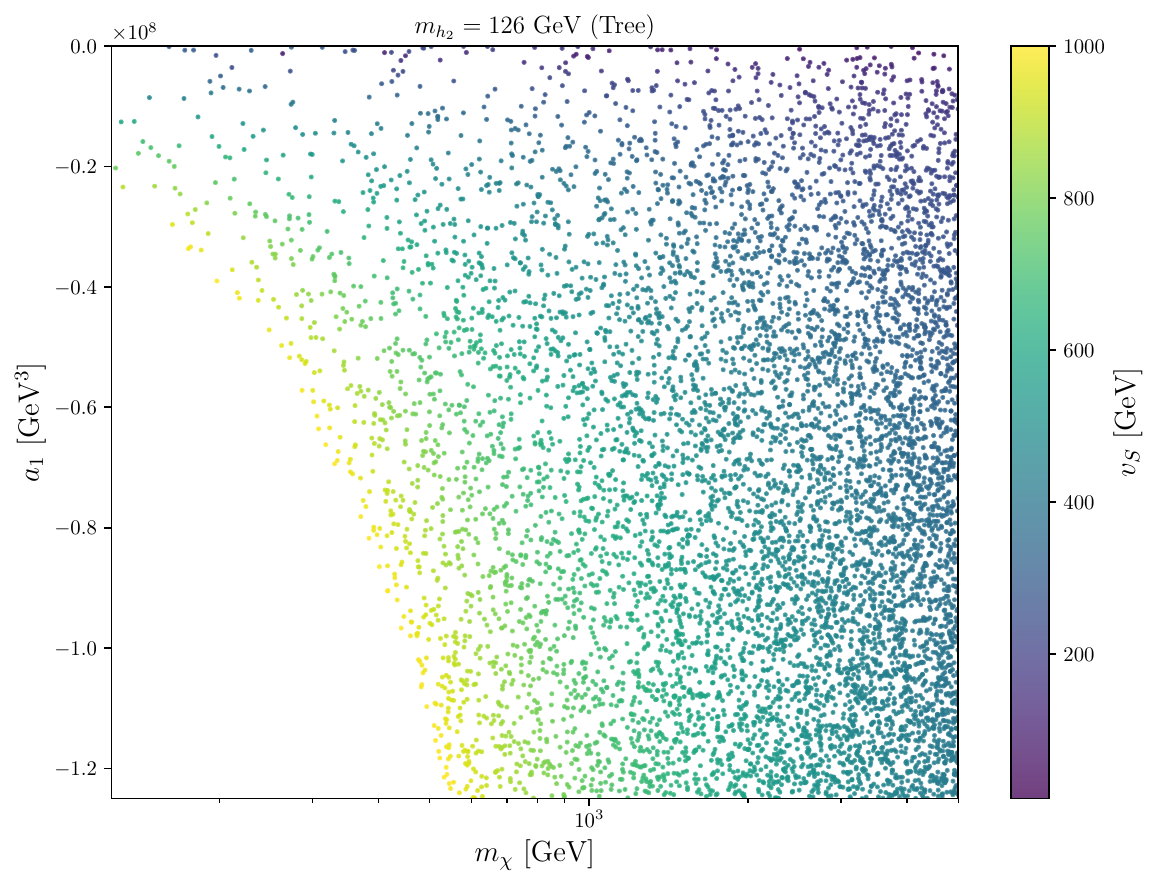}
  \end{minipage}
  \hspace{5mm}
  \begin{minipage}{0.38\columnwidth}
      \centering
    \includegraphics[width=\columnwidth]{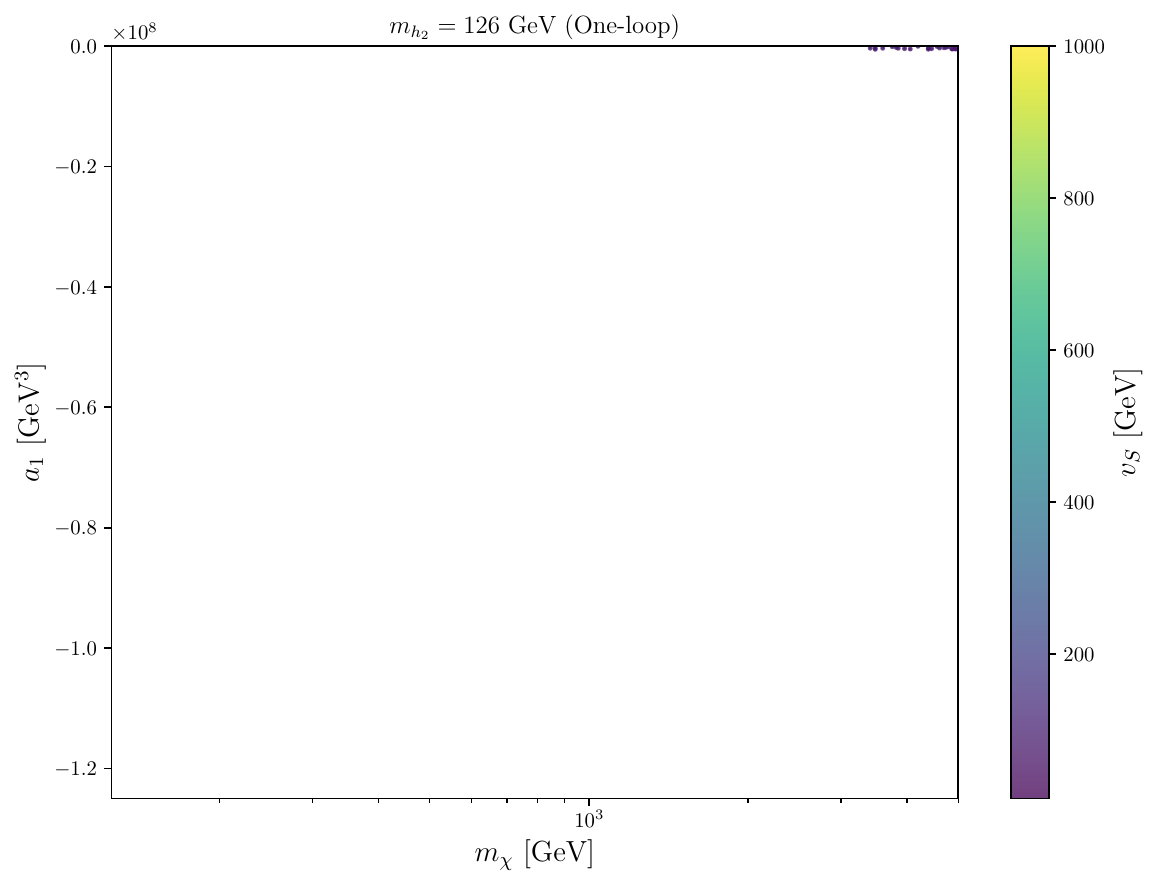}
  \end{minipage}
  
  \begin{minipage}{0.38\columnwidth}
    \centering
    \includegraphics[width=\columnwidth]{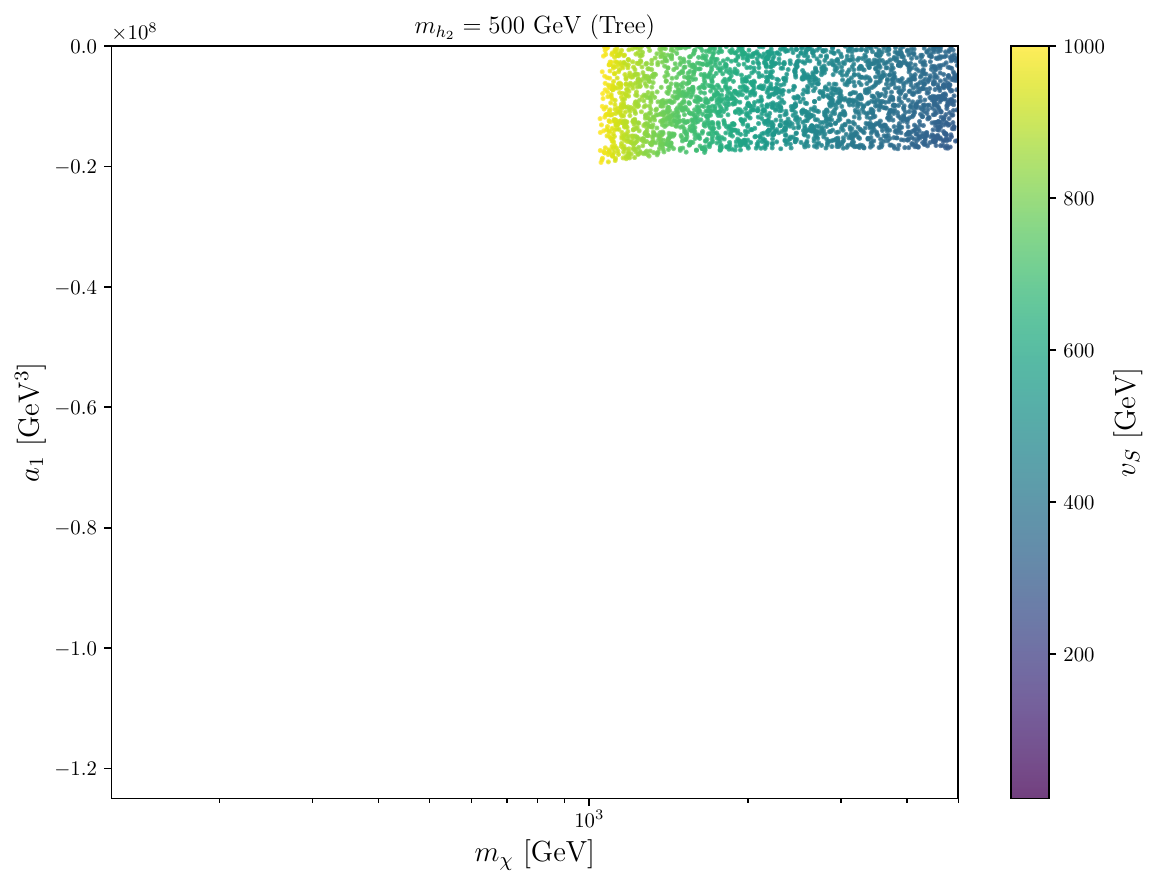}
  \end{minipage}
  \hspace{5mm}
  \begin{minipage}{0.38\columnwidth}
    \centering
    \includegraphics[width=\columnwidth]{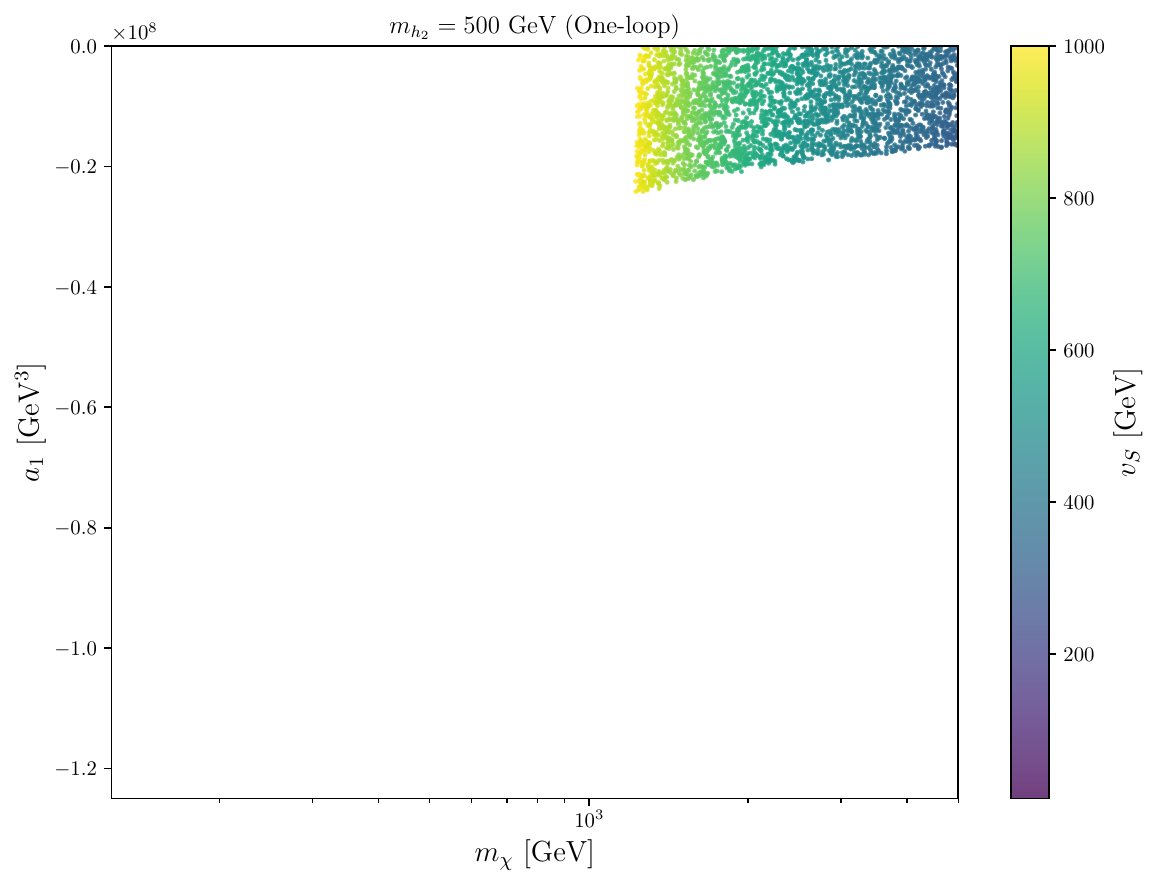}
  \end{minipage}
  
  \begin{minipage}{0.38\columnwidth}
    \centering
    \includegraphics[width=\columnwidth]{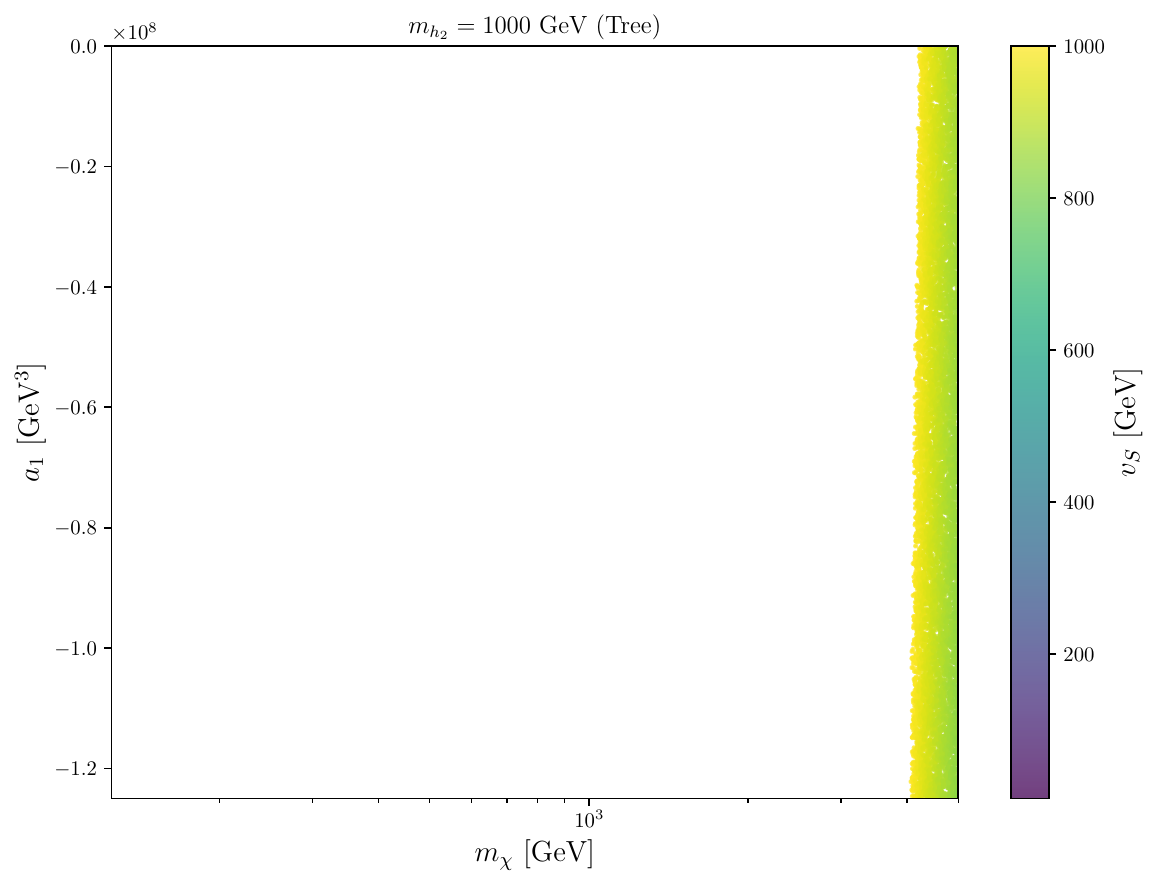}
  \end{minipage}
  \hspace{5mm}
  \begin{minipage}{0.38\columnwidth}
    \centering
    \includegraphics[width=\columnwidth]{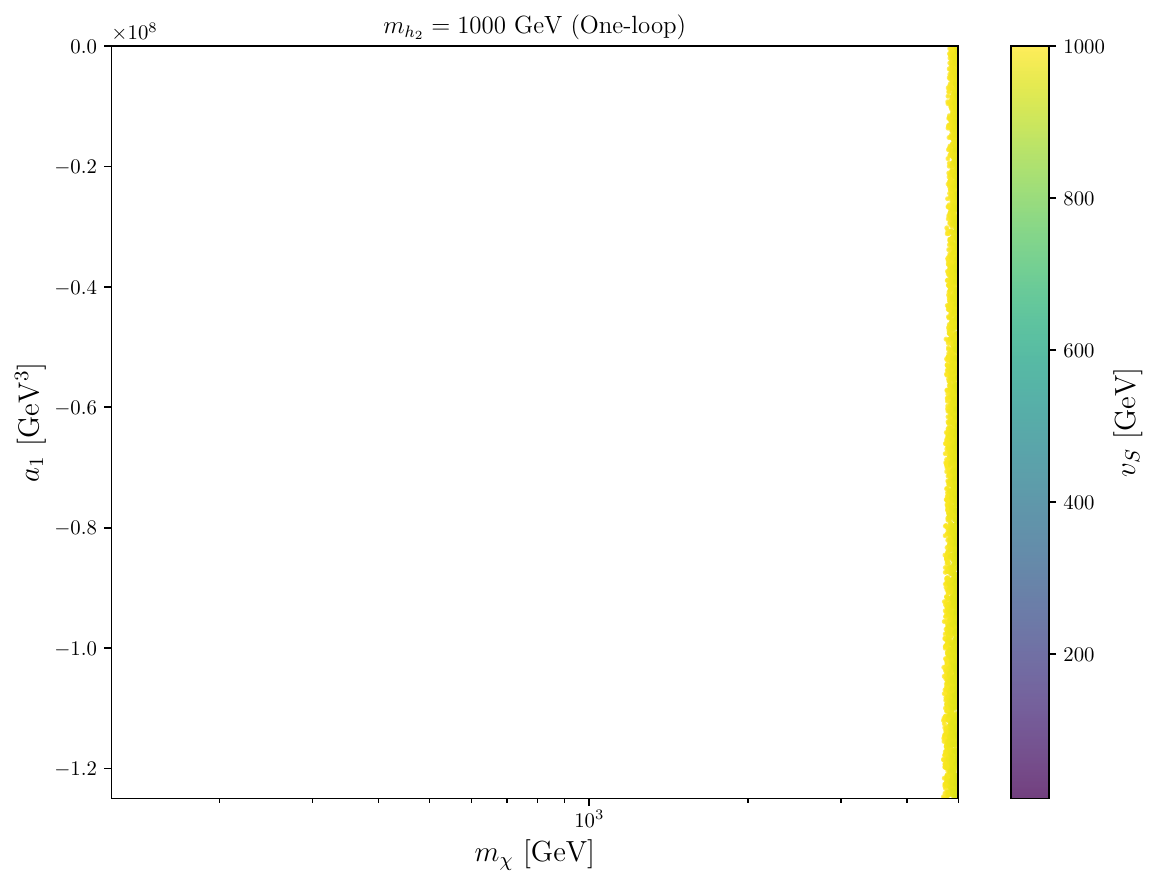}
  \end{minipage}
  
  \begin{minipage}{0.38\columnwidth}
      \centering
    \includegraphics[width=\columnwidth]{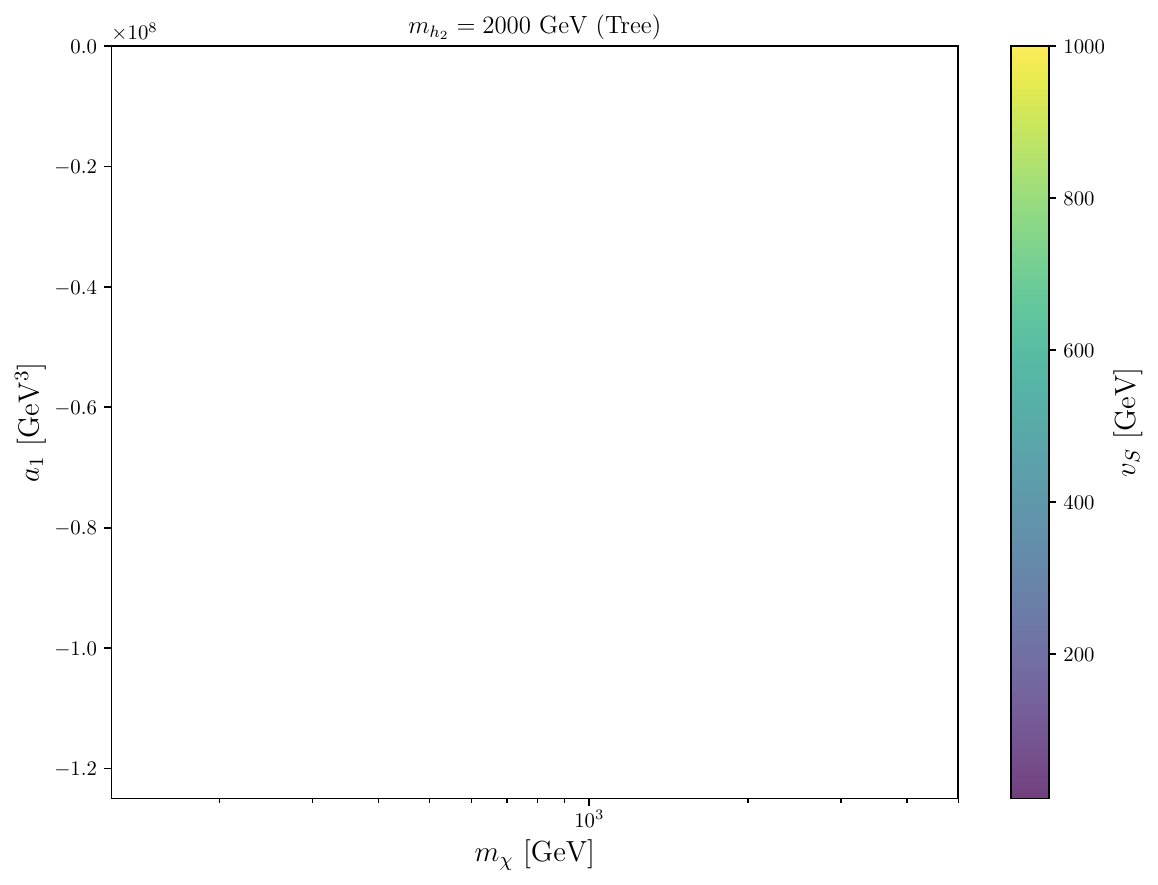}
  \end{minipage}
  \hspace{5mm}
  \begin{minipage}{0.38\columnwidth}
    \centering
    \includegraphics[width=\columnwidth]{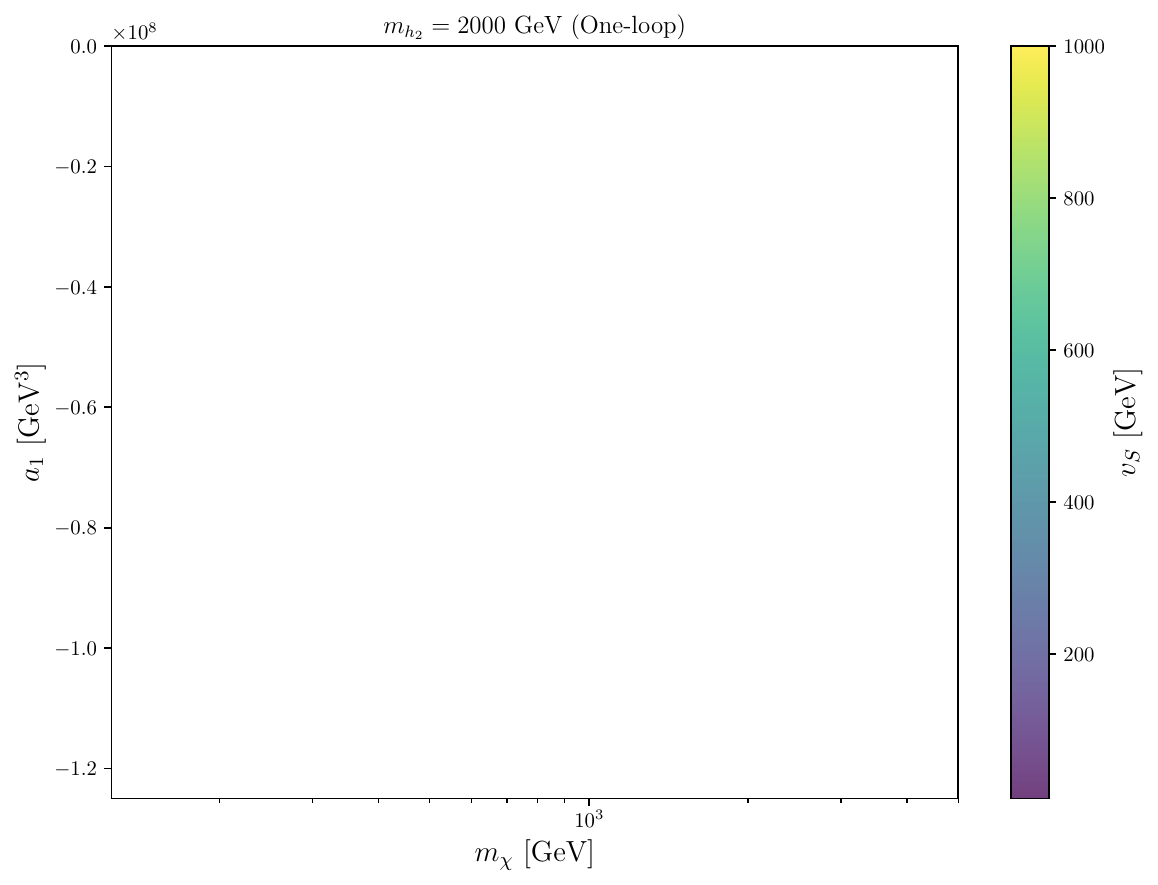}
  \end{minipage}
  \caption{%
   \setlength{\baselineskip}{14pt}
  The parameter region that satisfies (1) the results of the LZ experiment and, (2) the observed relic abundance within the 2$\sigma$ level. We set $\alpha=-\pi/14$ and $m_{h_2}=126,500,1000,2000$ GeV. The left four panels show the allowed parameter regions from the tree-level analysis, while the right ones depict those obtained by including the one-loop corrections.}
    \label{fig:all}
\end{figure}

%-------------------------------------------------------------%

Finally, Fig.~\ref{fig:all} presents the parameter regions that satisfy (1) the constraints from the LZ experiment, and (2) the observed relic abundance within the 2$\sigma$ level. In the tree-level analysis, the largest allowed region is obtained for $m_{h_2}=126$ GeV.
On the other hand, in the one-loop analysis, the allowed region shrinks drastically since far fewer parameter points satisfy the direct detection bound than at the tree level.
For $m_{h_2}=500,1000$ GeV, the allowed region still remains in both the tree-level and one-loop analyses.

%%%%%%%%%%%%%%%%%%%%%%%%%%%%%%%%%%%%%%%%%%%%%%%%
%				Summary 
%%%%%%%%%%%%%%%%%%%%%%%%%%%%%%%%%%%%%%%%%%%%%%%%

\section{Summary}\label{sec:sum}

We have analyzed the parameter space of the CxSM, a simple extension that includes a WIMP DM candidate. Our analysis explicitly incorporates the effects of the one-loop corrections, enabling a precise evaluation of theoretical and phenomenological constraints on the model. We have taken into account not only the counterterm contributions but also the derivatives of the one-loop corrections, and have considered two key DM experimental/observational constraints: the stringent bounds on the DM-quark scattering from the DM direct detection experiments, and the requirement to reproduce the observed DM relic abundance. The effects of the one-loop corrections to the vertices involving the $\chi$-particle and the counterterms for the effective couplings appear in the vertices relevant to both the scattering and annihilation processes.

We have performed a parameter space analysis of the CxSM using $m_\chi$, $a_1$, and $v_S$ as variables, while setting $m_{h_2}$ to four benchmark values. We found a larger $m_{h_2}$ enhances the effective couplings, which tend to reduce the allowed region from the DM direct detection experiments, while increasing the parameter space consistent with the observed relic density. For $m_{h_2} = 126$ GeV, we observed that a sizable allowed region already exists at the tree level, which reflects the cancellation mechanism in the degenerate scalar scenario, in which the mediator particles of the DM-quark scattering have nearly degenerate masses.

Combining these results, we found that after imposing all constraints, in the one-loop analysis, the allowed region becomes somewhat smaller compared to the tree-level case, and in the specific case of $m_{h_2}=$126 GeV it is drastically reduced.
Yet the degenerate scalar scenario, which works most effectively for $m_{h_2}=$126 GeV at the tree level, may become equally effective for another choice of $m_{h_2}$ with the radiatively corrected couplings.
Therefore, once the one-loop effects are taken into account, it becomes rather challenging to find parameter regions that satisfies all the conditions. To address this issue, introducing scalar cubic terms is expected to be effective. For example, by extending the scalar potential to include parameters ${c_1} H^{\dagger} H(S+\text { h.c.})$, ${c_2}|S|^2(S+\text { h.c.})$ and ${c_3}(S+\text { h.c.})^3$, one can treat the parameter $b_2 - b_1$ and the effective vertices of $\chi$-$\chi$-$h_i$, $D_i$, (associated with the DM direct detection experiments and the relic density and specific expressions are given by Eqs.~\eqref{DD1} and \eqref{DD2}) as free parameters, which would make it easier to find regions that satisfy all the required conditions. Moreover, the cubic terms are also known to be effective in realizing a strong first-order electroweak phase transition, which is essential for electroweak baryogenesis and gravitational wave production. A detailed analysis of the CxSM with cubic terms is left for future work.

%---------------------------------------------------------------------------------------------------
\begin{figure}[htpb]
\center
\includegraphics[width=8cm]{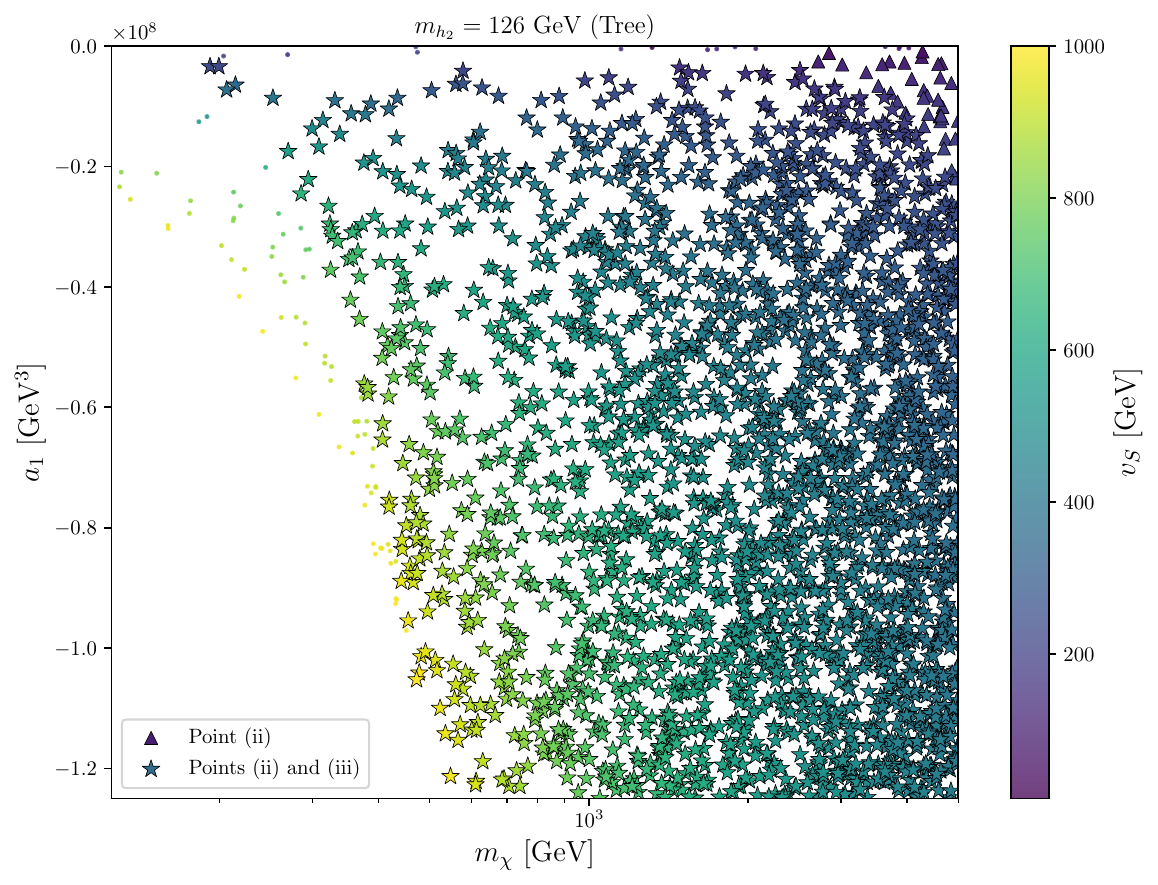}
\caption{For $m_{h_2}=126$ GeV, the stationary points appear within the allowed region of the parameter space shown in the left top panel of Fig.~\ref{fig:all}. The triangle markers correspond to the occurrence of Point (ii) \eqref{eq:branch_v0}, while the star markers correspond to the occurrence of both Points (ii) \eqref{eq:branch_v0} and (iii) \eqref{eq:branch_vneq0}. }
\label{fig:stationary}
\end{figure}
%---------------------------------------------------------------------------------------------------

In addition, as a future direction, it is important to investigate the vacuum structure more systematically in order to identify parameter regions where the discrete symmetry ensuring the stability of DM is not broken. At the tree level, when $m_{h_2}=500,1000$ GeV, no stationary point with $\varphi_\chi \neq 0$ exists for all the allowed parameters (see Eqs.~\eqref{eq:branch_v0} and \eqref{eq:branch_vneq0} for the precise conditions.). In contrast, for $m_{h_2}=$126 GeV, a lot of such stationary points appear as shown in Fig.~\ref{fig:stationary}. We explicitly compared the potential values at the stationary points with that of the desired vacuum $(v, v_{S}, 0)$ and found that $V_0(v, v_{S}, 0)$ is always the global minimum in the parameter space considered, thereby ensuring that the $\mathbb{Z}_2$ symmetry is preserved.

As another important direction for future work, we are extending the present analysis toward more complete loop calculations for both direct detection and DM annihilation. For direct detection, we are improving the present treatment by incorporating the full set of loop-induced contributions to the DM-quark and DM-gluon interactions, including self-energy, vertex, box, and triangle contributions, together with a systematic matching onto the effective operators relevant at the hadronic scale. For the annihilation processes, the full momentum dependence of the one-loop amplitudes should be retained, including both one-particle-irreducible vertex corrections and one-particle-reducible contributions involving the scalar propagators. A consistent treatment also requires radiative corrections associated with the renormalization of the mixing angle $\alpha$ and the wave-function renormalization factors. These improvements will allow a more precise and consistent evaluation of both the spin-independent DM-nucleon scattering cross section and the relic abundance. A detailed analysis including these effects is currently in progress.

%%%%%%%%%%%%%%%%%%%%%%%%%%%%%%%%%%%%%%%%%%%%%%%%
%			    	Appendix
%%%%%%%%%%%%%%%%%%%%%%%%%%%%%%%%%%%%%%%%%%%%%%%%

\appendix
\section{Derivatives of the one-loop correction to the effective potential} \label{app:deriv}

The first and second derivatives of the one-loop corrections are taken with respect to the background field and are written as follows:
\begin{align}
\frac{\partial V_{\mathrm{CW}}}{\partial \varphi_a} & =\sum_i n_i \frac{\partial \bar{m}_i^2}{\partial \varphi_a} \frac{\bar{m}_i^2}{32 \pi^2}\left(\log \frac{\bar{m}_i^2}{\bar{\mu}^2}-\gamma_i+\frac{1}{2}\right), \\
\frac{\partial^2 V_{\mathrm{CW}}}{\partial \varphi_a \partial \varphi_b} & =\sum_i n_i\left[\frac{\partial^2 \bar{m}_i^2}{\partial \varphi_a \partial \varphi_b} \frac{\bar{m}_i^2}{32 \pi^2}\left(\log \frac{\bar{m}_i^2}{\bar{\mu}^2}-\gamma_i+\frac{1}{2}\right)+\frac{\partial \bar{m}_i^2}{\partial \varphi_a} \frac{\partial \bar{m}_i^2}{\partial \varphi_b} \frac{1}{32 \pi^2}\left(\log \frac{\bar{m}_i^2}{\bar{\mu}^2}-\gamma_i+\frac{3}{2}\right)\right],
\end{align}
where $\bar{m}_i$ denotes the field-dependent mass of particle $i$. $\gamma_i=\frac{3}{2}$ for the scalars and fermions while $\gamma_i=\frac{5}{6}$ for the gauge bosons.
The first derivatives with respect to $\varphi$ and $\varphi_S$ for $\varphi_\chi=0$ are explicitly given by
\begin{align}
\frac{\partial V_{\mathrm{CW}}}{\partial \varphi} =\ & \frac{1}{\varphi} \sum_{\substack{i = t, b, c, \tau \\ W, Z}} \frac{n_i m_i^4}{16\pi^2} \left( \log \frac{m_i^2}{\bar{\mu}^2} - \gamma_i + \frac{1}{2} \right)
+ \sum_{i= h_2, \chi}\frac{\partial \bar{m}_i^2}{\partial \varphi}\cdot \frac{m_i^2}{32 \pi^2}\left(\log \frac{m_i^2}{\bar{\mu}^2}-1\right), \\
\frac{\partial V_{\mathrm{CW}}}{\partial \varphi_S}=\ &\sum_{i=h_2, \chi}\frac{\partial \bar{m}_i^2}{\partial \varphi_S}\cdot \frac{m_i^2}{32 \pi^2}\left(\log \frac{m_i^2}{\bar{\mu}^2}-1\right),\\
\frac{\partial^2 V_{\mathrm{CW}}}{\partial \varphi^2} =\ & \frac{1}{\varphi^2} \sum_{\substack{i = t, b, c, \tau \\ W, Z}} \frac{3 n_i m_i^4}{16\pi^2} \left( \log \frac{m_i^2}{\bar{\mu}^2} - \gamma_i + \frac{7}{6} \right) \\
& +\sum_{i=h_2, \chi}\left[\frac{\partial^2 \bar{m}_i^2}{\partial \varphi^2}\cdot \frac{m_i^2}{32 \pi^2}\left(\log \frac{m_i^2}{\bar{\mu}^2}-1\right)+\left(\frac{\partial \bar{m}_i^2}{\partial \varphi}\right)^2\cdot \frac{1}{32 \pi^2}\log \frac{m_i^2}{\bar{\mu}^2}\right]  ,\\
\frac{\partial^2 V_{\mathrm{CW}}}{\partial \varphi_S^2}
= \ & 
\sum_{i=h_2, \chi}\left[\frac{\partial^2 \bar{m}_i^2}{\partial \varphi_S^2}\cdot \frac{m_i^2}{32 \pi^2}\left(\log \frac{m_i^2}{\bar{\mu}^2}-1\right)+\left(\frac{\partial \bar{m}_i^2}{\partial \varphi_S}\right)^2\cdot \frac{1}{32 \pi^2}\log \frac{m_i^2}{\bar{\mu}^2}\right], \\
\frac{\partial^2 V_{\mathrm{CW}}}{\partial \varphi \, \partial \varphi_S}
= \ & 
\sum_{i=h_2, \chi}\left[ \frac{\partial^2 \bar{m}_i^2}{\partial \varphi \, \partial \varphi_S} \cdot \frac{m_i^2}{32\pi^2} \left( \log \frac{m_i^2}{\bar{\mu}^2} - 1 \right)
+ \frac{\partial \bar{m}_i^2}{\partial \varphi} \cdot \frac{\partial \bar{m}_i^2}{\partial \varphi_S} \cdot \frac{1}{32\pi^2} \log \frac{m_i^2}{\bar{\mu}^2} \right]
, \\
\frac{\partial V_{\mathrm{CW}}}{\partial \varphi_\chi}=\ &  \sum_{i=h_2, \chi}\frac{\partial \bar{m}_i^2}{\partial \varphi_\chi} \cdot \frac{m_i^2}{32\pi^2}
\left( \log \frac{m_i^2}{\bar{\mu}^2} - 1 \right)= 0,\\
\frac{\partial^2 V_{\mathrm{CW}}}{\partial \varphi_\chi^2}
= \ &  \sum_{i=h_2, \chi} \frac{\partial^2 \bar{m}_i^2}{\partial \varphi_\chi^2} \cdot \frac{m_i^2}{32\pi^2} \left( \log \frac{m_i^2}{\bar{\mu}^2} - 1 \right), \\
\frac{\partial^2 V_{\mathrm{CW}}}{\partial \varphi \, \partial \varphi_\chi}
= \ & \frac{\partial^2 V_{\mathrm{CW}}}{\partial \varphi_S \, \partial \varphi_\chi} = 0,
\end{align}
where the contributions from $h_1$-loop is omitted since the field-dependent mass squared of $h_1$ is negative for some region in ($\varphi$, $\varphi_S$)-plane.

In this model, the field-dependent mass-squared of the SM particles depend only on $\varphi$:
\begin{align}
\bar{m}_W^2&=\frac{g_2^2}{4} \varphi^2,\\
\quad \bar{m}_Z^2&=\frac{g_2^2}{4 \cos ^2 \theta_W} \varphi^2,\\
\quad \bar{m}_f^2&=\frac{y_f^2}{2} \varphi^2.
\end{align}
The symbol $f$ denotes the SM fermion, and in this context, we consider the top, bottom, charm quark, and the tau lepton.
$g_2$ denotes the SU(2)$_L$ gauge coupling constant, $\theta_W$ is the weak mixing angle, and $y_f$ represents the Yukawa coupling of the SM fermion $f$.
The field-dependent mass-squared of the scalar bosons are given by
\begin{align}
\bar{\mathcal{M}}_\phi^2=\left(\begin{array}{ccc}
\frac{m^2}{2}+\frac{3 \lambda}{4} \varphi^2+\frac{\delta_2}{4} \varphi_{S}^2 & \frac{\delta_2}{2} \varphi \varphi_S & 0 \\
\frac{\delta_2}{2} \varphi \varphi_S & \frac{b_2+b_1}{2}+\frac{3 d_2}{4} \varphi_S^2+\frac{\delta_2}{4} \varphi^2 & 0 \\
0 & 0 & \frac{b_2-b_1}{2}+\frac{d_2}{4} \varphi_S^2+\frac{\delta_2}{4} \varphi^2
\end{array}\right).
\end{align}
We can clearly see that
\begin{align}
\bar{m}_{h_1, h_2}^2= & \frac{1}{2}\left[\frac{m^2+b_2+b_1}{2}+\frac{3 \lambda+\delta_2}{4} \varphi^2+\frac{3 d_2+\delta_2}{4} \varphi_S^2\right. \nonumber \\
& \left. \pm \sqrt{\left(\frac{m^2-b_2-b_1}{2}+\frac{3 \lambda-\delta_2}{4} \varphi^2+\frac{\delta_2-3 d_2}{4} \varphi_S^2\right)^2+\left(\delta_2 \varphi \varphi_S\right)^2}\right],\\
\bar{m}_\chi^2=&\frac{b_2-b_1}{2}+\frac{d_2}{4} \varphi_S^2+\frac{\delta_2}{4} \varphi^2.
\end{align}

%%%%%%

\section{DM annihilation} \label{app:crosssection}
In this section, we present the scattering amplitudes and the final-state momentum integrals of the DM annihilation cross sections for each channel.

\subsection{$\chi\chi\to VV$}
%----------------------------------------------------------------------------------------------------------------------------------
\begin{figure}[htpb]
\center
\includegraphics[width=15cm]{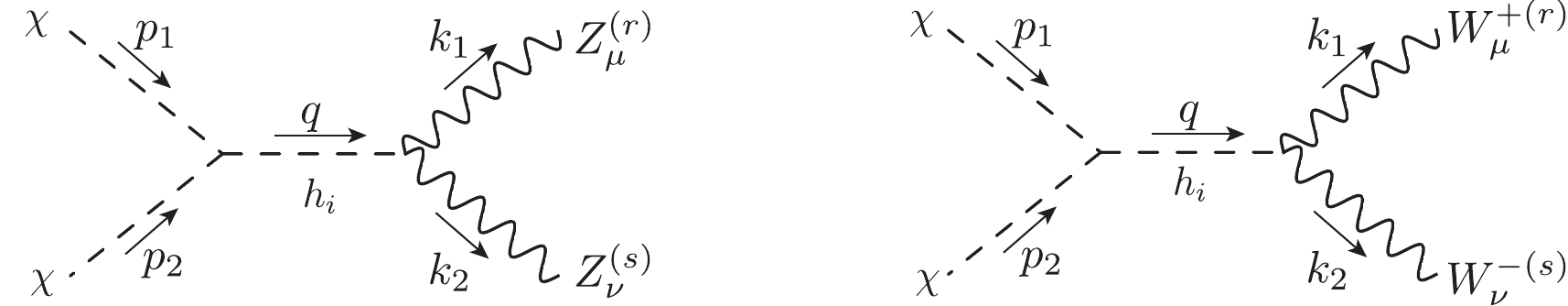}
\caption{Feynman diagram of the scattering process $\chi \chi \to VV $. }
\label{fig:chichiVV}
\end{figure}
%----------------------------------------------------------------------------------------------------------------------------------
The annihilation into a pair of gauge bosons $Z,W$ is an $s$-channel process mediated by $h_i$ as shown in Fig.~\ref{fig:chichiVV}. The amplitudes are given by
\begin{align}
i \mathcal{M}{(\chi \chi \rightarrow Z Z)} & =\frac{m_Z^2 \varphi^2}{v^2}\left(i \Delta_{h_1}\left(p_1+p_2\right) {D}_1 \cos \alpha-i \Delta_{h_2}\left(p_1+p_2\right) {D}_2 \sin \alpha\right) g^{\mu \nu} \epsilon_\mu^{(r)}\left(k_1\right) \epsilon_\mu^{(s)}\left(k_2\right), \\
i \mathcal{M}{(\chi \chi \rightarrow W W)} & =\frac{m_W^2 \varphi^2}{v^2}\left(i \Delta_{h_1}\left(p_1+p_2\right) {D}_1 \cos \alpha-i \Delta_{h_2}\left(p_1+p_2\right) {D}_2 \sin \alpha\right) g^{\mu \nu} \epsilon_\mu^{(r)}\left(k_1\right) \epsilon_\mu^{(s)}\left(k_2\right).
\end{align}
${D}_1$ and ${D}_2$ correspond to the vertices of $\chi\text{-}\chi\text{-}h_1$ and $\chi\text{-}\chi\text{-}h_2$ defined as
\begin{align}
{D}_1 &\equiv \delta_2 \cos\alpha+\frac{d_2 \varphi_S}{\varphi} \sin\alpha,\label{DD1}\\
{D}_2 &\equiv-\delta_2 \sin\alpha+\frac{d_2 \varphi_S}{\varphi} \cos\alpha.\label{DD2}
\end{align}
For later use, we also define
\begin{align}
{L}_1 &\equiv \lambda \cos\alpha+\frac{\delta_2 \varphi_S}{\varphi} \sin\alpha,\\
{L}_2 &\equiv-\lambda \sin\alpha+\frac{\delta_2 \varphi_S}{\varphi} \cos\alpha.
\end{align}
In addition, a propagator of this process can be expressed as
\begin{align}
\Delta_{h_i}(p)=\frac{1}{p^2-{m}_{h_i}^2+i {m}_{h_i} {\Gamma}_{h_i}},
\end{align}
with ${\Gamma}_{h_i}$ being a width of $h_i$. $g^{\mu\nu}$ is the metric tensor and $\epsilon_\mu(k_i)$ is the polarization vectors of the final-state vector bosons with momenta $k_i$.

We now proceed to compute the phase-space integral of the squared amplitude, which is denoted by the function $\mathcal{F}(s)$. We can obtain
\begin{align}
\mathcal{F}^{Z Z}(s)&=\frac{1}{2} \int \frac{d^3 \boldsymbol{k}_1}{(2 \pi)^3 2 E_1} \frac{d^3 \boldsymbol{k}_2}{(2 \pi)^3 2 E_2}(2 \pi)^4 \delta^4\left(p_1+p_2-k_1-k_2\right)|\mathcal{M}(\chi \chi \rightarrow Z Z)|^2\nonumber \\
&=\frac{s^2-4 s {m}_Z^2+12 {m}_Z^4}{8}\left|\Delta_{h_1}(\sqrt{s}) {D}_1 \cos \alpha-\Delta_{h_2}(\sqrt{s}) {D}_2 \sin \alpha\right|^2 \frac{1}{8 \pi} \sqrt{\frac{s-4 {m}_Z^2}{s}},\\
\mathcal{F}^{W W}(s)&=\frac{1}{2} \int \frac{d^3 \boldsymbol{k}_1}{(2 \pi)^3 2 E_1} \frac{d^3 \boldsymbol{k}_2}{(2 \pi)^3 2 E_2}(2 \pi)^4 \delta^4\left(p_1+p_2-k_1-k_2\right)|\mathcal{M}(\chi \chi \rightarrow W W)|^2\nonumber \\
&=\frac{s^2-4 s {m}_W^2+12 {m}_W^4}{4}\left|\Delta_{h_1}(\sqrt{s}) {D}_1 \cos \alpha-\Delta_{h_2}(\sqrt{s}) {D}_2 \sin \alpha\right|^2 \frac{1}{8 \pi} \sqrt{\frac{s-4 {m}_W^2}{s}}.
\end{align}

\subsection{$\chi\chi\to f\bar{f}$}
%----------------------------------------------------------------------------------------------------------------------------------
\begin{figure}[htpb]
\center
\includegraphics[width=6cm]{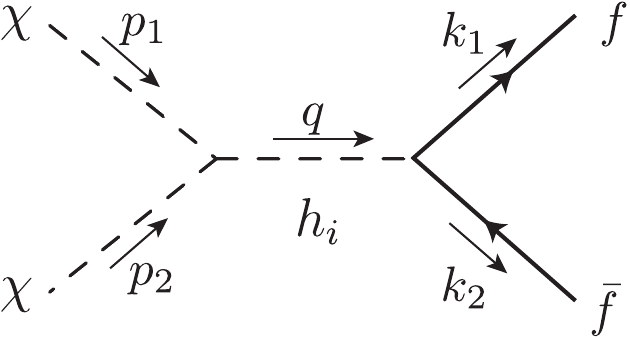}
\caption{Feynman diagram of the scattering process $\chi \chi \to f\bar{f} $. }
\label{fig:chichiff}
\end{figure}
%----------------------------------------------------------------------------------------------------------------------------------
The annihilation into a pair of fermion $f$ and antifermion $\bar{f}$ is an $s$-channel process mediated by $h_i$ as shown in Fig.~\ref{fig:chichiff}. The amplitude is given by
\begin{align}
i \mathcal{M}(\chi \chi \rightarrow f \bar{f})=-\frac{i}{2}\left(-i \frac{m_f}{v}\right) \varphi\left[i \Delta_{h_1}\left(p_1+p_2\right) {D}_1 \cos \alpha-i \Delta_{h_2}\left(p_1+p_2\right) {D}_2 \sin \alpha\right] \bar{u}^r\left(k_1\right) v^s\left(k_2\right),
\end{align}
where $\bar{u}(k_1)$, $v(k_2)$ are Dirac spinors for the outgoing fermion ($f$) and antifermion ($\bar{f}$) with momenta $k_1$ and $k_2$.

The phase-space integral of the squared amplitude becomes
\begin{align}
\mathcal{F}^{f \bar{f}}(s) & =N_C \int \frac{d^3 \boldsymbol{k}_1}{(2 \pi)^3 2 E_1} \frac{d^3 \boldsymbol{k}_2}{(2 \pi)^3 2 E_2}(2 \pi)^4 \delta\left(p_1+p_2-k_1-k_2\right)|\mathcal{M}(\chi \chi \rightarrow f \bar{f})|^2 \nonumber \\
& =\frac{N_C m_f^2 \varphi^2}{16 \pi v^2} \frac{\left(s-{m}_f^2\right)^{\frac{3}{2}}}{\sqrt{s}}\left|\Delta_{h_1}(\sqrt{s}) {D}_1 \cos \alpha-\Delta_{h_2}(\sqrt{s}) {D}_2 \sin \alpha\right|^2,
\end{align}
with $N_C=3$ for the quarks and $N_C=1$ for the leptons.

\subsection{$\chi\chi\to h_ih_j$}
%----------------------------------------------------------------------------------------------------------------------------------
\begin{figure}[htpb]
\centering
\includegraphics[width=17cm]{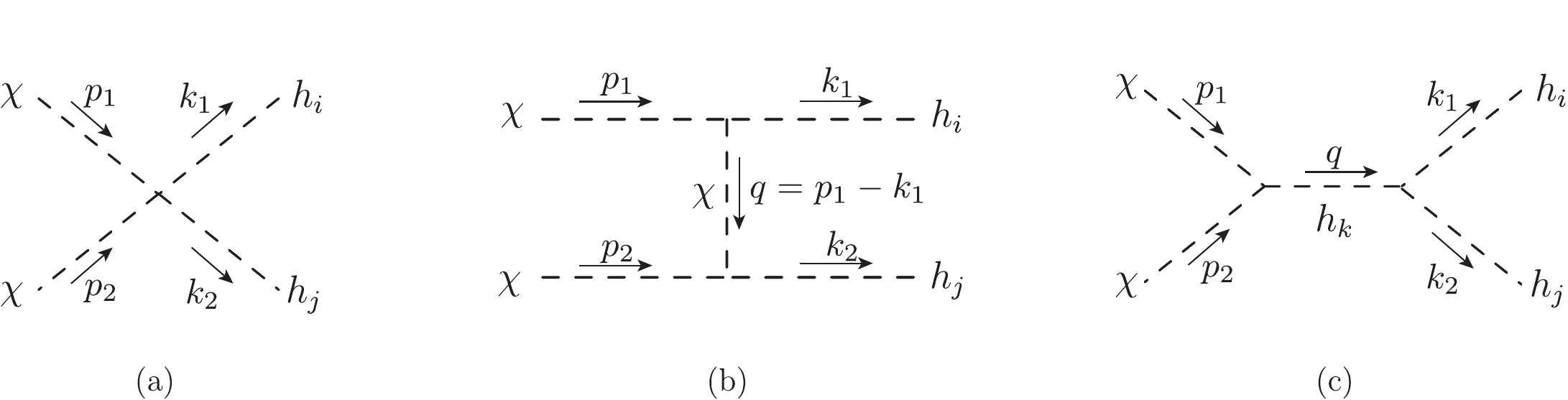}
\caption{Feynman diagram of the scattering process $\chi \chi \to h_ih_j $. }
\label{fig:chichihh}
\end{figure}
%----------------------------------------------------------------------------------------------------------------------------------
The annihilation into a pair of Higgs bosons $h_i$ is shown in Fig.~\ref{fig:chichihh}. The diagrams labeled (a), (b), and (c) correspond to the contact interaction, the $t$- and $u$-channel processes, and the $s$-channel process, respectively. Then, the amplitude is given by
\begin{align}
i \mathcal{M}\left(\chi \chi \rightarrow h_i h_j\right)=i \mathcal{M}_{ i j}^{(\mathrm{a})}+i \mathcal{M}_{ i j}^{(\mathrm{b})}+i \mathcal{M}_{ i j}^{(\mathrm{c})},
\end{align}
where
\begin{align}
& i \mathcal{M}_{11}^{(\mathrm{a})}=-i \frac{\delta_2 c_\alpha^2+d_2 s_\alpha^2}{2}, \\
& i \mathcal{M}_{11}^{(\mathrm{b})}=\left(-i \frac{{D}_1 \varphi}{2}\right)^2\left(i \Delta_\chi\left(p_1-k_1\right)+i \Delta_\chi\left(p_1-k_2\right)\right), \\
& i \mathcal{M}_{11}^{(\mathrm{c})}=-\frac{3}{4} {D}_1 \lambda_{111} \varphi^2 \cdot i \Delta_{h_1}\left(p_1+p_2\right)-\frac{1}{4} {D}_2 \lambda_{112} \varphi^2 \cdot i \Delta_{h_2}\left(p_1+p_2\right), \\
& i \mathcal{M}_{22}^{(\mathrm{a})}=-i \frac{\delta_2 s_\alpha^2+d_2 c_\alpha^2}{2}, \\
& i \mathcal{M}_{22}^{(\mathrm{b})}=\left(-i \frac{{D}_2 \varphi}{2}\right)^2\left(i \Delta_\chi\left(p_1-k_1\right)+i \Delta_\chi\left(p_1-k_2\right)\right), \\
& i \mathcal{M}_{22}^{(\mathrm{c})}=-\frac{1}{4} {D}_1 \lambda_{122} \varphi^2 \cdot i \Delta_{h_1}\left(p_1+p_2\right)-\frac{3}{4} {D}_2 \lambda_{222} \varphi^2 \cdot i \Delta_{h_2}\left(p_1+p_2\right), \\
& i \mathcal{M}_{12}^{(\mathrm{a})}=i \frac{\delta_2-d_2}{2} s_\alpha c_\alpha, \\
& i \mathcal{M}_{12}^{(\mathrm{b})}=\left(-i \frac{{D}_1 \varphi}{2}\right)\left(-i \frac{{D}_2 \varphi}{2}\right)\left(i \Delta_\chi\left(p_1-k_1\right)+i \Delta_\chi\left(p_1-k_2\right)\right), \\
& i \mathcal{M}_{12}^{(\mathrm{c})}=-\frac{1}{4} {D}_1 \lambda_{112} \varphi^2 \cdot i \Delta_{h_1}\left(p_1+p_2\right)-\frac{1}{4} {D}_2 \lambda_{122} \varphi^2 \cdot i \Delta_{h_2}\left(p_1+p_2\right) .
\end{align}
$\lambda_{111}$, $\lambda_{112}$, $\lambda_{122}$ and $\lambda_{222}$ correspond to the vertices of $h_1\text{-}h_1\text{-}h_1$, $h_1\text{-}h_1\text{-}h_2$, $h_1\text{-}h_2\text{-}h_2$ and $h_2\text{-}h_2\text{-}h_2$ defined as
\begin{align}
\lambda_{111} &\equiv {L}_1 \cos^2\alpha+{D}_1 \sin^2\alpha,\\
\lambda_{112} &\equiv {L}_2 \cos^2\alpha+{D}_2 \sin^2\alpha-2\left({L}_1-{D}_1\right) \cos\alpha \sin\alpha, \\
\lambda_{122} &\equiv {L}_1 \cos^2\alpha+{D}_1 \sin^2\alpha-2\left({L}_2-{D}_2\right) \cos\alpha \sin\alpha,\\
\lambda_{222} &\equiv {L}_2 \cos^2\alpha+{D}_2 \sin^2\alpha.
\end{align}

We compute the function $\mathcal{F}(s)$ separately for each final state: $h_1 h_1$, $h_2 h_2$, and $h_1 h_2$.
\underbar{$\chi\chi\to h_1 h_1$}
\begin{align}
\mathcal{F}^{h_1 h_1}(s)=&\frac{1}{2} \int \frac{d^3 k_1}{(2 \pi)^3 2 E_1} \frac{d^3 k_2}{(2 \pi)^3 2 E_2}(2 \pi)^4 \delta\left(p_1+p_2-k_1-k_2\right)\left|\mathcal{M}\left(\chi \chi \rightarrow h_1 h_1\right)\right|^2 \nonumber \\
= & \frac{1}{16 \pi} \sqrt{\frac{s-4 {m}_{h_1}^2}{s}}\left\{\frac{\left(\delta_2 \cos^2\alpha + d_2 \sin^2\alpha\right)^2}{4}+\left(\delta_2 \cos^2\alpha + d_2 \sin^2\alpha\right)\right. \nonumber \\
&\left[\frac{3}{4} \frac{{D}_1 \lambda_{111} \varphi^2\left(s-{m}_{h_1}^2\right)}{\left(s-\bar{m}_{h_1}^2\right)^2+{m}_{h_1}^2 {\Gamma}_{h_1}^2}+\frac{1}{4} \frac{{D}_2 \lambda_{112} \varphi^2\left(s-{m}_{h_2}^2\right)}{\left(s-{m}_{h_2}^2\right)^2+{m}_{h_2}^2 {\Gamma}_{h_2}^2}\right]  +\frac{{D}_1^4 \varphi^4}{8} \frac{1}{s {m}_\chi^2+{m}_{h_1}^2\left({m}_{h_1}^2-4 {m}_\chi^2\right)} \nonumber \\
& +\frac{{D}_1^2 \varphi^2}{4 p k}\left[\frac{{D}_1^2 \varphi^2}{4\left(s-2 {m}_{h_1}^2\right)}-\frac{\delta_2 \cos^2\alpha + d_2 \sin^2\alpha}{2}-\frac{3}{4} \frac{{D}_1 \lambda_{111} \varphi^2\left(s-{m}_{h_1}^2\right)}{\left(s-{m}_{h_1}^2\right)^2+{m}_{h_1}^2 {\Gamma}_{h_1}^2}\right. \nonumber \\
&\left.-\frac{1}{4} \frac{{D}_2 \lambda_{112} \varphi^2\left(s-{m}_{h_2}^2\right)}{\left(s-{m}_{h_2}^2\right)^2+{m}_{h_2}^2 {\Gamma}_{h_2}^2}\right] \log \left|\frac{s-2 {m}_{h_1}^2+4 p k}{s-2 {m}_{h_1}^2-4 p k}\right| \nonumber \\
& +\frac{9}{16} \frac{{D}_1^2 \lambda_{111}^2 \varphi^4}{\left(s-{m}_{h_1}^2\right)^2+{m}_{h_1}^2 {\Gamma}_{h_1}^2}+\frac{1}{16} \frac{{D}_2^2 \lambda_{112}^2 \varphi^4}{\left(s-{m}_{h_2}^2\right)^2+{m}_{h_2}^2 {\Gamma}_{h_2}^2} \nonumber \\
& \left.+\frac{3}{8} \frac{{D}_1 {D}_2 \lambda_{111} \lambda_{112} \varphi^4\left[\left(s-{m}_{h_1}^2\right)\left(s-{m}_{h_2}^2\right)+{m}_{h_1} {m}_{h_2} {\Gamma}_{h_1} {\Gamma}_{h_2}\right]}{\left[\left(s-{m}_{h_1}^2\right)\left(s-{m}_{h_2}^2\right)+{m}_{h_1} {m}_{h_2} {\Gamma}_{h_1} {\Gamma}_{h_2}\right]^2+\left[\left(s-{m}_{h_1}^2\right) {m}_{h_2} {\Gamma}_{h_2}-\left(s-{m}_{h_2}^2\right) {m}_{h_1} {\Gamma}_{h_1}\right]^2}\right\},
\end{align}
with $p=\sqrt{s / 4-{m}_\chi^2}$, $k=\sqrt{s / 4-{m}_{h_1}^2}$ and $D_{ \pm}=p^2+k^2 \pm 2 p k t+{m}_\chi^2=s / 2-{m}_{h_1}^2 \pm 2 p k t$ .
\\
\noindent\underbar{$\chi\chi\to h_2 h_2$}
\begin{align}
\mathcal{F}^{h_2 h_2}(s)= &\frac{1}{2} \int \frac{d^3 \boldsymbol{k}_1}{(2 \pi)^3 2 E_1} \frac{d^3 \boldsymbol{k}_2}{(2 \pi)^3 2 E_2}(2 \pi)^4 \delta\left(p_1+p_2-k_1-k_2\right)\left|\mathcal{M}\left(\chi \chi \rightarrow h_2 h_2\right)\right|^2 \nonumber \\
= & \frac{1}{16 \pi} \sqrt{\frac{s-4 {m}_{h_2}^2}{s}}\left\{\frac{\left(\delta_2 \sin^2\alpha+d_2 \cos^2\alpha\right)^2}{4}+\left(\delta_2 \sin^2\alpha+d_2 \cos^2\alpha\right)\right. \nonumber \\
&\left.\left[\frac{3}{4} \frac{{D}_2 \lambda_{222} \varphi^2\left(s-{m}_{h_2}^2\right)}{\left(s-{m}_{h_2}^2\right)^2+{m}_{h_2}^2 {\Gamma}_{h_2}^2}+\frac{1}{4} \frac{{D}_1 \lambda_{122} \varphi^2\left(s-{m}_{h_1}^2\right)}{\left(s-{m}_{h_1}^2\right)^2+{m}_{h_1}^2 {\Gamma}_{h_1}^2}\right]\right. +\frac{{D}_2^4 \varphi^4}{8} \frac{1}{s {m}_\chi^2+{m}_{h_2}^2\left({m}_{h_2}^2-4 {m}_\chi^2\right)} \nonumber \\
& +\frac{{D}_2^2 \varphi^2}{4 p k}\left[\frac{{D}_2^2 \varphi^2}{4\left(s-2 {m}_{h_2}^2\right)}-\frac{\delta_2 \sin^2\alpha+d_2 \cos^2\alpha}{2}-\frac{3}{4} \frac{{D}_2 \lambda_{222} \varphi^2\left(s-{m}_{h_2}^2\right)}{\left(s-{m}_{h_2}^2\right)^2+{m}_{h_2}^2 {\Gamma}_{h_2}^2}\right.\nonumber \\
&\left. -\frac{1}{4} \frac{{D}_1 \lambda_{122} \varphi^2\left(s-{m}_{h_1}^2\right)}{\left(s-{m}_{h_1}^2\right)^2+{m}_{h_1}^2 {\Gamma}_{h_1}^2}\right] \quad \times \log \left|\frac{s-2 {m}_{h_2}^2+4 p k}{s-2 {m}_{h_2}^2-4 p k}\right| \nonumber \\
& +\frac{9}{16} \frac{{D}_2^2 \lambda_{222}^2 \varphi^4}{\left(s-{m}_{h_2}^2\right)^2+{m}_{h_2}^2 {\Gamma}_{h_2}^2}+\frac{1}{16} \frac{{D}_1^2 \lambda_{122}^2 \varphi^4}{\left(s-{m}_{h_1}^2\right)^2+{m}_{h_1}^2 {\Gamma}_{h_1}^2} \nonumber \\
& \left.+\frac{3}{8} \frac{{D}_1 {D}_2 \lambda_{122} \lambda_{222} \varphi^4\left[\left(s-{m}_{h_1}^2\right)\left(s-{m}_{h_2}^2\right)+{m}_{h_1} {m}_{h_2} {\Gamma}_{h_1} {\Gamma}_{h_2}\right]}{\left[\left(s-{m}_{h_1}^2\right)\left(s-{m}_{h_2}^2\right)+{m}_{h_1} {m}_{h_2} {\Gamma}_{h_1} {\Gamma}_{h_2}\right]^2+\left[\left(s-{m}_{h_1}^2\right) {m}_{h_2} {\Gamma}_{h_2}-\left(s-{m}_{h_2}^2\right) {m}_{h_1} {\Gamma}_{h_1}\right]^2}\right\},
\end{align}
with $p=\sqrt{s / 4-{m}_\chi^2}$ and $k=\sqrt{s / 4-{m}_{h_2}^2}$.
\\
\noindent\underbar{$\chi\chi\to h_1 h_2$}
\begin{align}
\mathcal{F}_b^{h_1 h_2}(s)=&\int \frac{d^3 \boldsymbol{k}_1}{(2 \pi)^3 2 E_1} \frac{d^3 \boldsymbol{k}_2}{(2 \pi)^3 2 E_2}(2 \pi)^4 \delta\left(p_1+p_2-k_1-k_2\right)\left|\mathcal{M}\left(\chi \chi \rightarrow h_1 h_2\right)\right|^2 \nonumber \\
=& \frac{{k}(s)}{8 \pi \sqrt{s}}\left\{\frac{\left(d_2-\delta_2\right)^2 \cos^2\alpha \sin^2\alpha}{2}+\frac{\left(d_2-\delta_2\right) \cos\alpha \sin\alpha}{2} \right. \nonumber \\
&\left.\left[\frac{{D}_1 \lambda_{112} v^2\left(s-{m}_{h_1}^2\right)}{\left(s-{m}_{h_1}^2\right)^2+{m}_{h_1}^2 {\Gamma}_{h_1}^2}+\frac{{D}_2 \lambda_{122} \varphi^2\left(s-{m}_{h_2}^2\right)}{\left(s-{m}_{h_2}^2\right)^2+{m}_{h_2}^2 {\Gamma}_{h_2}^2}\right]\right. \nonumber \\
& +\frac{{D}_1 {D}_2 \varphi^2}{4}\left[\left(d_2-\delta_2\right) \cos\alpha \sin\alpha+\frac{1}{2} \frac{{D}_1 \lambda_{112} \varphi^2\left(s-{m}_{h_1}^2\right)}{\left(s-{m}_{h_1}^2\right)^2+{m}_{h_1}^2 {\Gamma}_{h_1}^2}+\frac{1}{2} \frac{{D}_2 \lambda_{122} \varphi^2\left(s-{m}_{h_2}^2\right)}{\left(s-{m}_{h_2}^2\right)^2+{m}_{h_2}^2 {\Gamma}_{h_2}^2}\right] \nonumber \\
& \quad \times \frac{1}{4 p k} \log \left|\frac{s-{m}_{h_1}^2-{m}_{h_2}^2+2 p k}{s-{m}_{h_1}^2-{m}_{h_2}^2-2 p k}\right| \nonumber \\
& +\frac{{D}_1^2 {D}_2^2 v^4}{32}\left[\frac{4}{\left(s / 2-\left({m}_{h_1}^2+{m}_{h_2}^2\right) / 2\right)^2-4 p^2 k^2}+\frac{1}{p k\left(s / 2-\left({m}_{h_1}^2+{m}_{h_2}^2\right) / 2\right)} \right. \nonumber \\
&\left. \log \left|\frac{s-{m}_{h_1}^2-{m}_{h_2}^2+2 p k}{s-{m}_{h_1}^2-{m}_{h_2}^2-2 p k}\right|\right] +\frac{1}{8} \frac{{D}_1^2 \lambda_{112}^2 \varphi^4}{\left(s-{m}_{h_1}^2\right)^2+{m}_{h_1}^2 {\Gamma}_{h_1}^2}+\frac{1}{8} \frac{{D}_{2122}^2 v^4}{\left(s-{m}_{h_2}^2\right)^2+{m}_{h_2}^2 {\Gamma}_{h_2}^2}\nonumber \\
&\left.+\frac{1}{4} \frac{{D}_1 {D}_2 \lambda_{112} \lambda_{122} v^4\left[\left(s-{m}_{h_1}^2\right)\left(s-{m}_{h_2}^2\right)+{m}_{h_1} {m}_{h_2} {\Gamma}_{h_1} {\Gamma}_{h_2}\right]}{\left[\left(s-{m}_{h_1}^2\right)\left(s-{m}_{h_2}^2\right)+{m}_{h_1} {m}_{h_2} {\Gamma}_{h_1} {\Gamma}_{h_2}\right]^2+\left[\left(s-{m}_{h_1}^2\right) {m}_{h_2} {\Gamma}_{h_2}-\left(s-{m}_{h_2}^2\right) {m}_{h_1} {\Gamma}_{h_1}\right]^2}\right\},
\end{align}
with $p=\sqrt{s / 4-{m}_\chi^2}$, $k={k}(s)=\sqrt{{\{\left(s-{m}_{h_1}^2-{m}_{h_2}^2\right)^2-4 {m}_{h_1}^2 {m}_{h_2}^2}\}/{4 s}}$ and $D_{ \pm}=p^2+k^2 \pm 2 p k t+{m}_\chi^2=\left(s-{m}_{h_1}^2-{m}_{h_2}^2\right) / 2 \pm 2 p k t$ .

%%%%%%

\section{One-loop corrections to the vertices including $\chi$-
particle} \label{app:chivertex}

The vertex functions are evaluated at the one-loop level, with the inclusion of the tree-level contributions that contain the finalized counterterms. These are employed in the one-loop analysis of the DM-quark scattering and the DM annihilation processes. For consistency with the finite counterterms, which are determined without the $h_1$-loop, we omit the contributions involving the $h_1$ propagator.
\\
\noindent\underbar{$\chi \chi h_i$-vertex}\\
%----------------------------------------------------------------------------------------------------------------------------------
\begin{figure}[htpb]
\centering
\includegraphics[width=17cm]{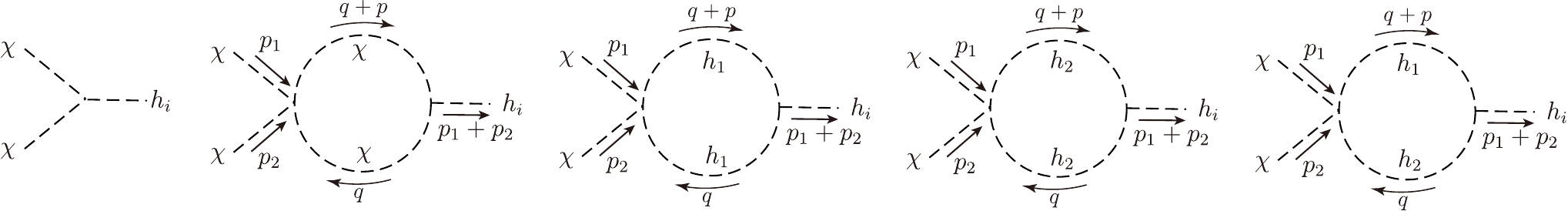}
\caption{The $\chi \chi h_i$-vertex at the one-loop level.}
\label{fig:chichihi}
\end{figure}
%----------------------------------------------------------------------------------------------------------------------------------
The $\chi \chi h_i$-vertex of Fig.~\ref{fig:chichihi} is expressed as
\begin{align}
& i \Gamma_{\chi \chi h_1}^\mathrm{basic}\left(p_1, p_2\right) \nonumber \\
& \stackrel{\overline{\mathrm{MS}}}{=}-i  \frac{v}{2}\left[D_1^{(1)}-\frac{3 d_2}{64 \pi^2} D_1 \varphi_2\left(p^2 ; m_\chi, m_\chi\right)-\frac{3}{64 \pi^2}\left(\delta_2 c_\alpha^2+d_2 s_\alpha^2\right) \lambda_{111} \varphi_2\left(p^2 ; m_{h_1}, m_{h_1}\right)\right. \nonumber \\
& \left.-\frac{1}{64 \pi^2}\left(\delta_2 s_\alpha^2+d_2 c_\alpha^2\right) \lambda_{122} \varphi_2\left(p^2 ; m_{h_2}, m_{h_2}\right)+\frac{\delta_2-d_2}{32 \pi^2} \lambda_{112} c_\alpha s_\alpha \varphi_2\left(p^2 ; m_{h_1}, m_{h_2}\right)\right],
\end{align}

\begin{align}
& i \Gamma_{\chi \chi h_2}^\mathrm{basic}\left(p_1, p_2\right) \nonumber \\
&\stackrel{\overline{\mathrm{MS}}}{=}-i \frac{v}{2}\left[D_2^{(1)}-\frac{3 d_2}{64 \pi^2} D_2 \varphi_2\left(p^2 ; m_\chi, m_\chi\right)-\frac{1}{64 \pi^2}\left(\delta_2 c_\alpha^2+d_2 s_\alpha^2\right) \lambda_{112} \varphi_2\left(p^2 ; m_{h_1}, m_{h_1}\right)\right. \nonumber \\
& \left.-\frac{3}{64 \pi^2}\left(\delta_2 s_\alpha^2+d_2 c_\alpha^2\right) \lambda_{222} \varphi_2\left(p^2 ; m_{h_2}, m_{h_2}\right)+\frac{\delta_2-d_2}{32 \pi^2} \lambda_{122} c_\alpha s_\alpha \varphi_2\left(p^2 ; m_{h_1}, m_{h_2}\right)\right],
\end{align}
where $p=p_1+p_2$ and $D_i^{(1)}(i=1,2)$ should be understood as including the finite counterterms. The loop function with two propagators $\varphi_2$ is defined below.

%----------------------------------------------------------------------------------------------------------------------------------
\begin{figure}[htpb]
\centering
\includegraphics[width=17cm]{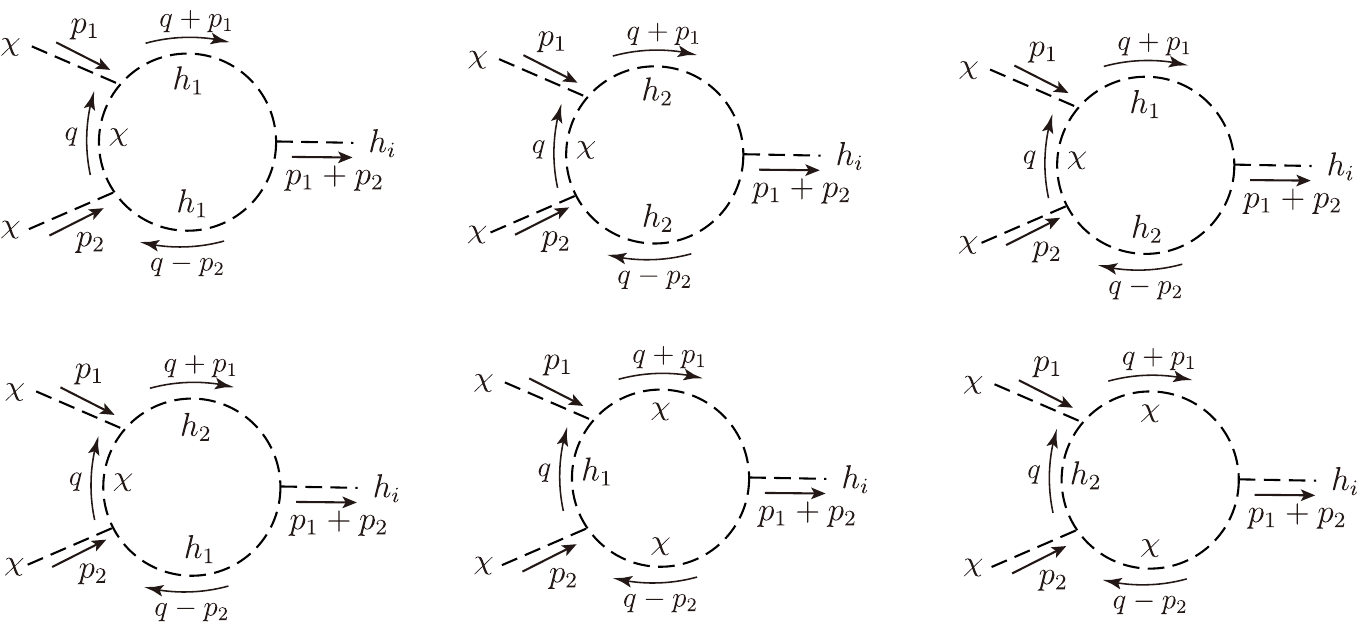}
\caption{Additional one-loop corrections to the $\chi \chi h_i$-vertex, suppressed by higher powers of the coupling constants and the VEV.}
\label{fig:chichihi2}
\end{figure}
%----------------------------------------------------------------------------------------------------------------------------------
Similarly, the $\chi \chi h_i$-vertex of Fig.~\ref{fig:chichihi2} is expressed as
\begin{align}
& i \Gamma_{\chi \chi h_1}^\mathrm{higher}\left(p_1, p_2\right) \nonumber \\
& =-i v^3 {\left[\frac{3 D_1^2 \lambda_{111}}{128 \pi^2} \varphi_3\left(p_1, p_2 ; m_{h_1}, m_{h_1}, m_\chi\right)+\frac{D_2^2 \lambda_{122}}{128 \pi^2} \varphi_3\left(p_1, p_2 ; m_{h_2}, m_{h_2}, m_\chi\right)\right.} \\
& +\frac{D_1 D_2 \lambda_{112}}{128 \pi^2}\left(\varphi_3\left(p_1, p_2 ; m_{h_1}, m_{h_2}, m_\chi\right)+\varphi_3\left(p_1, p_2 ; m_{h_2}, m_{h_1}, m_\chi\right)\right) \\
& \left.+\frac{D_1^3}{128 \pi^2} \varphi_3\left(p_1, p_2 ; m_\chi, m_\chi, m_{h_1}\right)+\frac{D_1 D_2^2}{128 \pi^2} \varphi_3\left(p_1, p_2 ; m_\chi, m_\chi, m_{h_2}\right)\right],
\end{align}
\begin{align}
& i \Gamma_{\chi \chi h_2}^\mathrm{higher}\left(p_1, p_2\right) \nonumber \\
& =-i v^3\left[\frac{D_1^2 \lambda_{112}}{128 \pi^2} \varphi_3\left(p_1, p_2 ; m_{h_1}, m_{h_1}, m_\chi\right)+\frac{3 D_2^2 \lambda_{222}}{128 \pi^2} \varphi_3\left(p_1, p_2 ; m_{h_2}, m_{h_2}, m_\chi\right)\right.\nonumber \\
& +\frac{D_1 D_2 \lambda_{122}}{128 \pi^2}\left(\varphi_3\left(p_1, p_2 ; m_{h_1}, m_{h_2}, m_\chi\right)+\varphi_3\left(p_1, p_2 ; m_{h_2}, m_{h_1}, m_\chi\right)\right) \\
& \left.\left.+\frac{D_1^2 D_2}{128 \pi^2} \varphi_3\left(p_1, p_2 ; m_\chi, m_\chi, m_{h_1}\right)+\frac{D_2^3}{128 \pi^2} \varphi_3\left(p_1, p_2 ; m_\chi, m_\chi, m_{h_2}\right)\right)\right].
\end{align}
The loop function with three propagators $\varphi_3$ is defined below.

\noindent\underbar{$\chi \chi h_ih_j$-vertex}\\
%----------------------------------------------------------------------------------------------------------------------------------
\begin{figure}[htpb]
\centering
\includegraphics[width=17cm]{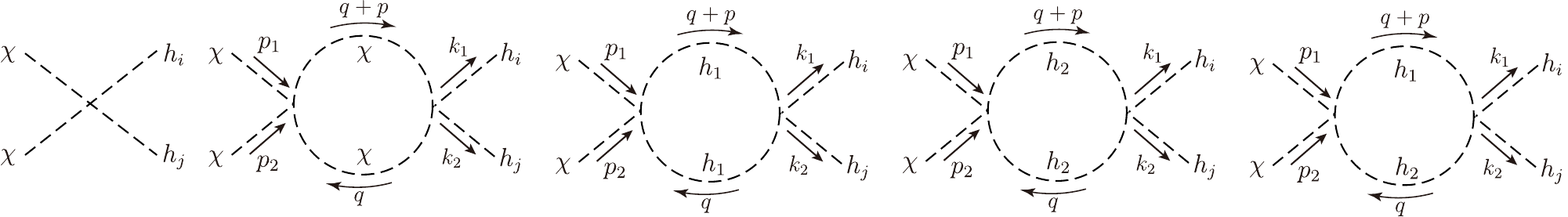}
\caption{The $\chi \chi h_i h_j$-vertex at the one-loop level.}
\label{fig:chichihihj}
\end{figure}
%----------------------------------------------------------------------------------------------------------------------------------
The $\chi \chi h_i h_j$-vertex of Fig.~\ref{fig:chichihihj} is expressed as
\begin{align}
& i \Gamma_{\chi \chi h_1 h_1}^\mathrm{basic}\left(p_1, p_2\right) \nonumber \\
&\stackrel{\overline{\mathrm{MS}}}{=}-i\left\{\frac{\left(\delta_2+\delta_2^{(1)}\right) c_\alpha^2+\left(d_2+d_2^{(1)}\right) s_\alpha^2}{2}-\frac{3 d_2\left(\delta_2 c_\alpha^2+d_2 s_\alpha^2\right)}{128 \pi^2} \varphi_2\left(p^2 ; m_\chi, m_\chi\right)\right. \nonumber\\
&-\frac{3\left(\delta_2 c_\alpha^2+d_2 s_\alpha^2\right)}{128 \pi^2}\left(\lambda s_\alpha^4+d_2 c_\alpha^4+2 \delta_2 s_\alpha^2 c_\alpha^2\right) \varphi_2\left(p^2 ; m_{h_1}, m_{h_1}\right) \nonumber\\
& -\frac{\delta_2 s_\alpha^2+d_2 c_\alpha^2}{128 \pi^2}\left[3\left(\lambda+d_2\right) c_\alpha^2 s_\alpha^2+\delta_2\left(1-6 c_a^2 s_a^2\right)\right] \varphi_2\left(p^2 ; m_{h_2}, m_{h_2}\right) \nonumber\\
& \left.+\frac{3\left(\delta_2-d_2\right) c_\alpha^2 s_\alpha^2}{64 \pi^2}\left[\left(\delta_2-\lambda\right) c_\alpha^2+\left(d_2-\delta_2\right) s_\alpha^2\right] \varphi_2\left(p^2 ; m_{h_1}, m_{h_2}\right)\right\},
\end{align}

\begin{align}
& i \Gamma_{\chi \chi h_2 h_2}^\mathrm{basic}\left(p_1, p_2\right) \nonumber \\
&\stackrel{\overline{\mathrm{MS}}}{=}-i\{ \frac{\left(\delta_2+\delta_2^{(1)}\right) s_\alpha^2+\left(d_2+d_2^{(1)}\right) c_\alpha^2}{2}-\frac{3 d_2\left(\delta_2 s_\alpha^2+d_2 c_\alpha^2\right)}{128 \pi^2} \varphi_2\left(p^2 ; m_\chi, m_\chi\right)\nonumber \\
& -\frac{\delta_2 c_\alpha^2+d_2 s_\alpha^2}{128 \pi^2}\left[3\left(\lambda+d_2\right) c_\alpha^2 s_\alpha^2+\delta_2\left(1-6 c_a^2 s_a^2\right)\right] \varphi_2\left(p^2 ; m_{h_1}, m_{h_1}\right)\nonumber \\
& -\frac{3\left(\delta_2 s_\alpha^2+d_2 c_\alpha^2\right)}{128 \pi^2}\left(\lambda s_\alpha^4+d_2 c_\alpha^4+2 \delta_2 s_\alpha^2 c_\alpha^2\right) \varphi_2\left(p^2 ; m_{h_2}, m_{h_2}\right)\nonumber \\
& \left.+\frac{3\left(\delta_2-d_2\right) c_\alpha^2 s_\alpha^2}{64 \pi^2}\left[\left(\delta_2-\lambda\right) s_\alpha^2+\left(d_2-\delta_2\right) c_\alpha^2\right] \varphi_2\left(p^2 ; m_{h_1}, m_{h_2}\right)\right\},
\end{align}

\begin{align}
& i \Gamma_{\chi \chi h_1 h_2}^\mathrm{basic}\left(p_1, p_2\right) \nonumber \\
&\stackrel{\overline{\mathrm{MS}}}{=} i c_\alpha s_\alpha\{ \frac{\delta_2+\delta_2^{(1)}-d_2-d_2^{(1)}}{2}-\frac{3 d_2\left(\delta_2-d_2\right)}{128 \pi^2} \varphi_2\left(p^2 ; m_\chi, m_\chi\right)\nonumber \\
& +\frac{3\left(\delta_2 c_\alpha^2+d_2 s_\alpha^2\right)}{128 \pi^2}\left[\left(\delta_2-\lambda\right) c_\alpha^2+\left(d_2-\delta_2\right) s_\alpha^2\right] \varphi_2\left(p^2 ; m_{h_1}, m_{h_1}\right)\nonumber \\
& +\frac{3\left(\delta_2 s_\alpha^2+d_2 c_\alpha^2\right)}{128 \pi^2}\left[\left(\delta_2-\lambda\right) s_\alpha^2+\left(d_2-\delta_2\right) c_\alpha^2\right] \varphi_2\left(p^2 ; m_{h_2}, m_{h_2}\right)\nonumber \\
& \left.-\frac{\delta_2-d_2}{64 \pi^2}\left[3\left(\lambda+d_2\right) c_\alpha^2 s_\alpha^2+\delta_2\left(1-6 c_a^2 s_a^2\right)\right] \varphi_2\left(p^2 ; m_{h_1}, m_{h_2}\right)\right\}.
\end{align}

%----------------------------------------------------------------------------------------------------------------------------------
\begin{figure}[htpb]
\centering
\includegraphics[width=17cm]{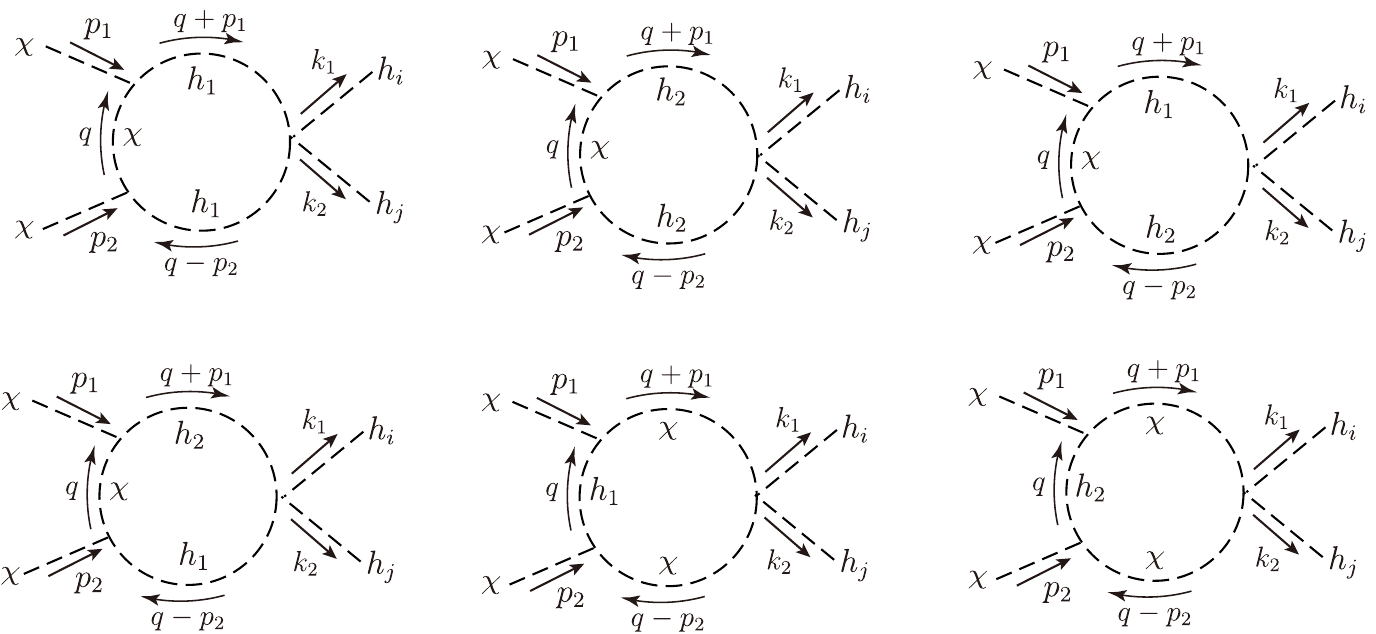}
\caption{Additional one-loop corrections to the $\chi \chi h_i h_j$-vertex, suppressed by higher powers of the coupling constants and the VEV.}
\label{fig:chichihihj2}
\end{figure}
%----------------------------------------------------------------------------------------------------------------------------------
The $\chi \chi h_i h_j$-vertex of Fig.~\ref{fig:chichihihj2} is expressed as
\begin{align}
& i \Gamma_{\chi \chi h_1 h_1}^\mathrm{higher}\left(p_1, p_2\right) \nonumber \\
&=-i v^2\left\{\frac{3 D_1^2}{128 \pi^2}\left(\lambda c_\alpha^4+d_2 s_\alpha^4+2 \delta_2 c_\alpha^2 s_\alpha^2\right) \varphi_3\left(p_1, p_2 ; m_{h_1}, m_{h_1}, m_\chi\right)\right.\nonumber \\
& +\frac{D_2^2}{128 \pi^2}\left[3\left(\lambda+d_2\right) c_\alpha^2 s_\alpha^2+\delta_2\left(1-6 c_\alpha^2 s_\alpha^2\right)\right] \varphi_3\left(p_1, p_2 ; m_{h_2}, m_{h_2}, m_\chi\right)\nonumber \\
& +\frac{3 D_1 D_2 c_\alpha s_\alpha}{128 \pi^2}\left[\left(\delta_2-\lambda\right) c_\alpha^2+\left(d_2-\delta_2\right) s_\alpha^2\right]\left(\varphi_3\left(p_1, p_2 ; m_{h_1}, m_{h_2}, m_\chi\right)+\varphi_3\left(p_1, p_2 ; m_{h_2}, m_{h_1}, m_\chi\right)\right)\nonumber \\
& \left.+\frac{D_1^2\left(\delta_2 c_\alpha^2+d_2 s_\alpha^2\right)}{128 \pi^2} \varphi_3\left(p_1, p_2 ; m_\chi, m_\chi, m_{h_1}\right)+\frac{D_2^2\left(\delta_2 c_\alpha^2+d_2 s_\alpha^2\right)}{128 \pi^2} \varphi_3\left(p_1, p_2 ; m_\chi, m_\chi, m_{h_2}\right)\right\},
\end{align}

\begin{align}
& i \Gamma_{\chi \chi h_2 h_2}^\mathrm{higher}\left(p_1, p_2\right) \nonumber \\
&=-i v^2\left\{\frac{D_1^2}{128 \pi^2}\left[3\left(\lambda+d_2\right) c_\alpha^2 s_\alpha^2+\delta_2\left(1-6 c_\alpha^2 s_\alpha^2\right)\right] \varphi_3\left(p_1, p_2 ; m_{h_1}, m_{h_1}, m_\chi\right)\right.\nonumber \\
& +\frac{3 D_2^2}{128 \pi^2}\left(\lambda s_\alpha^4+d_2 c_\alpha^4+2 \delta_2 c_\alpha^2 s_\alpha^2\right) \varphi_3\left(p_1, p_2 ; m_{h_2}, m_{h_2}, m_\chi\right)\nonumber  \\
& +\frac{3 D_1 D_2 c_\alpha s_\alpha}{128 \pi^2}\left[\left(\delta_2-\lambda\right) s_\alpha^2+\left(d_2-\delta_2\right) c_\alpha^2\right]\left(\varphi_3\left(p_1, p_2 ; m_{h_1}, m_{h_2}, m_\chi\right)+\varphi_3\left(p_1, p_2 ; m_{h_2}, m_{h_1}, m_\chi\right)\right) \nonumber \\
&\left.+\frac{D_1^2\left(\delta_2 s_\alpha^2+d_2 c_\alpha^2\right)}{128 \pi^2} \varphi_3\left(p_1, p_2 ; m_x, m_x, m_{h_1}\right)+\frac{D_2^2\left(\delta_2 s_\alpha^2+d_2 c_\alpha^2\right)}{128 \pi^2} \varphi_3\left(p_1, p_2 ; m_x, m_x, m_{h_2}\right)\right\},
\end{align}

\begin{align}
& i \Gamma_{\chi \chi h_1 h_2}^\mathrm{higher}\left(p_1, p_2\right) \nonumber \\
&=-i v^2\left\{\frac{3 D_1^2 c_\alpha s_\alpha}{128 \pi^2}\left[\left(\delta_2-\lambda\right) c_\alpha^2+\left(d_2-\delta_2\right) s_\alpha^2\right] \varphi_3\left(p_1, p_2 ; m_{h_1}, m_{h_1}, m_\chi\right)\right. \nonumber \\
& +\frac{3 D_2^2 c_\alpha s_\alpha}{128 \pi^2}\left[\left(\delta_2-\lambda\right) s_\alpha^2+\left(d_2-\delta_2\right) c_\alpha^2\right] \varphi_3\left(p_1, p_2 ; m_{h_2}, m_{h_2}, m_\chi\right) \nonumber \\
& +\frac{D_1 D_2}{128 \pi^2}\left[3\left(\lambda+d_2\right) c_\alpha^2 s_\alpha^2+\delta_2\left(1-6 c_\alpha^2 s_\alpha^2\right)\right]\left(\varphi_3\left(p_1, p_2 ; m_{h_1}, m_{h_2}, m_\chi\right)+\varphi_3\left(p_1, p_2 ; m_{h_2}, m_{h_1}, m_\chi\right)\right) \nonumber \\
& \left.-\frac{D_1^2\left(\delta_2-d_2\right) c_\alpha s_\alpha}{128 \pi^2} \varphi_3\left(p_1, p_2 ; m_\chi, m_\chi, m_{h_1}\right)-\frac{D_2^2\left(\delta_2-d_2\right) c_\alpha s_\alpha}{128 \pi^2} \varphi_3\left(p_1, p_2 ; m_\chi, m_\chi, m_{h_2}\right)\right\}.
\end{align}

The loop functions are defined as follows. First, the loop function with two propagators is defined as
\begin{align}
\int_q \Delta_1(q+p) \Delta_2(q) & =\mu^\epsilon \int \frac{d^d q}{(2 \pi)^d} \frac{1}{(q+p)^2-m_1^2} \frac{1}{q^2-m_2^2}=\mu^\epsilon \int \frac{d^d q}{(2 \pi)^d} \int_0^1 \frac{d x}{\left(q^2+2 x p q-M^2\right)^2} \nonumber \\
& =\frac{i}{16 \pi^2} \Gamma\left(\frac{\epsilon}{2}\right) \int_0^1 d x\left(\frac{M^2+x^2 p^2}{4 \pi \mu^2}\right)^{-\frac{\epsilon}{2}}\nonumber \\
&=\frac{i}{16 \pi^2}\left[\frac{2}{\epsilon}-\int_0^1 d x \log \frac{M^2+x^2 p^2}{4 \pi e^{-\gamma_E} \mu^2}+O(\epsilon)\right]\nonumber \\
& =\frac{i}{16 \pi^2}\left[\frac{2}{\epsilon}-\varphi_2\left(p^2 ; m_1, m_2\right)\right],
\end{align}
where $M^2(x) \equiv x m_1^2+(1-x) m_2^2-x p^2$.
\begin{align}
\varphi_2\left(p^2 ; m_1, m_2\right) &\equiv \int_0^1 d x \log \frac{x m_1^2+(1-x) m_2^2-\left(x-x^2\right) p^2}{\bar{\mu}^2}\nonumber \\
& =\left\{\begin{array}{lc}
\frac{1}{m_1^2-m_2^2}\left(m_1^2 \log \frac{m_1^2}{\bar{\mu}^2}-m_1^2 \log \frac{m_1^2}{\bar{\mu}^2}\right)-1 & \left(p=0, m_1 \neq m_2\right) \\
\log \frac{m_1^2}{\bar{\mu}^2} & \left(p=0, m_1=m_2\right).
\end{array}\right.
\end{align}
Here, $\varphi_2\left(p^2 ; m_1, m_2\right)=\varphi_2\left(p^2 ; m_2, m_1\right)$.
The loop function with three propagators is defined as
\begin{align}
\int_q & \Delta_1\left(q+p_1\right) \Delta_2\left(q-p_2\right) \Delta(q)=\int \frac{d^4 q}{(2 \pi)^4} \frac{1}{\left(q+p_1\right)^2-m_1^2} \frac{1}{\left(q-p_2\right)-m_2^2} \frac{1}{q^2-m_3^2} \\
& =\int \frac{d^4 q}{(2 \pi)^4} 2 \int_0^1 d x \int_0^{1-x} d y \frac{1}{\left[q^2+2\left(x p_1-y p_2\right) q-(1-x-y) m_3^2-x m_1^2-y m_2^2\right]^3} \\
& =-\frac{i}{16 \pi^2} \int_0^1 d x \int_0^{1-x} d y \frac{1}{M^2(x, y)+\left(x p_1-y p_2\right)^2} \\
& \equiv-\frac{i}{16 \pi^2} \varphi_3\left(p_1, p_2 ; m_1, m_2, m_3\right),
\end{align}
where $M^2(x, y) \equiv(1-x-y) m_3^2+x m_1^2+y m_2^2$. In the case of vanishing momenta of the incoming particles, it reduces to
\begin{align}
&\varphi_3\left(0 ; m_1, m_2, m_3\right)\nonumber \\
&\equiv \int_0^1 d x \int_0^{1-x} d y \frac{1}{(1-x-y) m_3^2+x m_1^2+y m_2^2}=\varphi_3\left(0 ; m_2, m_1, m_3\right)\nonumber \\
&=\left\{\begin{array}{ll}
\frac{m_1^2 \log \left(m_1^2 / \nu^2\right)}{\left(m_1^2-m_2^2\right)\left(m_1^2-m_3^2\right)}+\frac{m_2^2 \log \left(m_2^2 / \nu^2\right)}{\left(m_2^2-m_1^2\right)\left(m_2^2-m_3^2\right)}+\frac{m_3^2 \log \left(m_3^2 / \nu^2\right)}{\left(m_3^2-m_1^2\right)\left(m_3^2-m_2^2\right)} & \left(m_1 \neq m_2 \neq m_3 \neq m_1\right. \\
\frac{m_1^2}{\left(m_1^2-m_2^2\right)^2} \log \frac{m_1^2}{m_2^2}-\frac{1}{m_1^2-m_2^2} & \left(m_1 \neq m_2=m_3\right) \\
-\frac{m_3^2}{\left(m_1^2-m_3^2\right)^2} \log \frac{m_1^2}{m_3^2}+\frac{1}{m_1^2-m_3^2} & \left(m_1=m_2 \neq m_3\right) \\
\frac{1}{2 m_1^2} & \left(m_1=m_2=m_3\right).
\end{array}\right.
\end{align}
$\nu$ is an arbitrary mass scale.

For this model, ignoring the loops containing the $h_1$-propagator, the finite counterterms are given by
\begin{align}
\delta_2^{(1)}=& -\frac{2}{v v_{S}}\frac{\partial^2 V_{\mathrm{CW}}}{\partial \varphi \partial \varphi_S} \nonumber \\
=& -\frac{d_2 \delta_2}{64 \pi^2} \log \frac{m_\chi^2}{\bar{\mu}^2}\nonumber \\
& -\frac{2}{v v_{S}}\left[\left\langle\frac{\partial^2 \bar{m}_{h_2}^2}{\partial \varphi \partial \varphi_S}\right\rangle \frac{m_{h_2}^2}{32 \pi^2}\left(\log \frac{m_{h_2}^2}{\bar{\mu}^2}-1\right)+\left\langle\frac{\partial \bar{m}_{h_2}^2}{\partial \varphi}\right\rangle\left\langle\frac{\partial \bar{m}_{h_2}^2}{\partial \varphi_S}\right\rangle \frac{1}{32 \pi^2} \log \frac{m_{h_2}^2}{\bar{\mu}^2}\right], \\
d_2^{(1)}=&-\frac{2}{v_{S}^2}\left(\frac{\partial^2 V_{\mathrm{CW}}}{\partial \varphi_S^2}-\frac{1}{v_{S}}\frac{\partial V_{\mathrm{CW}}}{\partial \varphi_s}\right),\nonumber \\
= & -\frac{d_2^2}{64 \pi^2} \log \frac{m_\chi^2}{\bar{\mu}^2}\nonumber \\
& -\frac{2}{v_{S}^2}\left[\left(\left\langle\frac{\partial^2 \bar{m}_{h_2}^2}{\partial \varphi_S^2}\right\rangle-\frac{1}{v_{S}}\left\langle\frac{\partial \bar{m}_{h_2}^2}{\partial \varphi_S}\right\rangle\right) \frac{m_{h_2}^2}{32 \pi^2}\left(\log \frac{m_{h_2}^2}{\bar{\mu}^2}-1\right)+\left\langle\frac{\partial \bar{m}_{h_2}^2}{\partial \varphi_S}\right\rangle^2 \frac{1}{32 \pi^2} \log \frac{m_{h_2}^2}{\bar{\mu}^2}\right].
\end{align}
The counterterms of the effective couplings $D_1$ and $D_2$ are given by
\begin{align}
D_1^{(1)}&=D_1\left(1-\frac{d_2}{64 \pi^2} \log \frac{m_\chi^2}{\bar{\mu}^2}\right)\nonumber \\
&-\frac{2}{v v_{S}}\left[\left(\left\langle\frac{\partial^2 \bar{m}_{h_2}^2}{\partial \varphi \partial \varphi_S}\right\rangle c_\alpha+\left\langle\frac{\partial^2 \bar{m}_{h_2}^2}{\partial \varphi_S^2}\right\rangle s_\alpha-\frac{s_\alpha}{v_{S}}\left\langle\frac{\partial \bar{m}_{h_2}^2}{\partial \varphi_S}\right\rangle\right) \frac{m_{h_2}^2}{32 \pi^2}\left(\log \frac{m_{h_2}^2}{\bar{\mu}^2}-1\right)\right.\nonumber \\
&\left.+\left(\left\langle\frac{\partial \bar{m}_{h_2}^2}{\partial \varphi}\right\rangle c_\alpha+\left\langle\frac{\partial \bar{m}_{h_2}^2}{\partial \varphi_S}\right\rangle s_\alpha\right)\left\langle\frac{\partial \bar{m}_{h_2}^2}{\partial \varphi_S}\right\rangle \frac{1}{32 \pi^2} \log \frac{m_{h_2}^2}{\bar{\mu}^2}\right], \label{D1corr}\\
D_2^{(1)}&=D_2\left(1-\frac{d_2}{64 \pi^2} \log \frac{m_\chi^2}{\bar{\mu}^2}\right)\nonumber \\
&-\frac{2}{v v_{S}}\left[\left(-\left\langle\frac{\partial^2 \bar{m}_{h_2}^2}{\partial \varphi \partial \varphi_S}\right\rangle s_\alpha+\left\langle\frac{\partial^2 \bar{m}_{h_2}^2}{\partial \varphi_S^2}\right\rangle c_\alpha-\frac{c_\alpha}{v_{S}}\left\langle\frac{\partial \bar{m}_{h_2}^2}{\partial \varphi_S}\right\rangle\right) \frac{m_{h_2}^2}{32 \pi^2}\left(\log \frac{m_{h_2}^2}{\bar{\mu}^2}-1\right)\right.\nonumber \\
&\left.+\left(-\left\langle\frac{\partial \bar{m}_{h_2}^2}{\partial \varphi}\right\rangle s_\alpha+\left\langle\frac{\partial \bar{m}_{h_2}^2}{\partial \varphi_S}\right\rangle c_\alpha\right)\left\langle\frac{\partial \bar{m}_{h_2}^2}{\partial \varphi_S}\right\rangle \frac{1}{32 \pi^2} \log \frac{m_{h_2}^2}{\bar{\mu}^2}\right].\label{D2corr}
\end{align}

For both the DM direct detection and the relic abundance, the relevant amplitudes are dominated by contributions with vanishing incoming momenta. In addition, we omit the contribution from the diagrams containing the $h_1$-propagator here. Accordingly, the one-loop level vertex functions are given as follows. The three-point functions are
\begin{align}
&\Gamma_{\chi \chi h_1}(0)=\Gamma_{\chi \chi h_1}^\mathrm{basic}(0)+\Gamma_{\chi \chi h_1}^\mathrm{higher}(0)\nonumber \\
&=-\frac{v}{2}\left\{D_1+\delta_2^{(1)} c_\alpha^2+d_2^{(1)} s_\alpha^2-\frac{3 D_1 d_2}{64 \pi^2} \log \frac{m_\chi^2}{\bar{\mu}^2}-\frac{\left(\delta_2 s_\alpha^2+d_2 c_\alpha^2\right) \lambda_{122}}{64 \pi^2} \log \frac{m_{h_2}^2}{\bar{\mu}^2}\right.\nonumber \\
&\left.-\frac{D_2 v^2}{64 \pi^2}\left[\frac{D_2 \lambda_{122} m_\chi^2-D_1^2 m_{h_2}^2}{\left(m_{h_2}^2-m_\chi^2\right)^2} \log \frac{m_{h_2}^2}{m_\chi^2}-\frac{D_2 \lambda_{122}-D_1^2}{m_{h_2}^2-m_\chi^2}\right]\right\},
\label{Ghichi1_0}
\end{align}

\begin{align}
&\Gamma_{\chi \chi h_2}(0)=\Gamma_{\chi \chi h_2}^\mathrm{basic}(0)+\Gamma_{\chi \chi h_2}^\mathrm{higher}(0)\nonumber \\
&=-\frac{v}{2}\left\{D_2+\delta_2^{(1)} s_\alpha^2+d_2^{(1)} c_\alpha^2-\frac{3 D_2 d_2}{64 \pi^2} \log \frac{m_\chi^2}{\bar{\mu}^2}-\frac{3\left(\delta_2 s_\alpha^2+d_2 c_\alpha^2\right) \lambda_{222}}{64 \pi^2} \log \frac{m_{h_2}^2}{\bar{\mu}^2}\right.\nonumber \\
&\left.-\frac{D_1^2 v^2}{64 \pi^2}\left[\frac{3 \lambda_{222} m_\chi^2-D_2 m_{h_2}^2}{\left(m_{h_2}^2-m_\chi^2\right)^2} \log \frac{m_{h_2}^2}{m_\chi^2}-\frac{3 \lambda_{222}-D_2}{m_{h_2}^2-m_\chi^2}\right]\right\}.
\label{Ghichi2_0}
\end{align}

The 4-point vertices are
\begin{align}
&\Gamma_{\chi \chi h_1 h_1}(0)=\Gamma_{\chi \chi h_1 h_1}^\mathrm{basic}(0)+\Gamma_{\chi \chi h_1 h_1}^\mathrm{higher}(0)\nonumber \\
&= -\frac{\left(\delta_2+\delta_2^{(1)}\right) c_\alpha^2+\left(d_2+d_2^{(1)}\right) s_\alpha^2}{2}+\frac{3 d_2\left(\delta_2 c_\alpha^2+d_2 s_\alpha^2\right)}{128 \pi^2} \log \frac{m_\chi^2}{\bar{\mu}^2}\nonumber \\
& \quad+\frac{\delta_2 s_\alpha^2+d_2 c_\alpha^2}{128 \pi^2}\left[3\left(\lambda+d_2-2 \delta_2\right) c_\alpha^2 s_\alpha^2+\delta_2\right] \log \frac{m_{h_2}^2}{\bar{\mu}^2}\nonumber \\
& \quad+\frac{D_2^2 v^2}{128 \pi^2}\left[\frac{\left(3\left(\lambda+d_2-2 \delta_2\right) c_\alpha^2 s_\alpha^2+\delta_2\right) m_\chi^2-\left(\delta_2 c_\alpha^2+d_2 s_\alpha^2\right) m_{h_2}^2}{\left(m_{h_2}^2-m_\chi^2\right)^2} \log \frac{m_{h_2}^2}{m_\chi^2}\right.\nonumber \\
& \left.\quad-\frac{3\left(\lambda+d_2-2 \delta_2\right) c_\alpha^2 s_\alpha^2+\left(\delta_2-d_2\right) s_\alpha^2}{m_{h_2}^2-m_\chi^2}\right],
\label{Ghichi11_0}
\end{align}

\begin{align}
&\Gamma_{\chi \chi h_2 h_2}(0)=\Gamma_{\chi \chi h_2 h_2}^\mathrm{basic}(0)+\Gamma_{\chi \chi h_2 h_2}^\mathrm{higher}(0)\nonumber \\
&= -\frac{\left(\delta_2+\delta_2^{(1)}\right) s_\alpha^2+\left(d_2+d_2^{(1)}\right) c_\alpha^2}{2}+\frac{3 d_2\left(\delta_2 s_\alpha^2+d_2 c_\alpha^2\right)}{128 \pi^2} \log \frac{m_\chi^2}{\bar{\mu}^2}\nonumber \\
& \quad+\frac{3\left(\delta_2 s_\alpha^2+d_2 c_\alpha^2\right)}{128 \pi^2}\left(\lambda s_\alpha^4+d_2 c_\alpha^4+2 \delta_2 s_\alpha^2 c_\alpha^2\right) \log \frac{m_{h_2}^2}{\bar{\mu}^2}\nonumber \\
& \quad+\frac{D_1^2 v^2}{128 \pi^2}\left[\frac{3\left(\lambda s_\alpha^4+d_2 c_\alpha^4+2 \delta_2 c_\alpha^2 s_\alpha^2\right) m_\chi^2-\left(\delta_2 s_\alpha^2+d_2 c_\alpha^2\right) m_{h_2}^2}{\left(m_{h_2}^2-m_\chi^2\right)^2} \log \frac{m_{h_2}^2}{m_\chi^2}\right.\nonumber \\
& \left.\quad-\frac{3\left(\lambda s_\alpha^4+d_2 c_\alpha^4+2 \delta_2 c_\alpha^2 s_\alpha^2\right)-\left(\delta_2 s_\alpha^2+d_2 c_\alpha^2\right)}{m_{h_2}^2-m_\chi^2}\right],
\label{Ghichi22_0}
\end{align}

\begin{align}
&\Gamma_{\chi \chi h_1 h_2}(0)=\Gamma_{\chi \chi h_1 h_2}^\mathrm{basic}(0)+\Gamma_{\chi \chi h_1 h_2}^\mathrm{higher}(0)\nonumber \\
&=c_\alpha s_\alpha\left\{\frac{\delta_2+\delta_2^{(1)}-d_2-d_2^{(1)}}{2}-\frac{3 d_2\left(\delta_2-d_2\right)}{128 \pi^2} \log \frac{m_\chi^2}{\bar{\mu}^2}\right.\nonumber \\
&\left.+\frac{3\left(\delta_2 s_\alpha^2+d_2 c_\alpha^2\right)}{128 \pi^2}\left[\left(\delta_2-\lambda\right) s_\alpha^2+\left(d_2-\delta_2\right) c_\alpha^2\right] \log \frac{m_{h_2}^2}{\bar{\mu}^2}\right.\nonumber\\
&+\frac{D_2^2 v^2}{128 \pi^2}\left[\frac{3\left(\left(\delta_2-\lambda\right) s_\alpha^2+\left(d_2-\delta_2\right) c_\alpha^2\right) m_\chi^2-\left(\delta_2-d_2\right) m_{h_2}^2}{\left(m_\chi^2-m_{h_2}^2\right)^2} \log \frac{m_{h_2}^2}{m_\chi^2}\right.\nonumber \\
&\left.\left.-\frac{3\left(\left(\delta_2-\lambda\right) s_\alpha^2+\left(d_2-\delta_2\right) c_\alpha^2\right)-\left(\delta_2-d_2\right)}{m_{h_2}^2-m_\chi^2}\right]\right\}.
\label{Ghichi12_0}
\end{align}

%-------------------------------------------------------------%

\begin{figure}[h!]
  \centering
  \begin{minipage}{0.39\columnwidth}
    \centering
    
    \includegraphics[width=\columnwidth]{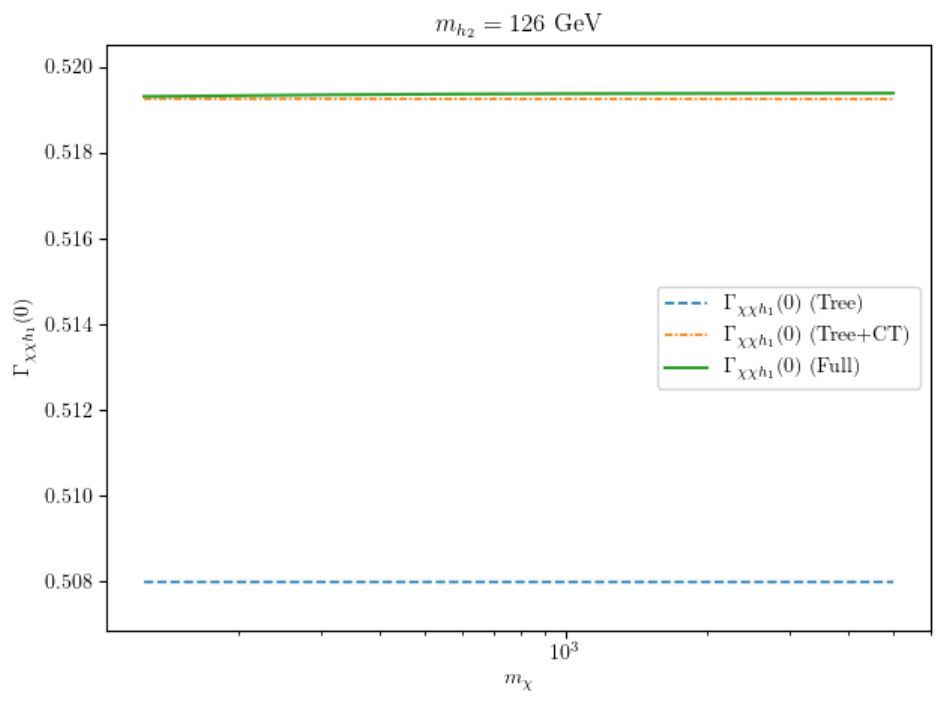}
  \end{minipage}
  \hspace{5mm}
  \begin{minipage}{0.39\columnwidth}
    \centering
    \includegraphics[width=\columnwidth]{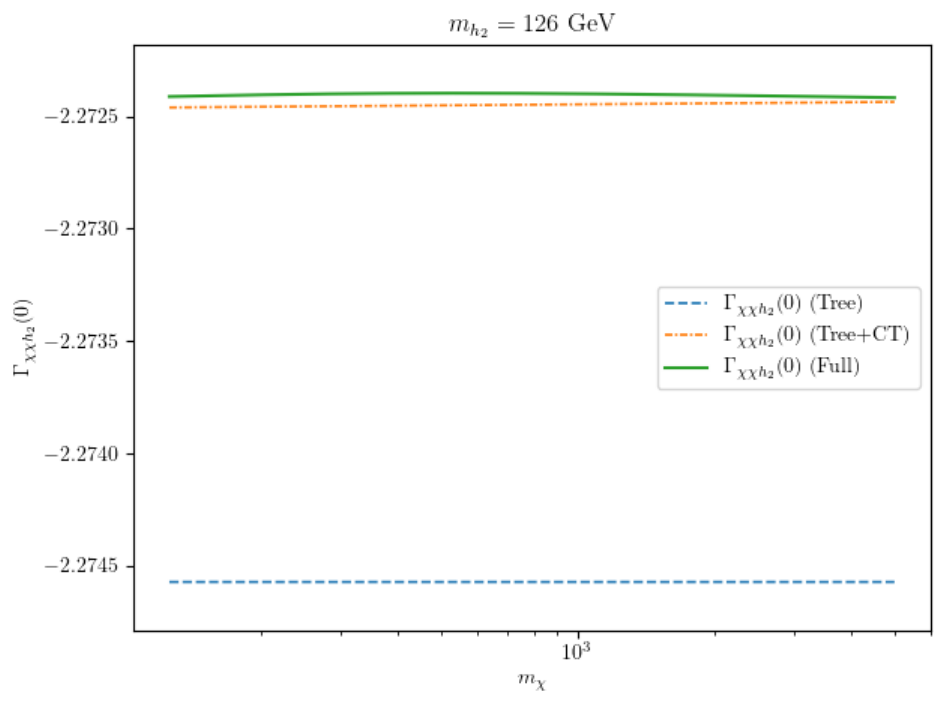}
  \end{minipage}
  
  \begin{minipage}{0.39\columnwidth}
    \centering
    \includegraphics[width=\columnwidth]{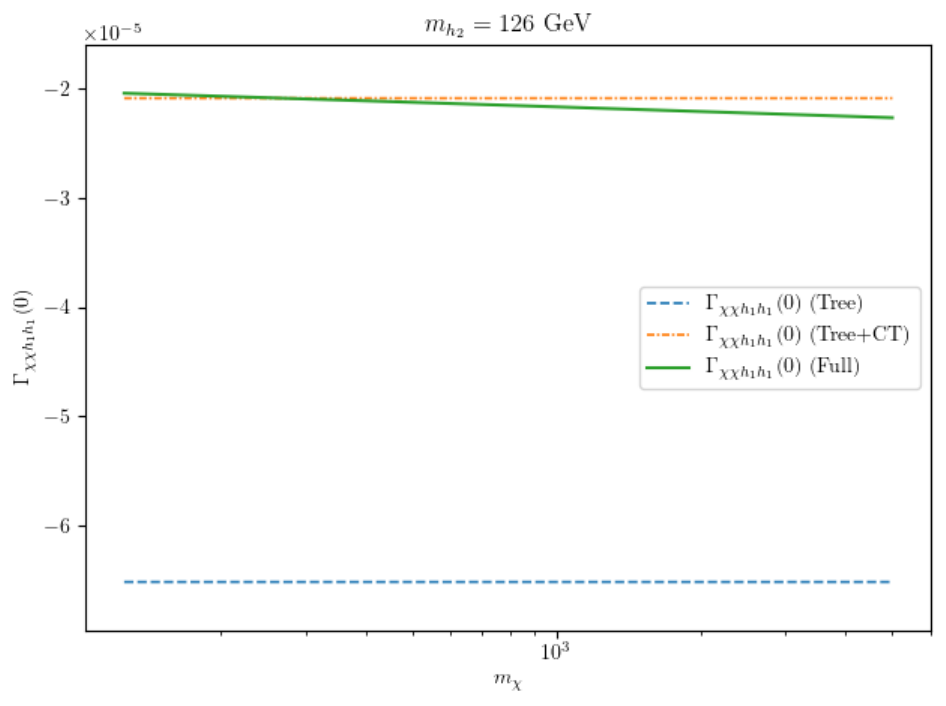}
  \end{minipage}
  \hspace{5mm}
  \begin{minipage}{0.39\columnwidth}
    \centering
    \includegraphics[width=\columnwidth]{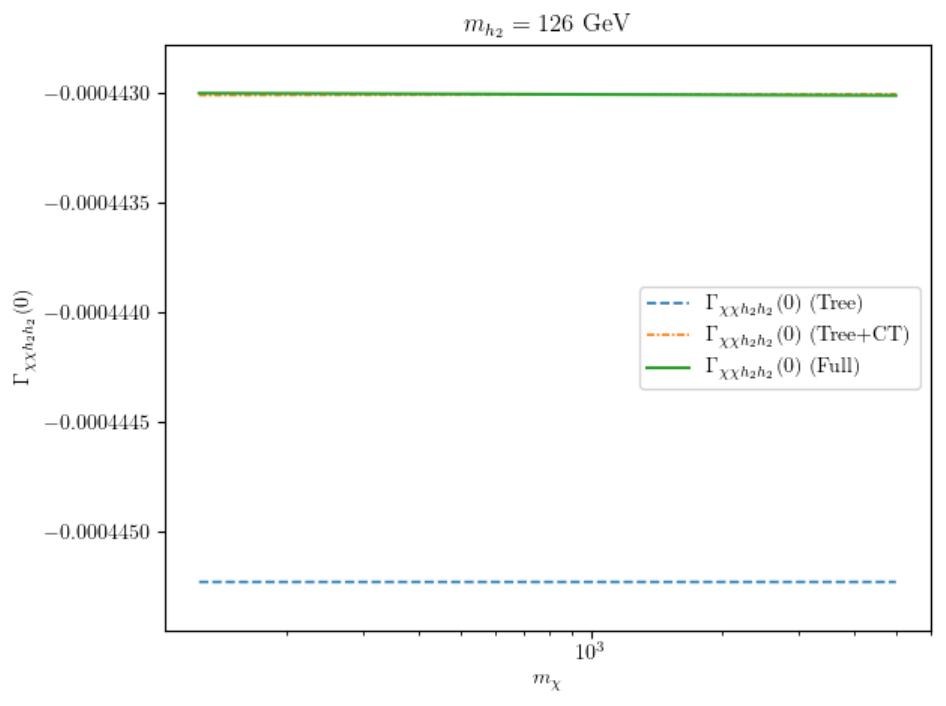}
  \end{minipage}
  
  \begin{minipage}{0.39\columnwidth}
    \centering
    \includegraphics[width=\columnwidth]{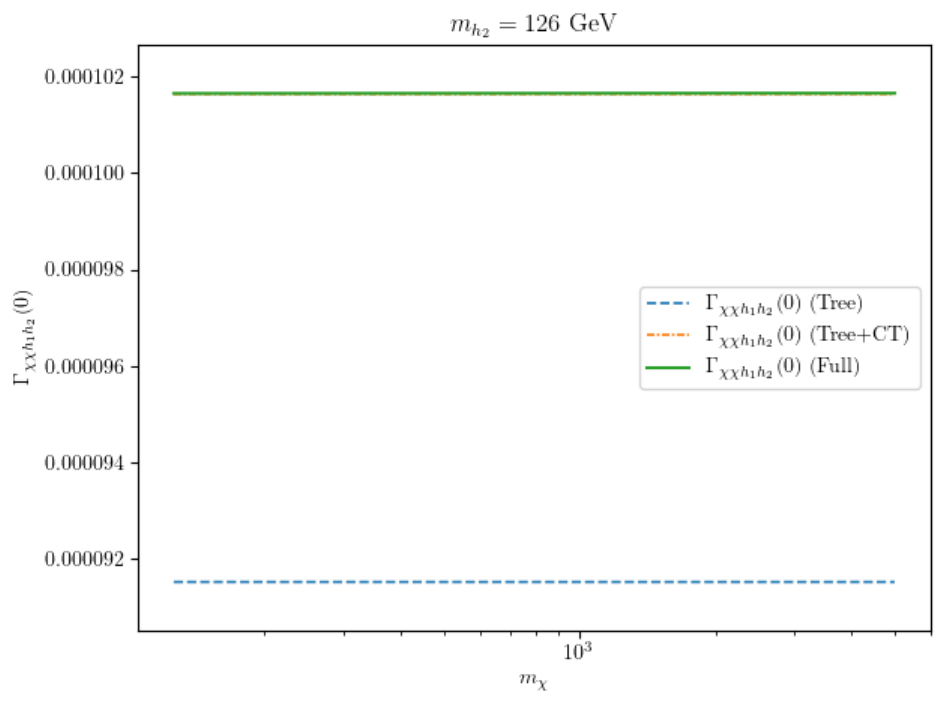}
  \end{minipage}
      \caption{Effective vertices associated with DM $\chi$ when $m_{h_2}=126$ GeV.}
    \label{fig:Gamma_126}
\end{figure}

%-------------------------------------------------------------%

%-------------------------------------------------------------%

\begin{figure}[h!]
  \centering
  \begin{minipage}{0.39\columnwidth}
    \centering
    
    \includegraphics[width=\columnwidth]{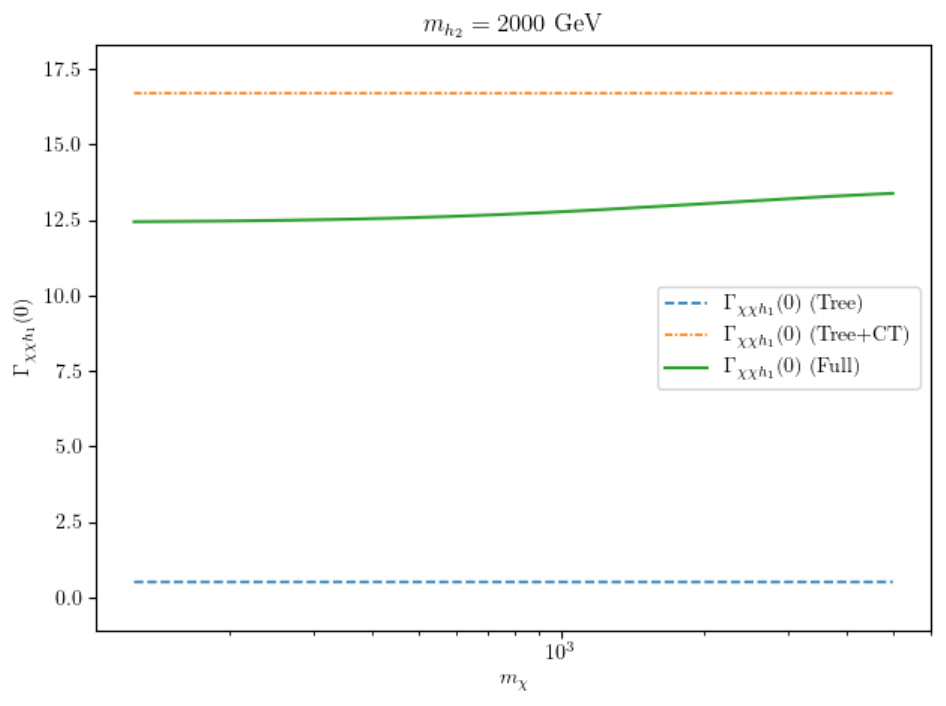}
  \end{minipage}
  \hspace{5mm}
  \begin{minipage}{0.39\columnwidth}
    \centering
    \includegraphics[width=\columnwidth]{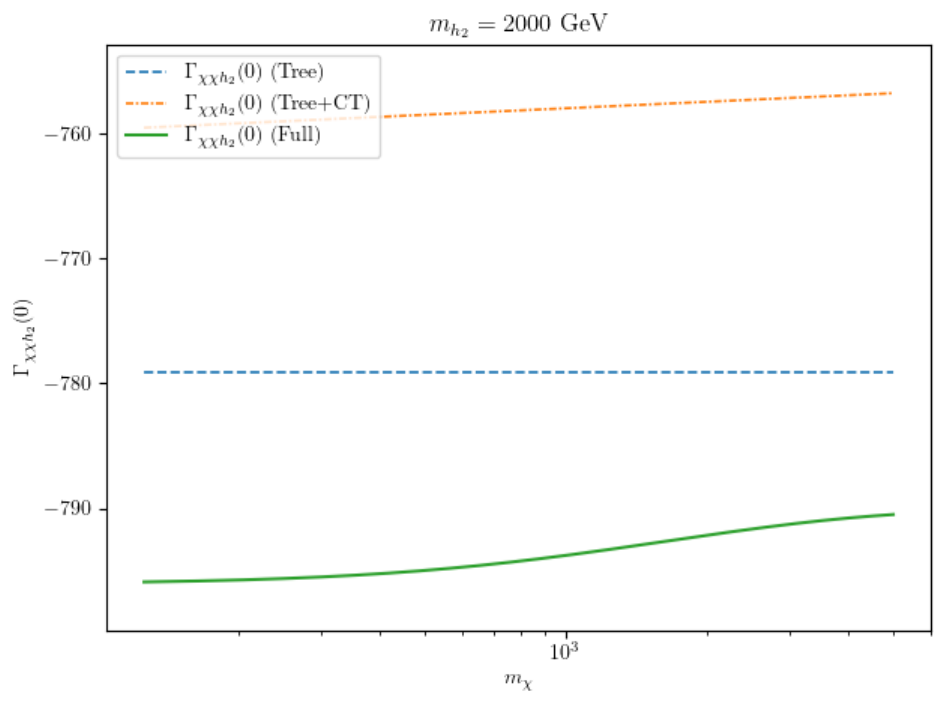}
  \end{minipage}
  
  \begin{minipage}{0.39\columnwidth}
    \centering
    \includegraphics[width=\columnwidth]{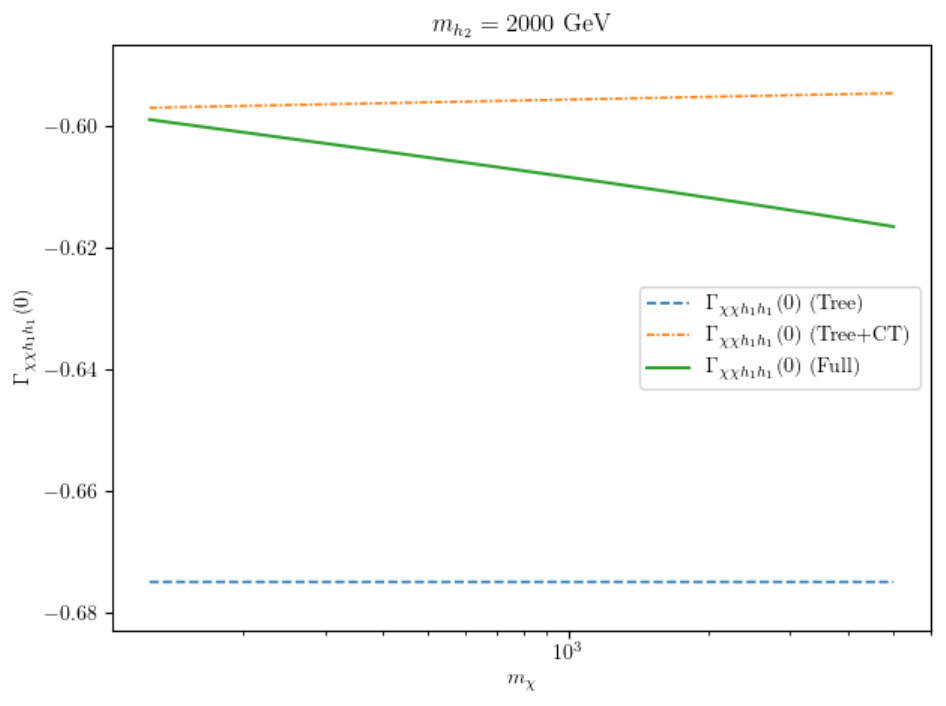}
  \end{minipage}
  \hspace{5mm}
  \begin{minipage}{0.39\columnwidth}
    \centering
    \includegraphics[width=\columnwidth]{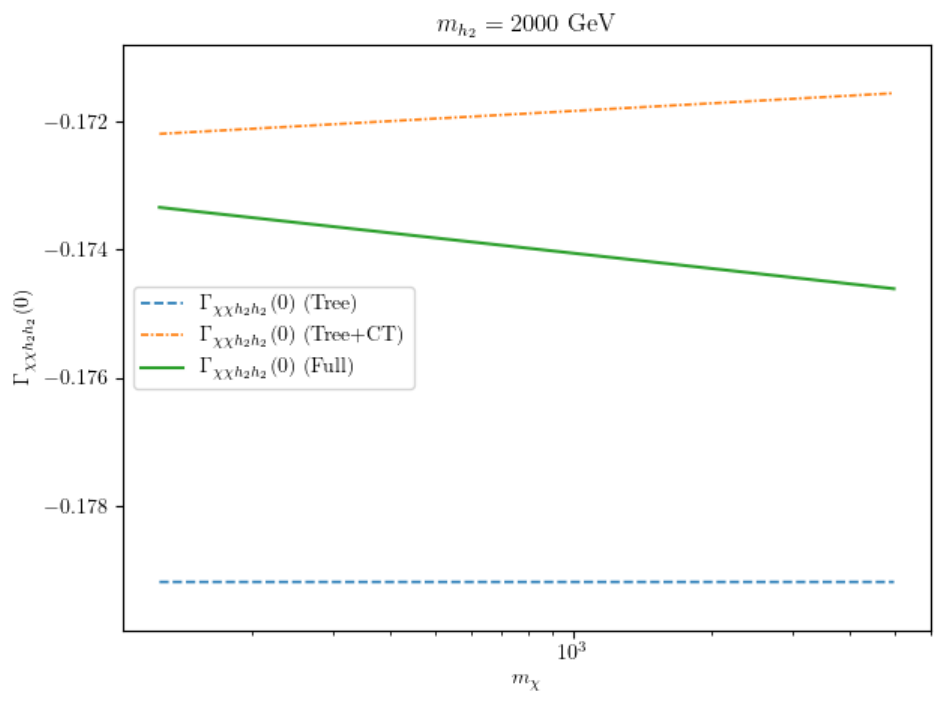}
  \end{minipage}
  
  \begin{minipage}{0.39\columnwidth}
    \centering
    \includegraphics[width=\columnwidth]{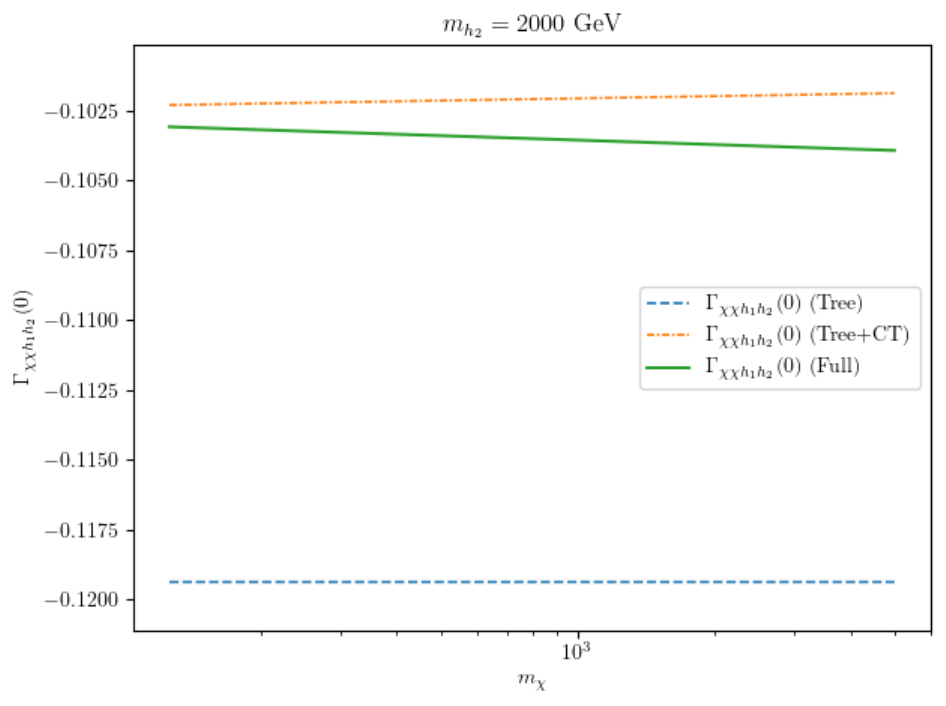}
  \end{minipage}
      \caption{Effective vertices associated with DM $\chi$ when $m_{h_2}=2000$ GeV.}
    \label{fig:Gamma_2000}
\end{figure}

%-------------------------------------------------------------%

As an example, the effective vertices \eqref{Ghichi1_0}-\eqref{Ghichi12_0} for $m_{h_2}$ = 126 GeV and $m_{h_2}$ = 2000 GeV are shown in Fig.~\ref{fig:Gamma_126} and \ref{fig:Gamma_2000}. One can see that while the one-loop corrections and the counterterms are present even for $m_{h_2}$ = 126 GeV, the couplings do not become so large as to invalidate perturbation theory even for $m_{h_2}$ = 2000 GeV.

%%%%%%%%%%%%%%%%%%%%%%%%%%%%%%%%%%%%%%%%%%%%%%%%
%					Reference
%%%%%%%%%%%%%%%%%%%%%%%%%%%%%%%%%%%%%%%%%%%%%%%%

\bibliography{biblist}

\end{document}